\documentclass[aps,pre,twocolumn,superscriptaddress,showpacs,floatfix,longbibliography]{revtex4-1}

\makeatletter
\def\@bibdataout@aps{%
\immediate\write\@bibdataout{%
@CONTROL{%
apsrev41Control%
\longbibliography@sw{%
    ,author="08",editor="1",pages="1",title="1",year="1"%
    }{%
    ,author="08",editor="1",pages="1",title="",year="1"%
    }%
  }%
}%
\if@filesw \immediate \write \@auxout {\string \citation {apsrev41Control}}\fi
}
\makeatother

\usepackage{graphicx}
\usepackage{dcolumn}
\usepackage{bm}

\usepackage{amssymb}
\usepackage{amsmath}
\usepackage{graphicx}
\usepackage{xcolor}

\usepackage[colorlinks,citecolor=blue,linkcolor=red,urlcolor=blue]{hyperref}

\begin{document}
\title{Universal Driven Critical Dynamics near the Boundary}
\author{Yu-Rong Shu}
\affiliation{School of Physics and Materials Science, Guangzhou University, Guangzhou 510006, China}

\author{Shuai Yin}
\email{yinsh6@mail.sysu.edu.cn}
\affiliation{School of Physics, Sun Yat-sen University, Guangzhou 510275, China}
\affiliation{Guangdong Provincial Key Laboratory of Magnetoelectric Physics and Devices, Sun Yat-sen University, Guangzhou 510275, China}

\date{\today}

\begin{abstract}
The celebrated Kibble-Zurek mechanism (KZM) describes the scaling of physical quantities when external parameters sweep through a critical point. Boundaries are ubiquitous in real systems, and critical behaviors near the boundary have attracted extensive research. Different boundary universality classes, including ordinary, special, extraordinary, and surface transitions, have been identified. However, the driven critical dynamics near boundaries remains unexplored. Here, we systematically investigate the driven critical dynamics in various boundary universality classes of the Ising model in both two and three dimensions, and discover a wealth of dynamic scaling behaviors. We find that for heating dynamics in all boundary universality classes, as well as for cooling dynamics in special, extraordinary, and surface transitions, the dynamic scaling behaviors of the order parameter can be described by a {\it normal} generalization of the KZM, called {\it boundary finite-time scaling} (BFTS). In contrast, for cooling dynamics in ordinary transition, we discover an {\it abnormal} logarithmic scaling on the driving rate. Moreover, for the special transition, in addition to temperature driving, we also consider the driven dynamics by driving the surface couplings. For increasing the surface coupling across the special transition point along the line of the ordinary transition, the prerequisite of the KZM, which requires that the correlation length/time in the initial state to be short-ranged, breaks down. We develop a generalized BFTS for a nonequilibrium initial state characterized by the waiting time, or the ``age", of the boundary. Possible generalizations are also discussed.
\end{abstract}

\maketitle

\section{Introduction}
The celebrated Kibble-Zurek mechanism (KZM) plays a crucial role in understanding nonequilibrium dynamics in phase transitions, with origins in both cosmology and condensed matter physics~\cite{Kibble1976,Zurek1985}. When a system is cooled at a finite rate $v$ across its critical point at $T_c$, starting from the symmetric disordered phase at $T>T_c$, the KZM divides the evolution into three stages according to the adiabatic-impulse scenario~\cite{Kibble1976,Zurek1985}. In the initial stage, the system evolves adiabatically, as the relaxation time is short and the dynamics can follow the external driving. As the system approaches the critical point, critical slowing down prevents adiabatic evolution, and the dynamics enter an impulse regime~\cite{Kibble1976,Zurek1985}. In the ordered side ($T<T_c$), uncorrelated local choices of ordered domains lead to the formation of topological defects~\cite{Kibble1976,Zurek1985}. Furthermore, the KZM asserts that the density of these topological defects follows a scaling relation determined by the cooling rate $v$~\cite{Kibble1976,Zurek1985}. Up to now, the KZM has aroused intensive investigations from both theoretical and experimental aspects, exerting far-reaching significance in both classical and quantum phase transitions~\cite{Kibble1976,Zurek1985,Zoller2005prl,Dziarmaga2005prl,PhysRevB.72.161201,Lin2014natphy,Ko2019,Keesling2019,PhysRevLett.129.227001,Ebadi2021,Du2023,sciadv.aba7292,science.abo6587,science.abq6753,PRXQuantum,Kang2023jap,Zeng2023prl,Weinberg2025arx}. 

As a generalization of the original KZM, a finite-time scaling (FTS) theory was proposed~\cite{Zhongf2005prb,Gong2010njp}. The FTS theory introduces a driving-induced time scale $\zeta_d\sim v^{-z/r}$, where $r=z+1/\nu$ with $z$ and $\nu$ being the dynamic and the correlation length critical exponents, respectively. It has been established that $\zeta_d$ effectively characterizes the critical dynamics in the vicinity of the critical point, providing a comprehensive framework to understand the universal dynamic scaling behaviors across the entire critical region~\cite{Zhongf2005prb,Zhong2006pre,Fan2007pre,Huangxz2010pre,Yin2014prb,Huang2014prb,feng2016prb,Fan2009pre}. The FTS theory reveals that driving-rate dependent scaling behaviors are not restricted to topological defects but also manifest in other macroscopic observables, such as the order parameter, correlation functions, and entanglement entropy~\cite{Zhongf2005prb,Zhong2006pre,Fan2007pre,Huangxz2010pre,Yin2014prb,Huang2014prb,feng2016prb,Fan2009pre,huqj2015prb,Cao2018prb}. Moreover, the FTS theory accommodates various driving protocols: from heating dynamics with ordered initial state~\cite{Zhongf2005prb,Zhong2006pre,Fan2007pre,Huangxz2010pre,Yin2014prb,Huang2014prb,feng2016prb,Fan2009pre,huqj2015prb,Cao2018prb} to driven dynamics under the symmetry-breaking field tuning~\cite{Gong2010njp}, and even temperature-driven processes near the quantum critical point~\cite{Yin2014prb}.
Recent works have further extended the FTS to phase transitions beyond the Landau paradigm~\cite{Shu2025nc,Huang2020prr,Zeng2025nc,Zeng2025prb}, critical dynamics with multiple length scales~\cite{Shu2025arx,Shu2024prb},
dynamical phase transitions~\cite{Liyh2019pre,Yinarxiv2024}, and integrated with relaxation critical dynamics, thereby extending the KZM beyond adiabatic initial conditions~\cite {Zhong2006pre,Huangyy2016prb,Yin2016prb}. In the work that comes later than Ref.~\cite{Zhongf2005prb}, full scaling forms analogous to those of FTS have been explored in alternative contexts~\cite{Deng2008,chandran2012prb,Grandi2011prb,Huse2012prl,Liu2014prb,Zurek2016prb}.

\begin{figure}[!htbp]
\centering
  \includegraphics[width=\linewidth,clip]{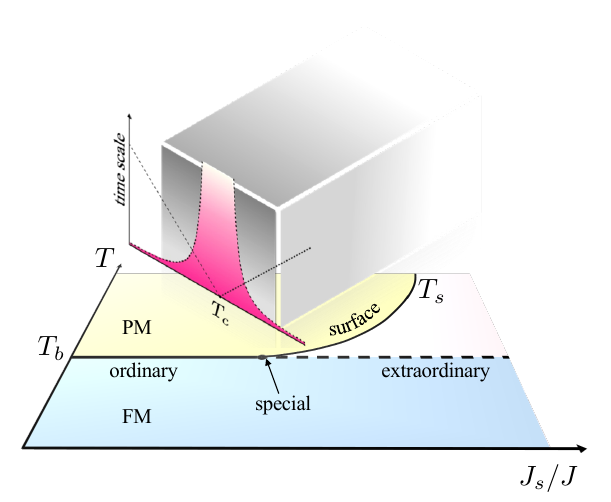}
  \vskip-3mm
  \caption{General phase diagram on the boundary of a system near its critical point and the Kibble-Zurek scheme. For a single bulk transition, different boundary universality classes, including ordinary, special, surface, and extraordinary transitions, can emerge. For each boundary universality class, we consider both heating and cooling dynamics across the critical point of the system to explore the dynamic scaling near the boundary. For the special transition, the driven dynamics of changing the surface coupling $J_s$ across the tricritical point along the line $T=T_b$ is also considered. For the $2$D Ising model, only the ordinary transition exists.
  }
  \label{fig:quench}
\end{figure}

Boundaries are ubiquitous in real materials. Numerous investigations have revealed that regions near boundaries harbor rich and intriguing critical phenomena, a subject that has sustained attention over decades~\cite{PTCP8,PTCP10,Diehl1997book,Pleimling2004jpa,Binder1972prb,Binder1984prl,Landau1990prb,Ruge1995prb,Deng2005pre,Deng2006pre}.
Notably, a single bulk universality class may be associated with several distinct boundary universality classes, including categories such as ordinary, special, extraordinary, and surface transitions, as illustrated in Fig.~\ref{fig:quench}. When the surface coupling $J_s<J_{sc}$, an ordinary transition occurs at $T_c=T_b$ on the boundary, which is ``passively'' induced by the coupling to the ordering transition in the bulk. For $J_s>J_{sc}$ and $d\geq 3$ ($d$ is the dimension of the system), a surface ordering phase transition can occur preemptively on the boundary at $T_s>T_b$. For $J_s>J_{sc}$ at $T_b$, the ordered boundary of the system triggers an extraordinary transition near the boundary. The ordinary and surface transition lines meet at the special transition point $J_s=J_{sc}$, which is a tricritical point with both $(J_s-J_{sc})$ and $(T-T_c)$ as its relevant directions. Recently, a new boundary universality class---the extraordinary-log transition---has been uncovered in the $O(N)$ model~\cite{Metlitski2022sp}
and further developed in various systems~\cite{Toldin2021prl,Hu2021prl,Hu2025arx,Toldin2022prl,Padayasi2022sp,Sun2022prb,Zhang2022prb,Sun2022prb_log,Sun2025sp}.
Studies on boundary criticality have also been extended to quantum critical systems, particularly in connection with topological phase transitions and their novel edge properties~\cite{Zhang2017prl,Zhu2025arx,Wang2024prb,Liu2024prl,Liu2025arx,Toldin2025arx,Wu2020prb,Scaffidi2017prx,Ding2018prl,Weber2018prb,Verresen2021prx,Zhu2021prb,Yu2022prl}.

These fascinating equilibrium boundary critical properties naturally motivate investigations of nonequilibrium critical properties near boundaries. While relaxation critical dynamics in this context has been explored~\cite{Dietrich1983zpb,Kikuchi1985prl,Diehl1994prb,Ritschel1995prl,Pleimling2004prl,Pleimling2005prb,Lin2008pre},
to the best of our knowledge, driven critical dynamics of boundary criticality remains unaddressed. Given the theoretical significance and practical implications of the KZM and FTS, investigation on driven dynamics near the boundaries is of considerable importance.

Motivated by these considerations, we systematically investigate the driven critical dynamics across different boundary universality classes under various driving protocols, using both the two-dimensional ($2$D) and three-dimensional ($3$D) Ising models as examples. The main discoveries include:

\begin{itemize}
\item For heating dynamics starting from $T<T_b$ across the critical point in all boundary universality classes, a set of {\it boundary finite-time scaling} (BFTS) forms can be established to describe the driven critical dynamics. These BFTS forms are {\it normal} generalizations of the conventional FTS forms combined with boundary critical exponents.

\item For cooling dynamics starting from $T>T_b$ across the critical point in ordinary transitions, the order parameters in both $2$D and $3$D exhibit a logarithmic dependence on the driving rate. New scaling forms are thus proposed.

\item For cooling dynamics in special, surface, and extraordinary transitions, the BFTS forms remain applicable.

\item For the special transition, in addition to temperature driving, we also consider the driven dynamics for varying the surface coupling along the ordinary transition line across the tricritical point. When the coupling strength is increased, the adiabatic-impulse scenario of the KZM breaks down because the initial state is a critical state of the ordinary transition with a divergent relaxation time. In this case, we propose a generalized BFTS form incorporating nonequilibrium initial states characterized by the ``age" of the boundary.

\end{itemize}

The rest of the paper is organized as follows. In Sec.~\ref{model}, we introduce the static boundary critical properties of the $2$D and $3$D Ising models and the numerical method employed. In Sec.~\ref{revfts}, we briefly review the bulk FTS forms. In Sec.~\ref{ord}, we study the driven critical dynamics of ordinary transition under both heating (Sec.~\ref{ordh}) and cooling (Sec.~\ref{ordc}) in $2$D and $3$D Ising models. In subsequent sections, only $3$D Ising model is considered, since the $2$D case only exhibits the ordinary transition. Section~\ref{sur} presents the dynamic scaling for the surface transition. In Sec.~\ref{spe}, we focus on the driven critical dynamics in the special transition. In Sec.~\ref{speT}, the temperature is vaired linearly across the critical point, whereas in Sec.~\ref{speJ}, the surface coupling is tuned. Then, in Sec.~\ref{ext}, we examine the dynamic scaling properties of the extraordinary transition. Finally, Sec.~\ref{sum} provides a summary and discussion.

\section{\label{model}Model and Method}
The Hamiltonian of the Ising model studied here is defined as 
\begin{equation}
\label{eq:hamiltonian}
H=-J\sum_{\langle ij\rangle}\sigma_i\sigma_j-J_s\sum_{\langle ij\rangle'}\sigma_{i}\sigma_{j},
\end{equation}
in which $\sigma_i=\pm 1$ represents the classical spin at site $i$, $\langle\rangle$ denotes the nearest-neighbor pairs with interaction $J$ in the bulk, and ${\langle\rangle'}$ denotes the nearest-neighbor pairs with interaction $J_s$ at the boundary. In the following, $J$ is chosen as the unit of energy scale and $g\equiv T-T_c$ denotes the distance to the critical temperature.

The bulk critical properties of model~(\ref{eq:hamiltonian}) have been well studied. In $2$D, the exact solution gives the critical point $T_b=2/\ln(1+\sqrt{2})$, the correlation length exponent $\nu=1$, and the order parameter exponent $\beta=1/8$. The dynamic exponent is given by $z=2.1667(5)$~\cite{Nightingale2000prb}. In $3$D, numerical studies yield $T_b=4.5115233(1)$~\cite{Ferrenberg2018pre}, $\beta=0.32653(10)$~\cite{Campostrini2002pre}, $\nu=0.6299709(40)$~\cite{Simmons-Duffin2017}, and $z=2.0245(2)$~\cite{Hasenbusch2020pre}.

Near the boundary, due to the reduction of coordination number, distinct critical phenomena characterized by additional critical exponents can appear, including: 
(i) For the ordinary transition in $2$D, the boundary order parameter exponent $\beta_1=1/2$~\cite{Diehl1997book,McCoy1973book}.
(ii) In $3$D, for the ordinary transition for $J_s<J_{sc}=1.50243(9)$~\cite{Hasenbusch2011prb_3}, $\beta_1=0.8033(4)$~\cite{Hasenbusch2011prb_2}. Note that $\beta_1>\beta$ in ordinary transitions in both $2$D and $3$D, since the transition is induced by the bulk ordering and is thus more moderate compared with bulk criticality.
(iii) For the special transition in $3$D, $\beta_1=0.2227(4)$~\cite{Hasenbusch2011prb_3}. In addition, the distance of the surface coupling to its critical value $g_J\equiv J_s-J_{sc}$ has the dimension of $\phi\simeq 0.52$~\cite{Lin2008pre,Landau1990prb}.
(iv) For the surface transition in $3$D for $J_s>J_{sc}$, the universality class reduces to that of the $2$D universality class with $\beta_1=\beta_{2D}=1/8$.
(v) For the extraordinary transition, no singularity appears at the surface since the spins are already ordered at $T_b$. Note that anomalous dimensions associated with the boundary spins are also used in the literature~\cite{PTCP8,PTCP10}. However, these exponents are not independent ones and can be related to $\beta_1$ from the scaling laws~\cite{PTCP8,PTCP10}.

Here we study the universal driven dynamics of different boundary universality classes, as illustrated in Fig.~\ref{fig:quench}. For driven critical dynamics, there is no additional divergence in the time direction, and thus no new critical exponents need to be introduced~\cite{Zhongf2005prb,Gong2010njp}. Accordingly, it is expected that the driven dynamics should be described by the above critical exponents.

The nonequilibrium evolution is simulated using Monte Carlo method with standard Metropolis dynamics~\cite{binderbook}.
Except in the case of changing the coupling $J_s$ across the special transition, the simulations are performed in two stages: an initial equilibration stage followed by a driven stage. In the equilibration stage, the system is thermalized at a fixed initial temperature $T_0$ for a sufficiently long time to generate an equilibrium state that serves as the starting configuration. In the driving stage, the temperature is varied linearly in time according to $T(t)=T_0\pm vt$ , where $v$ is the driving rate and $t$ is the simulation time in unit of a full Monte Carlo sweep through the lattice. Measurements are taken at successive values of $T$, so that results for different temperatures can be obtained within a single run. 

Simulations of the case of changing coupling $J_s$ across the special transition point are carried out in a similar two-stage manner but with different initial setups. In the case of decreasing $J_s$, the boundary initial state is an equilibrium state with $T_0=T_b$ and $J_{s0}>J_{sc}$. For increasing $J_s$, we incorporate a nonequilibrium initial state relaxed from an ordered state to $T_0=T_b$ and $J_{s0}<J_{sc}$ with a waiting time $t_a$. The subscript ``a'' in $t_a$ stands for the ``age" of the boundary state. 
In the driving stage, the surface coupling follows $J_s=J_{s0}\pm vt$.

In the simulations, the system size is chosen as $2L\times L$ in $2$D and $2L\times L\times L$ in $3$D, respectively. Open boundary condition is applied in the $x$-direction, which has $2L$ sites, while periodic boundary condition is imposed in other directions.
It has been established that the Metropolis Monte Carlo dynamics belongs to the Model A universality class~\cite{Hohenberg1977rmp,Folk2006jpa,Tauber2014book}, and can also be directly implemented in experiments~\cite{Chae2012prl,Skjaervo2019prx,Griffin2012prx,Lin2014natphy,Meier2017prx}.

\section{\label{revfts}Brief review on the FTS forms in the bulk}

For linearly varying the distance to the critical point, $g=\pm vt$, the FTS demonstrates that the external driving imposes a characteristic time scale $\zeta_d\sim v^{-z/r}$~\cite{Gong2010njp,Huang2014prb,feng2016prb} to the system. Near the critical point, $\zeta_d$ becomes shorter than the intrinsic correlation time of the system and thus dominates the critical dynamics. This is analogous to the finite-size scaling, where the system size $L$ controls the critical behavior when $L<\xi$, with $\xi\propto |g|^{-\nu}$ being the correlation length.

\begin{figure*}[!htbp]
\centering
  \includegraphics[width=\linewidth,clip]{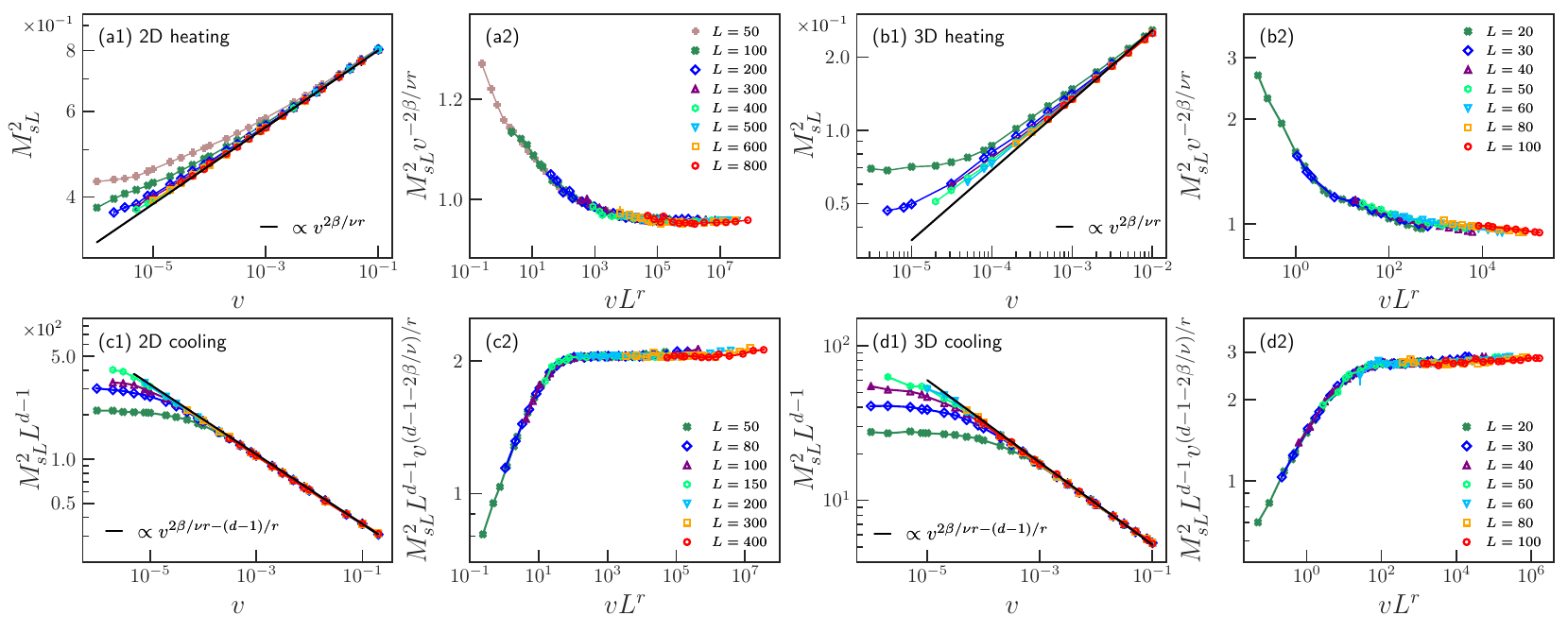}
  \vskip-3mm
  \caption{Driven dynamics of the order parameter in the bulk $M^{2}_{sL}$ at the bulk critical point $T=T_b$ and $J_s=1$. The initial states are prepared at $T_0=T_b\pm 2$ for cooling and heating processes, respectively.
    (a1) Dependence of $M^{2}_{sL}$ on $v$ for different system sizes during heating process in the $2$D Ising model. The solid line indicates a power-law dependence on $v$ with an exponent $2\beta/\nu r$.
    (a2) Data collapse after rescaling $M^{2}_{sL}$ and $v$ in (a1) as $M^{2}_{sL}v^{-2\beta/\nu r}$ and $vL^{r}$, verifying the FTS form in the bulk.
    Similarly, (b1) and (b2) represent the results for the $3$D Ising model during the heating process.
    (c1) Dependence of $M^{2}_{sL}L^{d-1}$ on $v$ during cooling process in the $2$D Ising model. The solid line indicates a power-law dependence on $v$ with an exponent $2\beta/\nu r-(d-1)/r$.
    (c2) Data collapse after rescaling $M^2_{sL}L^{d-1}$ and $v$ as $M^2_{sL}L^{d-1}v^{(d-1-2\beta/\nu)/r}$ and $vL^r$ confirms the FTS form for cooling.
    (d1) and (d2) show results for the $3$D Ising model, yielding consistent conclusion with the $2$D case. Log-log scales are used in all plots. Errorbars are smaller than the symbols.
  }
  \label{fig:xL}
\end{figure*}

In general, the FTS forms depend on the choice of the initial state. For driven dynamics starting from the ordered initial state far away from the critical point, the square of the bulk order parameter $M_b^2$ is defined as 
\begin{equation}
M_b^2=\frac{1}{L^{2d}} \sum_{i,j}\langle \sigma_i\sigma_j\rangle
\end{equation}
with $d$ being the spatial dimension of the system, obeys the FTS form~\cite{Zhongf2005prb,Huang2014prb,feng2016prb} 
\begin{equation}
\label{boporder}
M_b^2(g,L,v)=v^{2 \beta/\nu r}f_1(gv^{-1/\nu r},vL^{r}),
\end{equation}
in which $f_1$ (and other $f_i$ in the following) is the scaling function. Note that as a counterpart of $\zeta_d$, there is also a driving-induced length scale $\xi_d\propto \zeta_d^{1/z}\propto v^{-1/r}$. At $g=0$, when $v<L^{-r}$ (i.e., $\xi_d>L$), the usual finite-size scaling recovered and $M_b^2\propto L^{-2\beta/\nu}$. In contrast, when $v>L^{-r}$ (i.e., $\xi_d<L$), the finite-size effects can be neglected. In this case, $f_1$ tends to a constant and $M_b^2\propto v^{2\beta/\nu r}$.
Physically, the initial state lies in the symmetry-breaking ordered phase, containing a dominant large domain in which spins tend to align in the same direction. For $v>L^{-r}$, the driving-induced length scale $\xi_d$ divides the system into regions with typical size $\xi_d$. Within each region, the spins largely retain a memory of the initial ordering since flipping them requires an energy cost. As a result, the total magnetization remains finite, scaling as $v^{\beta/\nu r}$.

In contrast, for driven dynamics from the disordered initial state far from the critical point, the FTS of $M_b^2$ is~\cite{Huang2014prb,feng2016prb,Liu2014prb}
\begin{equation}
  \label{bopdisorder}
  \begin{split}
    M_b^2(g,L,v)&=L^{-d}v^{2\beta/\nu r-d/r}\\
    &\times f_2(gv^{-1/\nu r},vL^{r}).
  \end{split}
\end{equation}
At the critical point $g=0$, for small $v$, $M_b^2\propto L^{-2\beta/\nu}$ consistent with the case with an ordered initial state. For large $v$, $M_b^2\propto L^{-d}v^{2\beta/\nu r-d/r}$. Since $(2\beta/\nu r-d/r)<0$, $M_b^2$ decreases as $v$ increases. Physically, the initial state is disordered, with spins randomly oriented. For $v>L^{-r}$, correlated regions of typical size $\xi_d$ are formed. Within each region, the magnetization scale as $v^{\beta/\nu r}$, but the orientations of different regions remain random. As a result, $M_b^2$ measures the variance of the magnetization among these regions. With the number of correlated regions given by $N_v = (L/\xi_d)^d$, one has $M^2\propto v^{2\beta/\nu r}/N_{v}$ which leads to the asymptotic scaling form preceding the scaling function $f_2$.

In this paper, we explore the boundary critical dynamics under external driving. Since the translation symmetry in the $x$-direction is broken, we define an $x$-dependent order parameter as
\begin{equation}
M_{sx}^2=\frac{1}{L^{2(d-1)}} \sum_{i_x,j_x}\langle \sigma_{i_x}\sigma_{j_x}\rangle,
\end{equation}
in which $i_x$ and $j_x$ denote spin indices within the profile (a line in $2$D or a slice in $3$D) at position $x$.

To compare with the dynamics of $M_{sx}^2$ near the boundary discussed later, we first investigate the dynamics of $M_{sx}^2$ at the center of the bulk of the system with $x=L$, which is far away from the boundaries for sufficiently large $L$. Accordingly, the driven dynamics of $M_{sL}^2$ can be described by the FTS using bulk exponents.
For driven dynamics from the symmetry-breaking ordered phase, one has
\begin{equation}
\label{sLoporder}
M_{sL}^2(g,L,v)=v^{2 \beta/\nu r}f_3(gv^{-1/\nu r},vL^{r}),
\end{equation}
which takes the same form as Eq.~(\ref{boporder}). At $g=0$, when $v<L^{-r}$ (i.e., $\xi_d>L$), one recovers $M_{sL}^2\propto L^{-2\beta/\nu}$. In contrast, when $v>L^{-r}$ (i.e., $\xi_d<L$), finite-size effects can be neglected and $M_{sL}^2\propto v^{2\beta/\nu r}$.
Physically, for the ordered initial state, the profile at $x=L$ contains a domain with a dominant spin direction. Under fast driving, the memory of the initial state persists up to the critical point,
such that the regions with size $\xi_d$ at $x=L$ tend to retain the same spin direction. This explains the similarity between Eqs.~(\ref{boporder}) and (\ref{sLoporder}).

Similarly, for dynamics starting from a disordered initial state far from the critical point, the FTS of $M_{sL}^2$ is
\begin{equation}
  \label{sLopdisorder}
  \begin{split}
M_{sL}^2(g,L,v)&=L^{-(d-1)}v^{2 \beta/\nu r-(d-1)/r}\\
&\times f_4(gv^{-1/\nu r},vL^{r}),
\end{split}
\end{equation}
in which $d$ in Eq.~(\ref{bopdisorder}) is replaced by the profile dimension $(d-1)$.
At the critical point, $M_{sL}^2\propto L^{-2\beta/\nu}$ for small $v$. For large $v$, $M_{sL}^2\propto L^{-(d-1)}v^{2 \beta/\nu r-(d-1)/r}$. With a disordered initial state, for $v>L^{-r}$, correlated regions with typical size $\xi_d$ emerge within the profile at $x=L$. The magnetization in each region scales as $v^{\beta/\nu r}$, but the orientations of different regions remain random.
Consequently, $M_{sL}^2$ measures the variance of the magnetization among these regions, yielding $M_{sL}^2\propto L^{-(d-1)}v^{2 \beta/\nu r-(d-1)/r}$. Since $2 \beta/\nu r-(d-1)/r<0$, $M_{sL}^2$ decreases as $v$ increases for both $2$D and $3$D cases.

The driven dynamics of $M^2_{sL}$ at the critical point are shown in Fig.~\ref{fig:xL}  for both $2$D and $3$D. 
In Figs.~\ref{fig:xL}(a1) and \ref{fig:xL}(b1), we present the heating dynamics of $M^2_{sL}$ starting from the ordered phase for $2$D and $3$D, respectively. In both dimensions, for large $v$, $M^2_{sL}\propto v^{2\beta/\nu r}$ and is nearly independent of $L$, while for small $v$, $M^2_{sL}$ approaches its finite-size equilibrium value. In Figs.~\ref{fig:xL}(a2) and \ref{fig:xL}(b2), after rescaling $M^2_{sL}$ and $v$ as $M^2_{sL}v^{-2\beta/\nu r}$ and $vL^r$, the rescaled curves collapse well, confirming the FTS scaling form in Eq.~(\ref{sLoporder}).

In Figs.~\ref{fig:xL}(c1) and \ref{fig:xL}(d1), we show the cooling dynamics of $M^2_{sL}$ starting from the disordered phase in $2$D and $3$D, respectively.
In both cases, for large $v$, we find $M^2_{sL}\propto L^{-(d-1)}v^{2\beta/\nu r-(d-1)/r}$, while for small $v$, $M^2_{sL}$ saturates to its finite-size equilibrium value. In Figs.~\ref{fig:xL}(c2) and \ref{fig:xL}(d2), after rescaling $M^2_{sL}$ and $v$ as $M^2_{sL}L^{d-1}v^{(d-1-2\beta/\nu)/r}$ and $vL^r$, respectively, the rescaled curves collapse successfully, confirming the FTS scaling form in Eq.~(\ref{sLopdisorder}).

\section{\label{ord}Driven dynamics in the ordinary transition}

In this section, we explore the driven critical dynamics in the ordinary transition.

In $2$D systems with short-range interactions, spontaneous phase transitions involving continuous symmetry breaking at the $1$D boundary are forbidden by the Mermin-Wagner theorem. In $3$D, when the surface coupling $J_s$ is weaker than a critical value $J_{sc}$ (with $J_{sc} > J$), the surface cannot undergo an independent phase transition due to the reduced coordination number at the boundary. However, in such cases, ordinary boundary transitions can still occur passively, as they are induced by the bulk phase transition. 

Physically, in equilibrium, when the bulk is in the symmetry-breaking phase, the bulk order parameter acts as an effective symmetry-breaking field on the boundary through bulk-boundary coupling, thereby inducing boundary order. Moreover, at the critical point, the spins on the boundary become effectively long-range coupled mediated by the bulk spins due to the divergent correlation length~\cite{Jian2021sp}. Thus, the ordinary transition can exhibit universal scaling properties in equilibrium. However, as ordinary transitions are strongly influenced by the bulk transition, a single new independent exponent is needed~\cite{PTCP8,PTCP10}. Note that, for convenience, different boundary critical exponents are often introduced to describe various observables; however, they are usually not independent but related through scaling laws.
 
Here, we employ the dynamics of $M^2_{sx}$ to reveal the universal dynamic scaling properties. In Sec.~\ref{ordh}, we analyze the heating dynamics and develop the boundary FTS (BFTS) forms. In Sec.~\ref{ordc}, we study the cooling dynamics, where we uncover a logarithmic scaling relation on $v$ and propose new scaling forms.

\begin{figure}[!htbp]
\centering
  \includegraphics[width=\linewidth,clip]{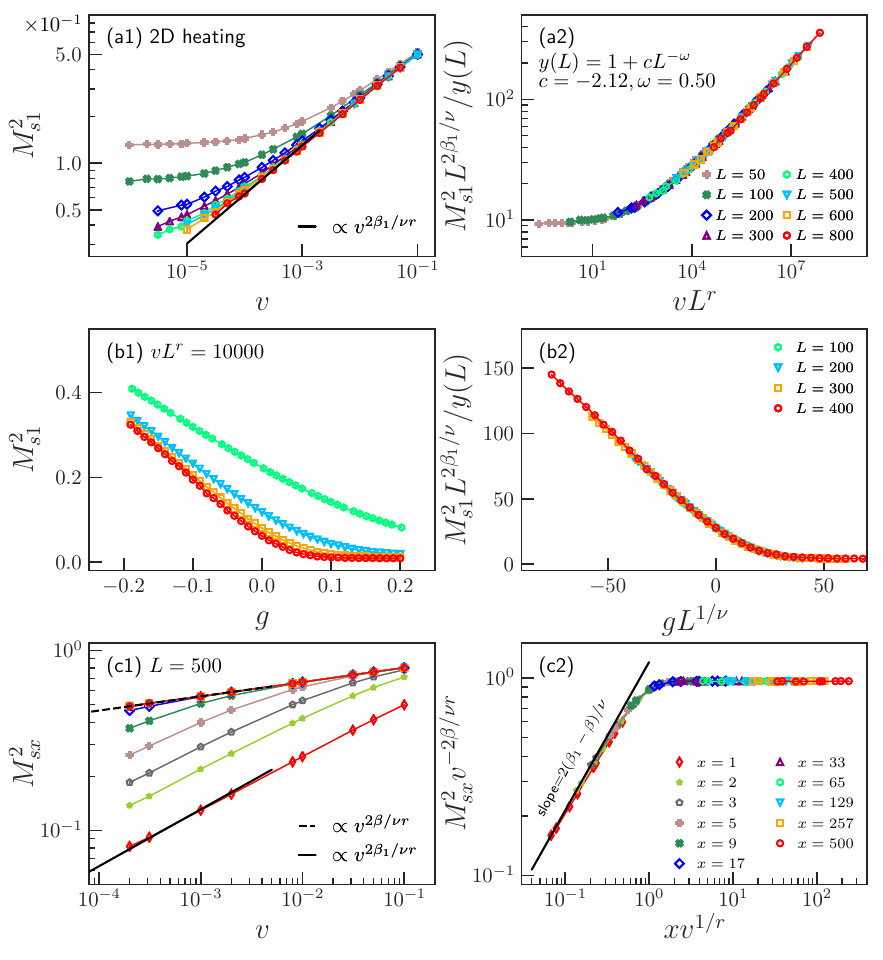}
  \vskip-3mm
  \caption{Heating dynamics of $M^{2}_{s1}$ in $2$D ordinary transition at $T=T_b$ and $J_s=1$. The initial state is prepared at $T_0=T_b-2$.
    (a1) Dependence of $M^{2}_{s1}$ on $v$. The solid line represents a power-law behavior with the exponent $2\beta_1/\nu r$.
    (a2) Rescaling $M^{2}_{s1}$ and $v$ as $M^{2}_{s1}L^{2\beta_1/\nu}$ (with a reasonable finite-size correction term $y(L)=1+cL^{-\omega}$ applied) and $vL^{r}$, respectively, the good collapse verifies the BFTS form at $g=0$.
    (b1) Dependence of $M^{2}_{s1}$ on $g$ for an arbitrary fixed $vL^{r}$.
    (b2) After rescaling the y-axis as in (a2), and $g$ as $gL^{1/\nu}$, all curves collapse well, confirming the validity of the BFTS form for $g\neq 0$.
    (c1) Dependence of $M^{2}_{sx}$ on $v$ for different $x$ with $L=500$. Near the boundary (small $x$), the scaling follows $v^{2\beta_1/\nu r}$ (solid line), while in the bulk (large $x$), it follows $v^{2\beta/\nu r}$ (dashed line). A crossover between the two regimes is observed.
    (c2) Data collapse of $M^{2}_{sx}v^{-2\beta/\nu r}$ versus $xv^{1/r}$. The solid line indicates a power law of $(xv^{1/r})^{2(\beta_1-\beta)/\nu}$ predicted by full BFTS form. Linear scales are used in (b1) and (b2); all other plots use log-log scales.
  }
  \label{fig:fm2d}
\end{figure}

\begin{figure}[!htbp]
\centering
  \includegraphics[width=\linewidth,clip]{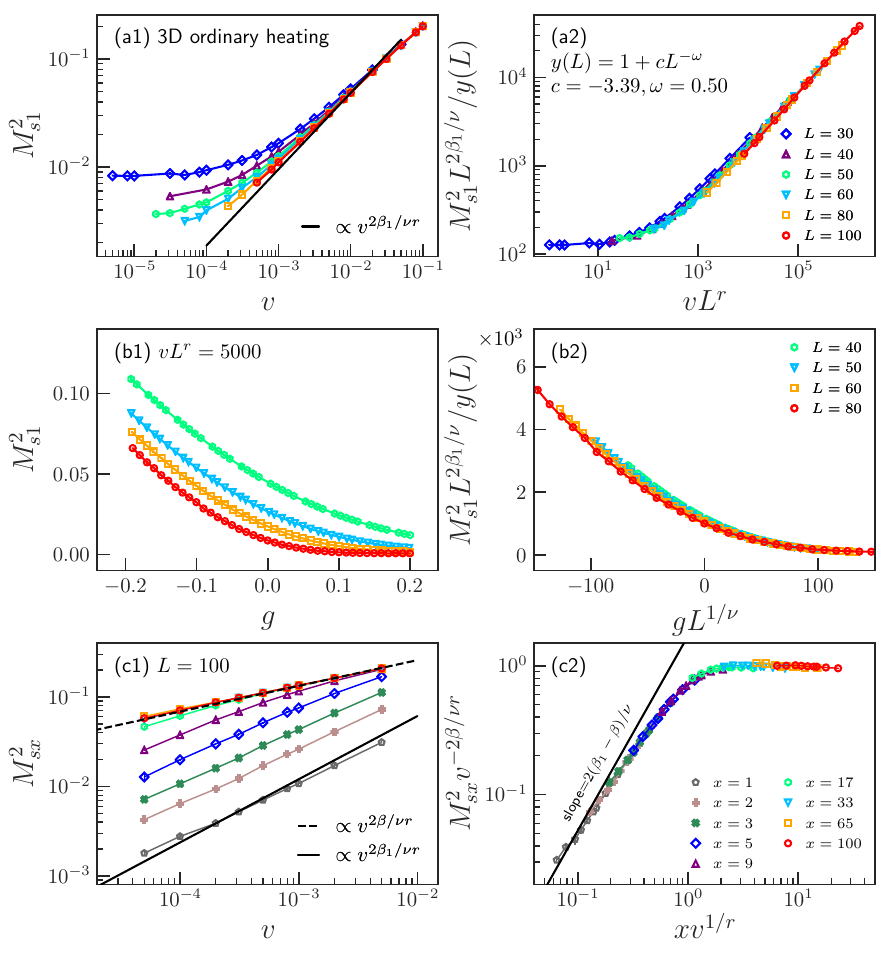}
  \vskip-3mm
  \caption{Heating dynamics in $3$D ordinary transition at $T=T_b$ and $J_s=1$. The initial state is set at $T_0=T_b-2$.
    (a1) Dependence of $M^{2}_{s1}$ on $v$. The solid line shows a power-law behavior with the exponent $2\beta_1/\nu r$.
    (a2) Rescaled curves of $M^{2}_{s1}L^{2\beta_1/\nu}$ (with finite-size correction) versus $vL^{r}$ to verify the BFTS form at $g=0$.
    (b1) Dependence of $M^{2}_{s1}$ on $g$ at fixed $vL^{r}$.
    (b2) After rescaling the y-axis as in (a2), and $g$ as $gL^{1/\nu}$, the good collapse confirms the BFTS form for $g\neq 0$.
    (c1) Dependence of $M^{2}_{sx}$ on $v$ for different $x$ with $L=100$.
    Near the boundary, $M^{2}_{sx}\propto v^{2\beta_1/\nu r}$ (small $x$, solid line), while in the bulk, $M^{2}_{sx}\propto v^{2\beta/\nu r}$ (large $x$, dashed line).
    Crossover behaviors in between are observed.
    (c2) Rescaled curves of $M^{2}_{sx}v^{-2\beta/\nu r}$ versus $xv^{1/r}$ for data collapse. The solid line is a power law of $(xv^{1/r})^{2(\beta_1-\beta)/\nu}$ predicted by full BFTS form.
    Linear scales are used in (b1) and (b2), while other plots use log-log scales.
  }  
  \label{fig:fm3dord}
\end{figure}

\subsection{\label{ordh}Heating dynamics}

In this section, we study the heating critical dynamics across the ordinary transition. For both $2$D and $3$D, the initial states are set at $T_0=T_b-2$ and $J_s=1$, and during the driving process, the temperature follows $T=T_0+vt$.

We at first explore $M^2_{sx}$ at the boundary profile $x=1$. In equilibrium, at the critical point one has $M^2_{s1}\propto L^{-2\beta_1/\nu}$ for a system of size $L$. Compared with the bulk case, where $M^2_{sL}\propto L^{-2\beta/\nu}$ as shown in Eq.~(\ref{bopdisorder}), the only difference is that the bulk order parameter exponent $\beta$ is replaced by the boundary exponent $\beta_1$. This analogy suggests that the dynamic scaling properties at the boundary may be obtained by a similar substitution.
For heating dynamics from the symmetry-breaking ordered phase, by analogy with Eq.~(\ref{sLoporder}), this {\it normal} generalization of the FTS forms leads to the BFTS form
\begin{equation}
\label{sLoporderx1}
\begin{split}
M_{s1}^2(g,L,v)&=v^{2 \beta_1/\nu r}\\
&\times f_5(gv^{-1/\nu r},vL^{r}),
\end{split}
\end{equation}
in which $\beta_1=1/2$~\cite{Diehl1997book,McCoy1973book} and $\beta_1=0.8033(4)$~\cite{Hasenbusch2011prb_2} for the $2$D and $3$D Ising model, respectively.

To verify Eq.~(\ref{sLoporderx1}), for $2$D case, we show the $v$ dependence of $M^2_{s1}$ in Fig.~\ref{fig:fm2d}(a1) at $g=0$.
For large $v$ and $L$, we find that $M^2_{s1}\propto v^{2\beta_1/\nu r}$. In this region, $M^2_{s1}$ is almost independent of $L$. For small $v$, $M^2_{s1}$ approaches its finite-size equilibrium value. These scaling behaviors are analogous to the bulk cases discussed below Eqs.~(\ref{boporder}) and (\ref{sLoporder}). 

At $g=0$, Eq.~(\ref{sLoporderx1}) can be converted to $M_{s1}^2(0,L,v)=L^{-2 \beta_1/\nu}f_{5'}(vL^r)$ as $f_5(0,vL^{r})=(vL^r)^{-2\beta_1\nu r}f_{5'}(vL^r)$.
After rescaling $M^2_{s1}$ and $v$ as $M^2_{s1}L^{2\beta_1/\nu}$ and $vL^r$, respectively, we find that the rescaled curves collapse well after applying a reasonable finite-size scaling correction, as shown in Fig.~\ref{fig:fm2d}(a2).
Here we use $L$ to rescale the quantities since the equilibrium finite-size scaling for $v\rightarrow 0$ is well established. In the vicinity of the critical point, for fixed $vL^{r}$, the dependence of $M^2_{s1}$ on $g$ is shown in Fig.~\ref{fig:fm2d}(b1). After rescaling $M^{2}_{s1}$ and $g$ as $M^2_{s1}L^{2\beta_1/\nu}$ (with the same finite-size correction) and $gL^{1/\nu}$, respectively, the rescaled curves in Fig.~\ref{fig:fm2d}(b2) also collapse well. These results confirm the BFTS form of Eq.~(\ref{sLoporderx1}).

We also study the $3$D case to verify the universality of Eq.~(\ref{sLoporderx1}). At $g=0$, Fig.~\ref{fig:fm3dord}(a1) shows that for large $v$ and $L$, $M^2_{s1}\propto v^{2\beta_1/\nu r}$, consistent with the $2$D case.
After rescaling $M^2_{s1}$ and $v$ as $M^2_{s1}L^{2\beta_1/\nu}$ and $vL^r$ (with a finite-size correction), the curves collapse well, as shown in Fig.~\ref{fig:fm3dord}(a2).
In addition, we study the behavior of $M^{2}_{s1}$ near the critical point. For fixed $vL^{r}$, after rescaling $M^{2}_{s1}$ and $g$ as $M^2_{s1}L^{2\beta_1/\nu}$ (with the same finite-size correction) and $gL^{1/\nu}$, respectively, successful collapse is observed, as suggested by Eq.~(\ref{sLoporderx1}).
These results confirm that the BFTS form of Eq.~(\ref{sLoporderx1}) can apply in the driven dynamics from the ordered phase in the ordinary transition.

To understand the origin of Eq.~(\ref{sLoporderx1}), we note that the heating critical dynamics combines the critical scaling properties with those of the symmetry-breaking ordered phase, as we discussed in Sec.~\ref{revfts}.
In the initial state, both the bulk and boundary are in the ordered phase, dominated by the symmetry-breaking domain. As the system is driven toward the critical point, correlated regions with typical size $\xi_d\propto v^{-1/r}$ emerge in the bulk, within which the spins preferentially orient along the original ordered direction.
Through bulk-boundary coupling, these ordered regions extend to the boundary. Consequently, the total boundary magnetization acquires a finite value $\propto v^{\beta_1/\nu r}$, accounting for the BFTS form of Eq.~(\ref{sLoporderx1}).

Next, we discuss the critical properties near the boundary, focusing on how boundary critical behavior penetrates into bulk critical properties under different driving rates. To this end, we consider a large system where finite-size effects are negligible, and the profile at $x=L$ effectively reflects the bulk limit $x\rightarrow \infty$.

In $2$D, Fig.~\ref{fig:fm2d}(c1) shows the curves of $M^2_{sx}$ versus $v$ for different $x$. For each $x$ in the bulk ($x>1$), $M^2_{sx}$ exhibits a crossover: for small $x$, $M^2_{sx}\propto v^{2 \beta_1/\nu r}$, while for large $x$, $M^2_{sx}\propto v^{2 \beta/\nu r}$. As $x$ increases, the crossover driving rate $v$ shifts to smaller values. Similar behaviors are also found in $3$D, as shown in Fig.~\ref{fig:fm3dord} (c1). These results demonstrate that for large $x$ and large $v$, the dominant ordered domain in the initial state primarily controls the behavior of $M^{2}_{sx}$; in contrast, for small $x$ and small $v$, boundary fluctuations become increasingly significant.

The results in both $2$D and $3$D suggest that for $L\rightarrow\infty$ and $g=0$, $M_{sx}^2$ should satisfy
\begin{equation}
\label{sLoporderx}
M_{sx}^2(v)=v^{2 \beta/\nu r}f_6(xv^{1/r}),
\end{equation}
in which for large $x$ and $v$, $f_6$ tends to a constant and $M_{sx}^2(v)\propto v^{2 \beta/\nu r}$; while for small $v$ and small $x$, $f_6(xv^{1/r})\propto (xv^{1/r})^{2(\beta_1-\beta)/\nu}$, leading to $M_{sx}^2(v)\propto v^{2 \beta_1/\nu r}$.

The scaling collapse in Figs.~\ref{fig:fm2d}(c2) and \ref{fig:fm3dord}(c2) for $2$D and $3$D, respectively, verifies Eq.~(\ref{sLoporderx}). For large $xv^{1/r}$, the rescaled curves flatten, consistent with the scaling function $f_6$ approaching a constant. For small $xv^{1/r}$, the rescaled curves follows $M^{2}v^{-2\beta/\nu r}\propto (xv^{1/r})^{2(\beta_1-\beta)/\nu}$, explaining the crossover behaviors.

Accordingly, a full FTS form for the heating dynamics in ordinary transition can be established, which not only incorporates the original bulk FTS and the BFTS but also captures crossover behaviors discussed above.
Combining the FTS forms of Eqs.~(\ref{sLoporder}), (\ref{sLoporderx1}), and (\ref{sLoporderx}), the full scaling FTS form for the heating dynamics near the boundary with ordinary transition is given by
\begin{equation}
\label{fullsLoporder}
M_{sx}^2(g,L,v)=v^{2 \beta/\nu r}f_7(gv^{-1/\nu r},vL^{r},xv^{1/r}).
\end{equation}
For large $xv^{1/r}$, Eq.~(\ref{fullsLoporder}) reduces to Eq.~(\ref{sLoporder}); while for small $xv^{1/r}$, Eq.~(\ref{fullsLoporder}) recovers the BFTS of Eq.~(\ref{sLoporderx1}).

\subsection{\label{ordc}Cooling dynamics}

Having extended the FTS to the BFTS in heating dynamics, we now turn to the cooling case in the ordinary transition.
To realize the cooling process, we prepare the initial state at $T_0=T_b+2$ and $J_s=1$ for both $2$D and $3$D, and during the driving, the temperature follows $T=T_0-vt$.

For the cooling dynamics, a central question is whether the replacement of the bulk exponent $\beta$ by the boundary exponent $\beta_1$, as discussed in Sec.~\ref{ordh}, also applies.
According to Eq.~(\ref{sLopdisorder}), for cooling dynamics from the disordered phase, this leads to the putative BFTS form
\begin{equation}
  \label{sLopdisorderx1}
  \begin{split}
    M_{s1}^2(g,L,v)&=L^{-(d-1)}v^{2 \beta_1/\nu r-(d-1)/r}\\
    &\times f_8(gv^{-1/\nu r},vL^{r}),
  \end{split}
\end{equation}
which expects that $M_{s1}^2\propto L^{-(d-1)}v^{2 \beta_1/\nu r-(d-1)/r}$ for large $v$ and $g=0$.
Based on the discussion below Eq.~(\ref{sLopdisorder}) and the scaling behaviors of $M_{s1}^2$, in particular, Eq.~(\ref{sLopdisorderx1}) predicts that for large $v$, $M_{s1}^2L^{d-1}$ should be independent of $v$ since $[2\beta_1/\nu-(d-1)]=0$ in $2$D, whereas in $3$D, $M_{s1}^2$ should increase as $v$ increases since $[2\beta_1/\nu-(d-1)]>0$. However, as we shall see below, both Eq.~(\ref{sLopdisorderx1}) and these deductions fail.

\begin{figure}[!htbp]
\centering
  \includegraphics[width=\linewidth,clip]{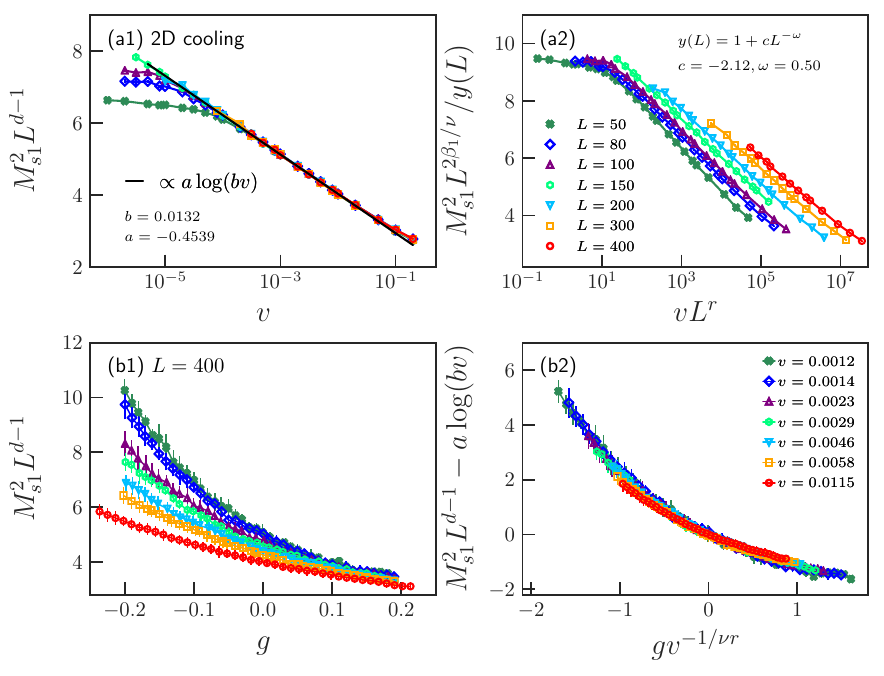}
  \vskip-3mm
  \caption{Cooling dynamics in $2$D ordinary transition at $T=T_b$ and $J_s=1$. The initial state is set at $T_0=T_b+2$.
    (a1) Dependence of $M^{2}_{s1}L^{d-1}$ on $v$. The solid line represents a logarithmic dependence of $a\log(bv)$.
    (a2) Rescaled curves of $M^{2}_{s1}L^{2\beta_1/\nu}$ (with finite-size correction) versus $vL^{r}$. Only data close to equilibrium collapses, indicating the breakdown of the normal BFTS form.
    (b1) Dependence of $M^{2}_{s1}L^{d-1}$ on $g$ for fixed $vL^{r}$ with $L=400$. 
    (b2) Rescaled curves of $M^{2}_{s1}L^{d-1}-a\log(bv)$ versus $gL^{1/\nu}$, the good collapse verifies Eq.~(\ref{oporderlog1}).
    Log-linear scales are used in (a1) and (a2), and linear scales are used in (b1) and (b2).
  }
  \label{fig:pm2d}
\end{figure}

In $2$D, Fig.~\ref{fig:pm2d}(a1) shows the curves of $M_{s1}^2L^{d-1}$ versus $v$. For small $v$, $M_{s1}^2L^{d-1}$ saturates to its finite-size scaling value due to finite-size effects, while for large $v$, it becomes almost independent of $L$. However, $M_{s1}^2L^{d-1}$ clearly decreases with increasing $v$, inconsistent with the prediction of Eq.~(\ref{sLopdisorderx1}). Moreover, transforming Eq.~(\ref{sLopdisorderx1}) at $g=0$ to $M_{s1}^2(L,v)=L^{-2\beta_1/\nu}f_{8'}(vL^r)$, we rescale $M_{s1}^2$ and $v$ as $M_{s1}^2L^{2\beta_1/\nu}$ and $vL^r$, respectively, and find that the rescaled curves only collapse in the small $v$ region, but deviate significantly for large $v$. This result again contradicts Eq.~(\ref{sLopdisorderx1}), as shown in Fig.~\ref{fig:pm2d}(a2).

\begin{figure}[!htbp]
\centering
  \includegraphics[width=\linewidth,clip]{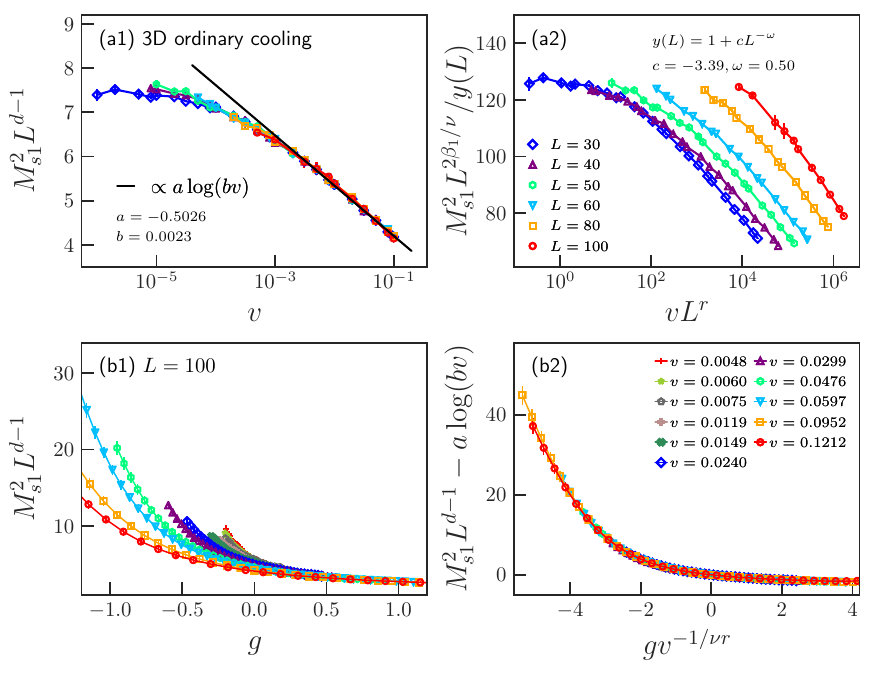}
  \vskip-3mm
  \caption{Cooling dynamics in $3$D ordinary transition at $T=T_b$ and $J_s=1$. The initial state is set at $T_0=T_b+2$.
    (a1) Dependence of $M^{2}_{s1}L^{d-1}$ on $v$. The solid line represents a logarithmic dependence of $a\log(bv)$.
    (a2) Rescaled curves of $M^{2}_{s1}L^{2\beta_1/\nu}$ (with finite-size correction) versus $vL^{r}$. Only data close to equilibrium collapses, indicating the breakdown of the normal BFTS form.
    (b1) Dependence of $M^{2}_{s1}L^{d-1}$ on $g$ for an arbitrary fixed $vL^{r}$ with $L=100$. 
    (b2) Rescaled curves of $M^{2}_{s1}L^{d-1}-a\log(bv)$ versus $gL^{1/\nu}$, the good collapse verifies Eq.~(\ref{oporderlog1}).
    Log-linear scales are used in (a1) and (a2), and linear scales are used in (b1) and (b2).
  }
  \label{fig:pm3dord}
\end{figure}

Similar violations of Eq.~(\ref{sLopdisorderx1}) for large $v$ are also observed in $3$D. Figure~\ref{fig:pm3dord}(a1) shows the curves of $M_{s1}^2L^{d-1}$ versus $v$. For small $v$, $M_{s1}^2L^{d-1}$ saturates to its finite-size value, while for large $R$, it becomes nearly independent of $L$. Moreover, $M_{s1}^2L^{d-1}$ clearly again deceases with increasing $v$, in contradiction with the prediction of Eq.~(\ref{sLopdisorderx1}). In addition, in Fig.~\ref{fig:pm3dord}(a2), no collapse is observed in the rescaled curves of $M_{s1}^2L^{2\beta_1/\nu}$ versus $vL^{r}$ for large $v$, further indicating the breakdown of Eq.~(\ref{sLopdisorderx1}).

Beyond the numerical evidence, the prediction of Eq.~(\ref{sLopdisorderx1}) also contradicts the intuitive physical picture. For the cooling dynamics starting from the disordered phase, the dynamic behaviors at the critical point should reflect a combination of the initial state and the universal critical properties. For larger driving rates, more initial state information is preserved, resulting in a smaller $M_{s1}^2$. Therefore, $M_{s1}^2$ must decrease with increasing $v$.

\begin{figure}[tbp]
\centering
  \includegraphics[width=0.8\linewidth,clip]{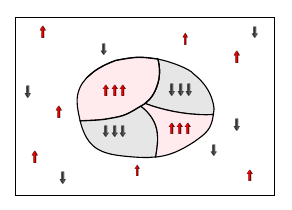}
  \vskip-3mm
  \caption{Illustration of suppressed domain formation near the boundary during cooling to $T_b$, where enhanced fluctuations and reduced coordination number prevent the emergence of correlated domains, resulting in the breakdown of the normal BFTS.   
  }
  \label{fig:illu_abnormal}
\end{figure}

In addition, as discussed above, the scaling relation $M_{sL}^2 \propto L^{-(d-1)}v^{2 \beta/\nu r-(d-1)/r}$ in Eq.~(\ref{sLopdisorder}) can be interpreted as the variance of the order parameter within domains of typical size $\xi_d$. At the boundary, however, the breakdown of the analogous scaling relation $M_{s1}^2 \propto L^{-(d-1)}v^{2 \beta_1/\nu r-(d-1)/r}$ suggests that such domains are not formed. A physical picture is illustrated in Fig.~\ref{fig:illu_abnormal}: although the random domains with typical size $\xi_d$ can develop in the bulk, their influence cannot extend to the boundary, where strong fluctuations--—enhanced by the reduced coordination number—--disrupt domain formation.

The absence of correlated domains at $T_c$ during cooling dynamics can also be understood from a different perspective. As discussed above, in equilibrium, the ordinary transition can be regarded as a phase transition in a system of one lower dimension with effective long-range interactions. Under cooling, however, the driving carries memory of the initial disordered state to the critical point. As a result, the boundary retains only finite-length couplings, which suppresses the formation of domains.

The failure of the BFTS form of Eq.~(\ref{sLopdisorderx1}) necessitates a new scaling relation to characterize the cooling dynamics of ordinary transitions. From Figs.~\ref{fig:pm2d}(a1) and \ref{fig:pm3dord}(a1), for large $v$, we find that the dependence of $M_{sL}^2 L^{(d-1)}$ on $v$ satisfies a logarithmic function:
\begin{equation}
\label{oporderlog}
M_{s1}^2 \propto aL^{-(d-1)}\log(bv), 
\end{equation}
in which $a$ and $b$ are dimension-dependent constants with $a<0$ and $b>0$.

We further generalize this scaling relation to the cooling process in the regime of large $L$ and large $v$. Since $M_{s1}^2(v)\propto L^{-(d-1)}$, we propose the following the scaling form:
\begin{equation}
\label{oporderlog1}
M_{s1}^2(v)=aL^{-(d-1)}\log(bv)+L^{-(d-1)}f_9(gv^{-1/\nu r}).
\end{equation}

Figures~\ref{fig:pm2d}(b1) and \ref{fig:pm3dord}(b1) shows the evolution of $M_{s1}^2$ versus $g$ with different $v$ for a large system size in $2$D and $3$D, respectively. After rescaling the data according to Eq.~(\ref{oporderlog1}), the rescaled curves collapse well, confirming the validity of Eq.~(\ref{oporderlog1}).

In contrast to the scaling form obtained from Eq.~(\ref{sLopdisorderx1}) $M_{s1}^2\propto L^{-(d-1)}v^{2 \beta_1/\nu r-(d-1)/r}$, which cannot capture the decreasing of $M_{s1}^2$ with increasing $v$ when $[2 \beta_1/\nu r-(d-1)/r]\geq0$, the logarithmic dependence on $v$ is the leading function to describe this behavior. Thus, the appearance of the logarithmic scaling of $M_{s1}^2$ on the driving rate seems plausible.

Taken together, for the ordinary transition, heating dynamics follows the BFTS as a natural extension of the bulk FTS, whereas cooling dynamics deviates from this extension and exhibits a logarithmic dependence on the driving rate. These results demonstrate a clear asymmetry between heating and cooling, arising from the enhanced boundary fluctuations that suppress correlated domain formation.

\section{\label{sur}Driven dynamics in the surface transition}
In this section, we investigate the driven critical dynamics of the surface transition in $3$D. When the surface coupling $J_s$ exceeds a critical value $J_{sc}$, the surface undergoes an ``autonomous'' phase transition as the temperature is tuned. Different from the ordinary transition, the surface transition in Model~(\ref{eq:hamiltonian}) belongs to the usual $2$D Ising universality class, and its driven dynamics is therefore expected to follow the behavior dictated by the $2$D Ising critical exponents. For the heating dynamics, the BFTS form is
\begin{equation}
  \label{heatingsur}
    \begin{split}
      M_{s1}^2(g,L,v)&=v^{2 \beta_{2D}/\nu_{2D} r_{2D}}\\
      &\times f_{10}(gv^{-1/\nu_{2D} r_{2D}},vL^{r_{2D}}),
    \end{split}
\end{equation}
while for the cooling dynamics, the BFTS form is
\begin{equation}
  \label{coolingsur}
  \begin{split}
    M_{s1}^2(g,L,v)=&L^{-(d-1)}v^{2 \beta_{2D}/\nu_{2D} r_{2D}-(d-1)/r_{2D}}\\
    &\times f_{11}(gv^{-1/\nu_{2D} r_{2D}},vL^{r_{2D}}).
  \end{split}
\end{equation}

\begin{figure}[!htbp]
\centering
\includegraphics[width=\linewidth,clip]{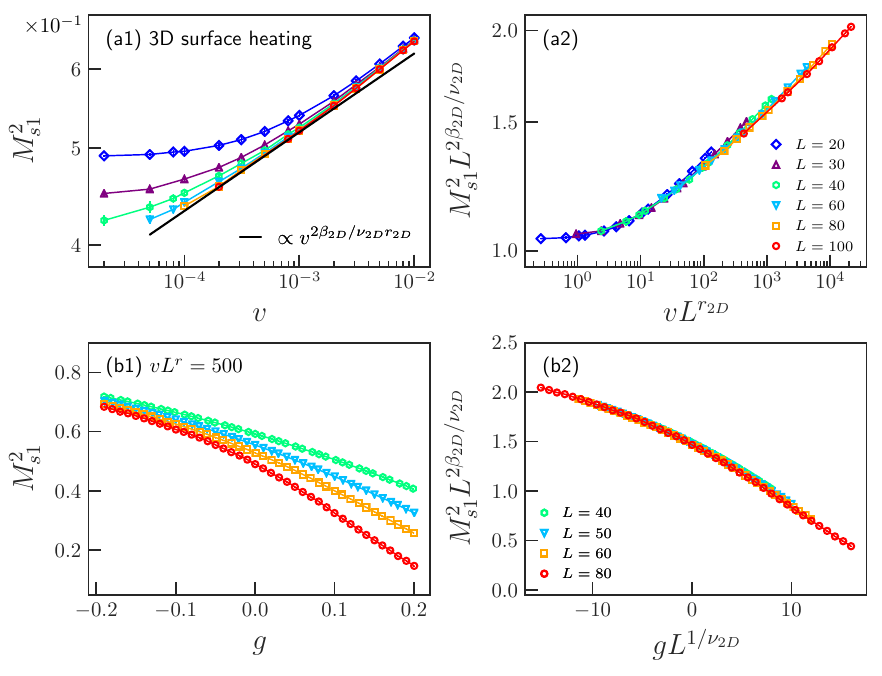}
  \vskip-3mm
  \caption{Heating dynamics in $3$D surface transition at $T=T_s$ and $J_s=2$. The initial state is set at $T_0=T_s-2$.
    (a1) Dependence of $M^{2}_{s1}$ on $v$. The solid line represents a power law of $v^{2\beta_{2D}/\nu_{2D} r_{2D}}$ with the $2$D Ising exponents.
    (a2) Rescaled curves of $M^{2}_{s1}L^{2\beta_{2D}/\nu_{2D}}$ versus $vL^{r_{2D}}$, the good collapse verifies the BFTS form at $g=0$.
    (b1) Dependence of $M^{2}_{s1}$ on $g$ for fixed $vL^{r_{2D}}$.
    (b2) After rescaling the y-axis as in (a2), and $g$ as $gL^{1/\nu_{2D}}$, the good collapse confirms the BFTS form for $g\neq 0$.
    Log-log scales are used in (a1) and (a2), while (b1) and (b2) use linear scales.
  }
  \label{fig:fm3dsurf}
\end{figure}

\begin{figure}[!htbp]
\centering
  \includegraphics[width=\linewidth,clip]{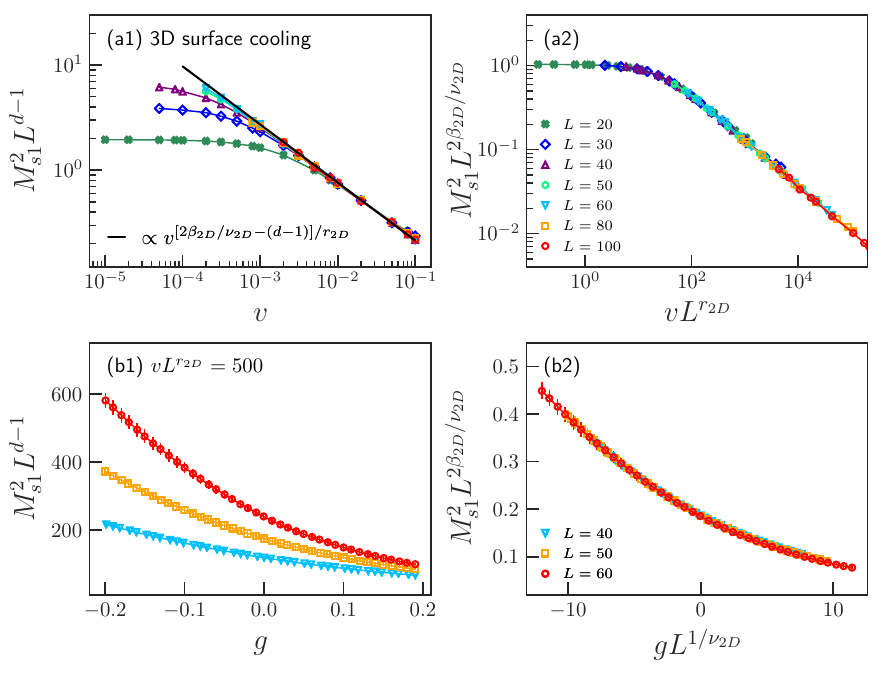}
  \vskip-3mm
  \caption{Cooling dynamics in $3$D surface transition at $T=T_s$ and $J_s=2$. The initial state is set at $T_0=T_s+2$.
    (a1) Dependence of $M^{2}_{s1}L^{d-1}$ on $v$. The solid line represents a power law of $v^{[2\beta_{2D}/\nu_{2D}-(d-1)]/r_{2D}}$ with the $2$D Ising exponents.
    (a2) Rescaled curves of $M^{2}_{s1}L^{2\beta_{2D}/\nu_{2D}}$ versus $vL^{r_{2D}}$ to verify the BFTS form at $g=0$.
    (b1) Dependence of $M^{2}_{s1}$ on $g$ for an arbitrary fixed $vL^{r_{2D}}$.
    (b2) After rescaling the y-axis as in (a2), and $g$ as $gL^{1/\nu_{2D}}$, the good collapse confirms the BFTS form for $g\neq 0$.
    Log-log scales are used in (a1) and (a2), while (b1) and (b2) use linear scales.
  }
  \label{fig:pm3dsurf}
\end{figure}

For the $3$D Ising model, $J_{sc}=1.50243(9)$~\cite{Hasenbusch2011prb}. Here we set the surface coupling strength $J_{s}=2$, for which the surface transition temperature is $T_s=4.955$~\cite{Lin2008pre}.
For the heating dynamics, Fig.~\ref{fig:fm3dsurf}(a1) shows the dependence of $M^2_{s1}$ on $v$ for different $L$ at $T_s$. At large driving rates, $M^2_{s1}\propto v^{2\beta_{2D}/\nu_{2D} r_{2D}}$ and $M^2_{s1}$ is almost independent of $L$, while at small $v$, $M^2_{s1}$ recovers the equilibrium finite-size scaling. In Fig.~\ref{fig:fm3dsurf}(a2), after rescaling $M^2_{s1}$ and $v$ as $M^2_{sL}L^{-2\beta_{2D}/\nu_{2D}}$ and $vL^{r_{2D}}$, respectively, the rescaled curves collapse well, confirming Eq.~(\ref{heatingsur}) at the surface critical point for $J_s=2$. In addition, for fixed $vL^{r_{2D}}$, Fig.~\ref{fig:fm3dsurf}(b1) shows the evolution of $M^2_{s1}$ in the driven process. After rescaling $M^2_{s1}$ and $g$ according to Eq.~(\ref{heatingsur}), the rescaled curves also collapse successfully, as shown in Fig.~\ref{fig:fm3dsurf}(b2), confirming Eq.~(\ref{heatingsur}). 

For the cooling dynamics, Fig.~\ref{fig:pm3dsurf}(a1) shows the curves of $M^2_{s1}$ versus $v$ for different $L$ at $T_s=4.955$ and $J_s=2$~\cite{Lin2008pre}. For large $v$, the behavior of $M^2_{s1} L^{d-1}$ demonstrates a power-law dependence on $v$ with an exponent of $2 \beta_{2D}/\nu_{2D} r_{2D}-(d-1)/r_{2D}$, consistent with the BFTS form of Eq.~(\ref{coolingsur}). Moreover, $M^2_{s1}L^{d-1}$ is almost independent of $L$ in the large $v$ regime. After rescaling $M^2_{s1}$ and $v$ as $M^2_{sL}L^{2\beta_{2D}/\nu_{2D} r_{2D}}$ and $vL^{r_{2D}}$, respectively, all rescaled curves collapse well, confirming Eq.~(\ref{coolingsur}).
To further explore the dynamic scaling in the cooling process, we fix $vL^{r_{2D}}$ to an arbitrary constant and compute $M^2_{s1}$ for different $L$ and $g$, as shown in Fig.~\ref{fig:pm3dsurf}(b1). The successful collapse of the rescaled curves in Fig.~\ref{fig:pm3dsurf}(b2) again verifies Eq.~(\ref{coolingsur}).

These results show that both heating and cooling dynamics of the surface transition follow the BFTS forms. This confirms that the driven critical dynamics of the surface transition is fully governed by the $2$D Ising universality class.

\section{\label{spe}Driven dynamics in the special transition}

The special transition is a tricritical point that lies at the watershed between the ordinary and the special transition~\cite{PTCP8,PTCP10}. Near this point, there are two relevant directions: one associated with the temperature $T$ and the other goes along the surface coupling $J_s$. While the KZM mainly focused on the usual critical point, its generalization to the tricritical point in the bulk has recently been explored~\cite{Wangtl2025arx,Wanght2025arx}. Here, we explore the driven critical dynamics near this surface tricritical point. We first consider the driven dynamics by driving the temperature in Sec.~\ref{speT} and then turn to the case of varying the surface coupling~\ref{speJ}.

\subsection{\label{speT} Changing temperature}
\subsubsection{\label{speTin} Heating dynamics}
To study the temperature-driven heating dynamics of the special transition, we fix the boundary coupling $J_{sc}=1.50243$~\cite{Hasenbusch2011prb_3} and choose the initial temperature as $T_0=T_b-2$, corresponding to the symmetry-breaking ordered phase. In this setup, the {\it normal} generalization of the FTS forms leads to the BFTS form at the boundary with $x=1$,
\begin{equation}
\label{oporderspeT}
M_{s1}^2(g,L,v)=v^{2 \beta_1/\nu r}f_{12}(gv^{-1/\nu r},vL^{r}),
\end{equation}
in which $\beta_1=0.2227(4)$~\cite{Hasenbusch2011prb_3} for the $3$D Ising model. Note that this scaling form is similar to the corresponding BFTS in the ordinary transition.

\begin{figure}[htbp]
\centering
  \includegraphics[width=\linewidth,clip]{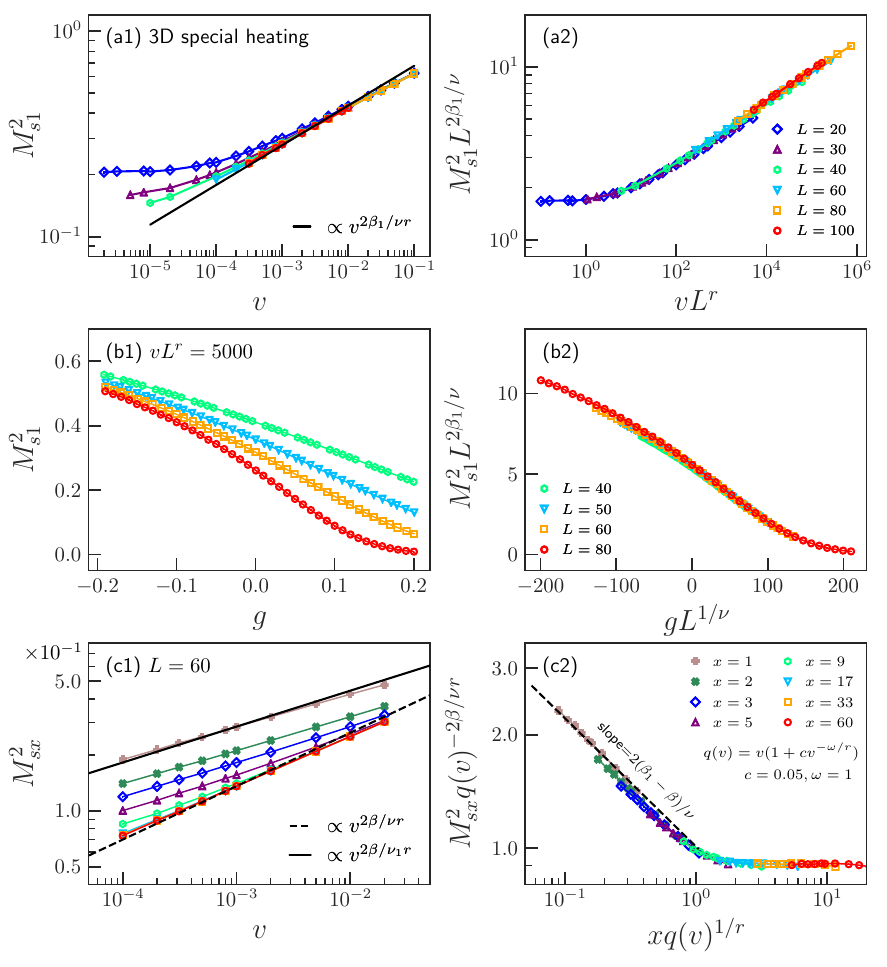}
  \vskip-3mm
  \caption{Heating dynamics in $3$D special transition at $T=T_b$ with $J_s=J_{sc}$. The initial state is set at $T_0=T_b-2$.
    (a1) Dependence of $M^{2}_{s1}$ on $v$. The solid line represents a power-law behavior with the exponent $2\beta_1/\nu r$, with $\beta_1=0.2227$ being the boundary order parameter exponent of the $3$D Ising special universality class.
    (a2) Rescaled curves of $M^{2}_{s1}L^{2\beta_1/\nu}$ versus $vL^{r}$ to verify the BFTS form at $g=0$.
    (b1) Dependence of $M^{2}_{s1}$ on $g$ for an arbitrary fixed $vL^{r}$.
    (b2) Rescaled curves of $M^{2}_{s1} L^{2\beta_1/\nu}$ versus $gL^{1/\nu}$ to verify the BFTS form for $g\neq 0$.
    (c1) Dependence of $M^{2}_{sx}$ on $v$ for different $x$ with $L=60$.
    The solid line indicates a power law of $v^{2\beta_1/\nu r}$ for small $x$ and the dashed line represents a power law of $v^{2\beta/\nu r}$ for large $x$. Crossover behaviors in between are observed.
    (c2) Rescaled curves of $M^{2}_{sx}q(v)^{-2\beta/\nu r}$ versus $xq(v)^{1/r}$ for data collapse, where $q(v)=v(1+cv^{-\omega/r})$ includes a driving-rate correction. The solid line is a power law of $[xq(v)^{1/r}]^{2(\beta_1-\beta)/\nu}$ expected by full BFTS form.
    Linear scales are used in (b1) and (b2), while other plots use log-log scales.
  }
  \label{fig:fm3dsp}
\end{figure}

To verify Eq.~(\ref{oporderspeT}), we show in Fig.~\ref{fig:fm3dsp}(a1) the curves of $M_{s1}^2$ versus $v$ for different $L$. For large $v$, $M^2_{s1}\propto v^{2\beta_1/\nu r}$, consistent with Eq.~(\ref{oporderspeT}). In Fig.~\ref{fig:fm3dsp}(a2), the good collapse of the rescaled curves of $M^2_{sL}L^{2\beta/\nu }$ and $vL^r$ confirms Eq.~(\ref{heatingsur}) at $g=0$. With fixed $vL^{r}$, the driven process near the critical point is shown in Fig.~\ref{fig:fm3dsp}(b1). After rescaling $M^2_{s1}$ and $g$ according to Eq.~(\ref{oporderspeT}), the curves collapse well, as shown in Fig.~\ref{fig:fm3dsurf}(b2), confirming the BFTS of Eq.~(\ref{heatingsur}).
The heating dynamics at the special transition resembles that of the ordinary transition.

In Fig.~\ref{fig:fm3dsp}(c1), we further show the dynamics scaling behaviors near the boundary with different $x$ at $g=0$ for large $L$. For small $x$, $M_{sx}^2\propto v^{2\beta_1/\nu r}$, while for large $x$, the behavior crosses over to $M_{sx}^2\propto v^{2\beta/\nu r}$. This crossover arises because, for large $x$ and large $v$, the dominant ordered domain in the bulk exerts a stronger influence, whereas for small $x$ and small $v$, the ordered domain at the boundary becomes more significant. This behavior is similar to that observed in ordinary transitions.

A key difference between the ordinary transition and the special transition, however, emerges in the $x$-dependence of $M^{2}_{sx}$: for a fixed $v$, in ordinary transition, $M_{sx}^2$ increases with increasing $x$ (the $x=1$ curve lies at the bottom in Fig.~\ref{fig:fm3dord}(c1)); while in special transition, it decreases with increasing $x$ (the $x=1$ curve lies at the top in Fig.~\ref{fig:fm3dsp}(c1)). This contrast arises from the relation between exponents: in the special transition, $\beta_1<\beta$, whereas in the ordinary transition, $\beta_1>\beta$.
Despite this difference, the scaling collapse in Fig.~\ref{fig:fm3dsp}(c2) indicates that for large $L$ the crossover behaviors can be described by
\begin{equation}
\label{sLoporderspe}
M_{sx}^2(v)=v^{2 \beta/\nu r}f_{13}(xv^{1/r}),
\end{equation}
analogous to Eq.~(\ref{sLoporderx}). For large $x$ and large $v$, $M^2_{sx}$ reduces to the bulk behavior $M_{sx}^2\propto v^{2 \beta/\nu r}$ and $f_{13}(xv^{1/r})$ tends to a constant, similar to the situation Eq.~(\ref{sLoporderx}). However, the asymptotic behavior for small $x$ and small $v$ is different.
As shown in Fig.~\ref{fig:fm3dsp}(c2), $f_{13}(\tilde{x})\propto \tilde{x}^{2(\beta_1-\beta)/\nu}$ decreases with $\tilde{x}$, in which $\tilde{x}=xq(v)^{1/r}$ and $q(v)=v(1+cv^{-\omega/r})$ with a driving-rate correction term introduced~\cite{Liu2025cpb}, while $f_6(xv^{1/r})$ increases with $xv^{1/r}$, as shown in Figs.~\ref{fig:fm2d}(c2) and \ref{fig:fm3dord}(c2).

Combining the FTS forms of Eqs.~(\ref{sLoporder}), (\ref{oporderspeT}), and (\ref{sLoporderspe}), the full scaling FTS form for the heating dynamics in special transition is
\begin{equation}
  \label{fullsLoporderspe}
  M_{sx}^2(g,L,v)=v^{2 \beta/\nu r}f_{14}(gv^{-1/\nu r},vL^{r},xv^{1/r}).
\end{equation}
For large $xv^{1/r}$, Eq.~(\ref{fullsLoporderspe}) reduces Eq.~(\ref{sLoporder}), while for small $xv^{1/r}$, it recovers the BFTS of Eq.~(\ref{oporderspeT}).

\subsubsection{\label{speTde} Cooling dynamics}

Given the similarity between the special transition and ordinary transition in heating dynamics, it is natural to ask whether the same applies to cooling.
For the cooling case, the boundary coupling is fixed at $J_{sc}=1.50243$~\cite{Hasenbusch2011prb_3} and the initial temperature is $T_0=T_b+2$, corresponding to the disordered phase.

We first focus on the case at $x=1$. Referring to Eq.~(\ref{sLopdisorder}), for the cooling dynamics from the disordered phase, the {\it normal} generalization by replacing $\beta$ of Eq.~(\ref{sLopdisorder}) as $\beta_1$ leads to the BFTS form
\begin{equation}
  \label{opdisorderspec}
  \begin{split}
    M_{s1}^2(g,L,v)&=L^{-(d-1)}v^{2 \beta_1/\nu r-(d-1)/r}\\
    &\times f_{15}(gv^{-1/\nu r},vL^{r}),
  \end{split}
\end{equation}
which appears to be similar to Eq.~(\ref{sLopdisorderx1}) in Sec.~\ref{ordc}.
\begin{figure}[tbp]
\centering
  \includegraphics[width=\linewidth,clip]{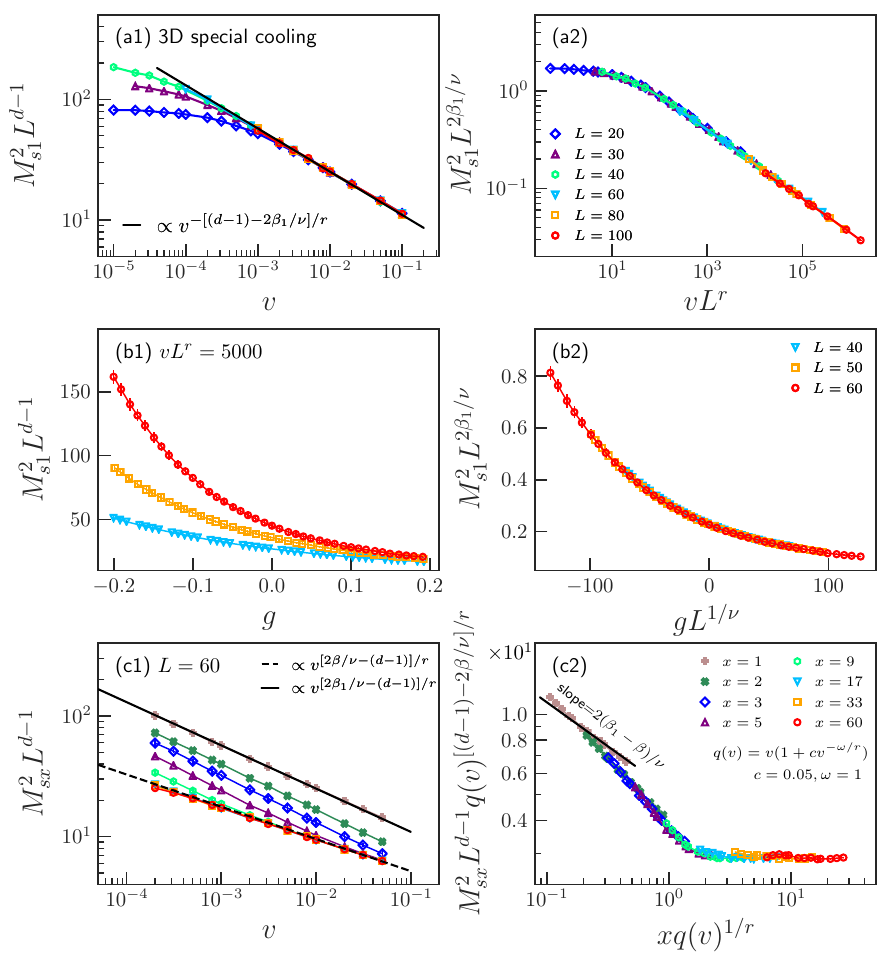}
  \vskip-3mm
  \caption{Cooling dynamics in $3$D special transition at $T=T_b$ with $J_s=J_{sc}$. The initial state is set at $T_0=T_b+2$.
      (a1) Dependence of $M^{2}_{s1}L^{d-1}$ on $v$. The solid line is a power law with the exponent $2\beta_1/\nu r$, with $\beta_1=0.2227$ being the boundary order parameter exponent of the $3$D Ising special universality class.
    (a2) Rescaled curves of $M^{2}_{s1}L^{2\beta_1/\nu}$ versus $vL^{r}$ to verify the BFTS form at $g=0$.
    (b1) Dependence of $M^{2}_{s1}$ on $g$ for an arbitrary fixed $vL^{r}$.
    (b2) Rescaled curves of $M^{2}_{s1} L^{2\beta_1/\nu}$ versus $gL^{1/\nu}$ to verify the BFTS form for $g\neq 0$.
    (c1) Dependence of $M^{2}_{sx}$ on $v$ for different $x$ with $L=60$.
    The solid line indicates a power law of $v^{[2\beta_1/\nu-(d-1)]/r}$ for small $x$ and the dashed line represents a power law $v^{[2\beta/\nu-(d-1)]/r}$ for large $x$. Crossover behaviors in between are observed.
    (c2) Rescaled curves of $M^{2}_{sx}L^{d-1}q(v)^{[(d-1)-2\beta/\nu]/r}$ versus $xq(v)^{1/r}$ for data collapse, where $q(v)=v(1+cv^{-\omega/r})$ includes a driving-rate correction. The solid line is a power law of $[xq(v)^{1/r}]^{2(\beta_1-\beta)/\nu}$ expected by full BFTS form.
    Linear scales are used in (b1) and (b2), while other plots use log-log scales.
  }
  \label{fig:pm3dsp}
\end{figure}
However, there is a key difference. As discussed in Sec.~\ref{ordc}, the counterpart of Eq.~(\ref{opdisorderspec}), namely Eq.~(\ref{sLopdisorderx1}), fails to capture the cooling dynamics in ordinary transition, since $2 \beta_1/\nu r-(d-1)/r\geq 0$ for the ordinary transition, which is contradictory to numerical evidence that $M^2_{x1}$ decreases with increasing $v$. Differently, here $2 \beta_1/\nu r-(d-1)/r< 0$ for the special transition, suggesting that the scaling form may be valid in this case.
 
To verify Eq.~(\ref{opdisorderspec}), we examine the dependence of $M_{s1}^2$ on $v$ for different $L$. For large $v$ and $g=0$, Fig.~\ref{fig:pm3dsp}(a1) shows that $M_{s1}^2L^{(d-1)}\propto v^{2 \beta_1/\nu r-(d-1)/r}$, consistent with Eq.~(\ref{opdisorderspec}). In addition, the rescaled curves collapse well following Eq.~(\ref{opdisorderspec}) with $g=0$, as shown in Fig.~\ref{fig:pm3dsp}(a2).
Around the critical point, fixing $vL^{r}$ to a constant, we further compute the dependence of $M_{s1}^2$ on $g$ for different $L$, as shown in Fig.~\ref{fig:pm3dsp}(b1). The successful collapse of the rescaled curves in Fig.~\ref{fig:pm3dsp}(b2) confirms the BFTS form Eq.~(\ref{opdisorderspec}) for the special transition.

\begin{figure}[!htbp]
\centering
  \includegraphics[width=0.8\linewidth,clip]{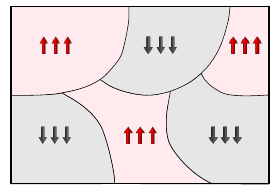}
  \vskip-3mm
  \caption{Illustration of boundary domain formation at the special transition, where the stronger coupling compensates the reduced coordination and the enhanced fluctuations, enabling bulk-comparable ordering.
  }
  \label{fig:illuspe}
\end{figure}

Comparing with the ordinary transition, the validity of Eq.~(\ref{opdisorderspec}) implies that domains can form at the boundary in the special transition, as illustrated in Fig.~\ref{fig:illuspe}. The special transition lies at the meeting point of the ordinary transition---where the boundary coupling is too weak to support domain formation during cooling, as discussed in Sec.~\ref{ordc}---and the surface transition---where the boundary coupling is strong enough to induce ordering before the bulk.
Accordingly, at the special transition, the ability to form domains at the boundary becomes comparable to that in the bulk, as the sufficiently strong boundary coupling compensates for the reduced coordination number.

The enhanced ordering ability at the boundary suggests that nontrivial scaling behavior may arise near the boundary.
To examine this, we compute $M_{sx}^2$ as a function of $v$ for different $x$ after cooling to the special transition point with a sufficiently large $L$. As shown in Fig.~\ref{fig:pm3dsp}(c1), for large $v$ and large $x$, $M_{sx}^2$ follows $M_{sx}^2\propto L^{-(d-1)}v^{2 \beta/\nu r-(d-1)/r}$, while for small $x$ and small $x$, it crosses over to $M_{sx}^2\propto L^{-(d-1)}v^{2 \beta_1/\nu r-(d-1)/r}$. This crossover can be captured by the scaling form
\begin{equation}
\label{opdisorderspecx}
M_{sx}^2(v)=L^{-(d-1)}v^{2 \beta/\nu r-(d-1)/r}f_{16}(xv^{1/r}).
\end{equation}
Note that here it is expected that $f_{16}(xv^{1/r})$ behaves as $(xv^{1/r})^{2(\beta_1-\beta)/\nu}$ for small $xv^{1/r}$ to fit the crossover between the scaling behaviors in the bulk and at the surface. However, a remarkable scaling correction needs to be considered as demonstrated by the deviation of the rescaling curve and the curve of $(xv^{1/r})^{2(\beta_1-\beta)/\nu}$. Nevertheless, with a driving-rate correction included, the good scaling collapse in Fig.~\ref{fig:pm3dsp}(c2) can confirm Eq.~(\ref{opdisorderspecx}).

Considering the three scaling forms Eqs.~(\ref{sLopdisorder}), (\ref{opdisorderspec}) and (\ref{opdisorderspecx}) together, we arrive at the full FTS form of the special transition:
\begin{equation}
  \label{fullsdisorderspe}
  \begin{split}
    M_{sx}^2(g,L,v)&=L^{-(d-1)}v^{2 \beta/\nu r-(d-1)/r}\\
    &\times f_{17}(gv^{-1/\nu r},vL^{r},xv^{1/r}).
  \end{split}
\end{equation}
For large $xv^{1/r}$, Eq.~(\ref{fullsdisorderspe}) reduces Eq.~(\ref{sLopdisorder}); while for small $xv^{1/r}$, it recovers the BFTS of Eq.~(\ref{opdisorderspec}).

\subsection{\label{speJ} Changing coupling}

\subsubsection{\label{speJde} Decreasing coupling}
Here, we consider the driven dynamics of the special transition point by linearly decreasing $J_s$ from $J_{s0}=J_{sc}+1$ across the critical value $J_{sc}$. The temperature is fixed at $T_b$ during the driving process.
In the initial state, the bulk is at the critical point, while the surface is in the ordered phase, since $T_b$ is below the surface transition point $T_s$.
This satisfies the original adiabatic-impulse scenario of the KZM. Accordingly, the BFTS form should be analogous to Eq.~(\ref{oporderspeT}), with the difference that the dimension of $g_J=J_s-J_{sc}$ differs from that of $g=T-T_c$. It has been shown that $\dim(g_J)=\phi \dim(g)$ with $\phi\simeq 0.52$~\cite{Lin2008pre,Landau1990prb}. As a result, the scaling dimension of $g_J$ is $\nu_J\equiv \nu/\phi$, and the driving rate $v$ has an effective dimension of $r_J=z+\phi/\nu$. These scaling analyses lead to the BFTS form for increasing $J$,
\begin{equation}
\label{oporderspeJ}
M_{s1}^2(g,L,v)=v^{2 \beta_1/\nu r_J}f_{12}(g_Jv^{-1/\nu_J r_J},vL^{r_{J}}).
\end{equation}
Note that the dimension of $M_{s1}^2$ at the special point is $\beta_1/\nu$. Thus $\nu$ in the exponent of $v$ before $f_{12}$ should not be replaced by $\nu_J$.

\begin{figure}[!htbp]
\centering
  \includegraphics[width=\linewidth,clip]{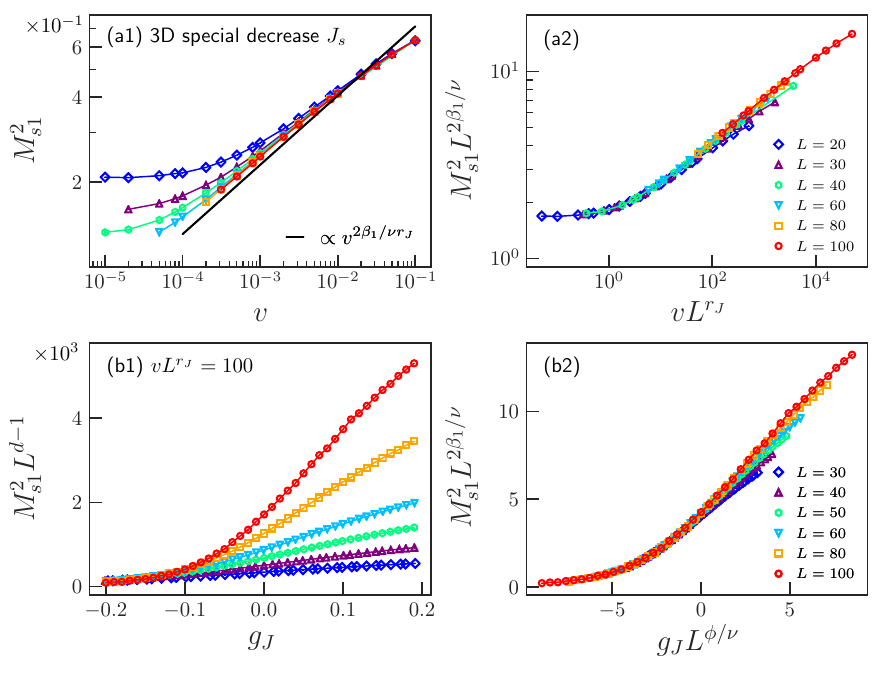}
  \vskip-3mm
  \caption{Driven dynamics in $3$D special transition at $T=T_b$ with decreasing $J_s$ from $J_{s0}=J_{sc}+1$ to $J_{sc}$.
    (a1) Dependence of $M^{2}_{s1}$ on $v$. The solid line represents a power law of $v^{2\beta_{1}/\nu r_{J}}$ where $r_J=z+\phi/\nu$ with $\phi\simeq 0.52$.
    (a2) Rescaled curves of $M^{2}_{s1}L^{2\beta/\nu}$ versus $vL^{r_{J}}$ to verify the BFTS form at $g_J=0$, with $g_J=J_{sc}-J_{s}$.
    (b1) Dependence of $M^{2}_{s1}$ on $g_J$ for an arbitrary fixed $vL^{r_{J}}$.
    (b2) After rescaling the y-axis as in (a2), and $g_J$ as $g_JL^{1/\nu}$, the good collapse confirms the BFTS form for $g_J\neq 0$.
    Log-log scales are used in (a1) and (a2), while (b1) and (b2) use linear scales.
  }
  \label{fig:fm3dspdeJ}
\end{figure}

The validity of Eq.~(\ref{oporderspeJ}) is shown in Fig.~\ref{fig:fm3dspdeJ}, similar to the case of the heating dynamics in Sec.~\ref{speTin}.
Figure~\ref{fig:fm3dspdeJ}(a1) shows that for large $v$ and $g_J=0$, $M_{s1}^2\propto v^{2 \beta_1/\nu r_J}$. The scaling collapse in Fig.~\ref{fig:fm3dspdeJ}(a2) confirms Eq.~(\ref{oporderspeJ}) at $g_J=0$.
To verify the case of $g_J\neq 0$, we show the dependence of $M^{2}_{s1}$ on $g$ with fixed $vL^{r}$ and perform rescaling in Fig.~\ref{fig:fm3dspdeJ}(b1) and (b2), respectively. The good collapse again supports Eq.~(\ref{oporderspeJ}).

\subsubsection{\label{speJin} Increasing coupling}
Here we study the case for increasing coupling from $J_s<J_{sc}$. In this case, the equilibrium states for the initial parameters with $J_{s0}<J_{sc}$ and $T=T_b$ are critical in both the bulk and the surface. In particular, the latter corresponds to the critical state of ordinary transition. Since both states have divergent correlation time scales, the prerequisite of the original KZM, which requires an adiabatic initial stage, breaks down.

In general, in the thermodynamic limit, the critical system cannot equilibrate as a result of the critical slowing down. Thus, it is instructive to consider the case for driving the system from a nonequilibrium state. 

Here, we focus on the driven dynamics from a nonequilibrium state, which corresponds to the state of the system that relaxes from a completely ordered state for a time $t_a$ under the parameters $T=T_b$ and $J_s=J_{s0}<J_{sc}$. $t_a$ is usually referred to as the waiting time, or the ``age" of the system. Then, $J_s$ is increased linearly across the special point. Apparently, the driven dynamics for this case is beyond the adiabatic-impulse scenario of the KZM.

We construct a general scaling theory for this case in the following way. At $g_J=0$, for large driving rate, $M_{s1}^2$ should retain memory of the initial state at $t_a$ and satisfy $M_{s1}^2\propto t^{-2\beta^{o}_{1}/\nu}$. Here, to avoid confusion, we use $\beta^o_1$ and $\beta_1$ to denote the boundary order parameter exponent of the ordinary transition and special transition, respectively. Meanwhile, $M_{s1}^2$ should also reflect the scaling dimension of the special point and depend on the driving rate $v$. Accordingly, we can assume the scaling form as
\begin{equation}
\label{opdisorderspeJ}
\begin{split}
M_{s1}^2(g,v,t_a)&=t_a^{-2\beta^{o}_{1}/\nu z}v^{2(\beta_1-\beta^{o}_{1})/\nu r_J}\\
&\times f_{15}(g_Jv^{-1/\nu_J r_J},vt_a^{r_J/z}),
\end{split}
\end{equation}
for systems with a large size $L$ where finite-size effects can be neglected.

\begin{figure}[!htbp]
\centering
  \includegraphics[width=\linewidth,clip]{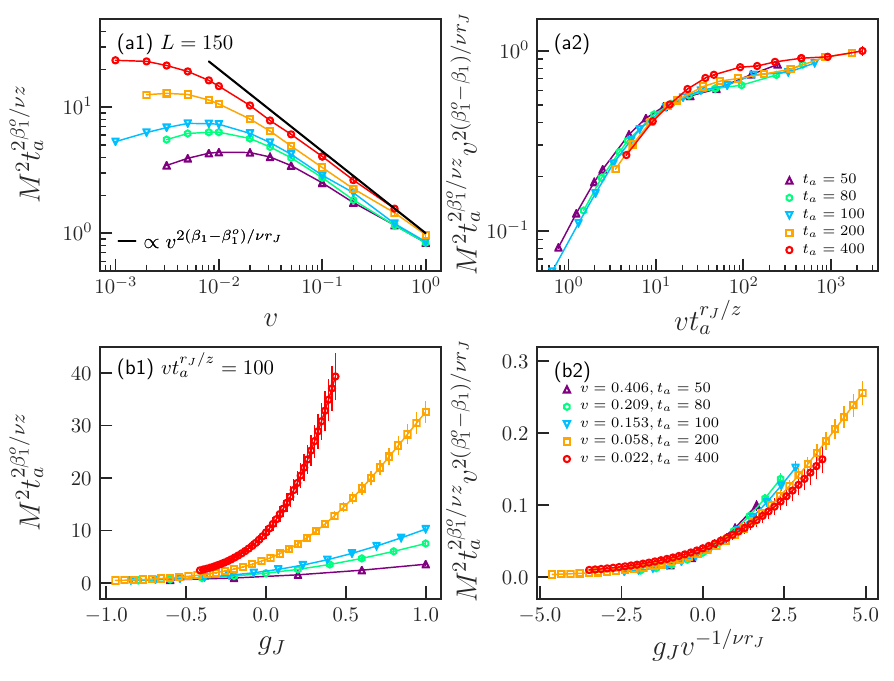}
  \vskip-3mm
  \caption{Driven dynamics in $3$D special transition at $T=T_b$ with increasing $J_s$ from $J_{s0}=J_{sc}-1$ to $J_{sc}$ for $L=150$. The initial state is chosen as the state after relaxing from the ordered state with time $t_a$.
    (a1) Dependence of $M^{2}_{s1}t_a^{2\beta^{o}_{1}/\nu z}$ on $v$ for different $t_a$. The solid line represents a power law of $v^{2(\beta_{1}-\beta^{o}_{1})/\nu r_{J}}$ where $r_J=z+\phi/\nu$ with $\phi\simeq 0.52$.
    (a2) Rescaled curves of $M^{2}_{s1}t_a^{2\beta^{o}_{1}/\nu z}v^{2(\beta^{o}_{1}-\beta_{1})/\nu r_{J}}$ versus $vt_a^{r_{J}/z}$ to verify the BFTS form at $g_J=0$, with $g_J=J_{sc}-J_{s}$.
    (b1) Dependence of $M^{2}_{s1}$ on $g_J$ for fixed $vt_a^{r_{J}/z}$ for different $t_a$.
    (b2) After rescaling the y-axis as in (a2), and $g_J$ as $g_Jv^{-1/\nu r_{J}}$, the good collapse confirms the BFTS form for $g_J\neq 0$.
    Log-log scales are used in (a1) and (a2), while (b1) and (b2) use linear scales.
  }
  \label{fig:inJ}
\end{figure}

To verify Eq.~(\ref{opdisorderspeJ}), we first focus on the case for $g_J=0$. The initial state is set as $J_{s0}=J_{sc}-1$ and $T=T_b$. As shown in Fig.~\ref{fig:inJ}(a1), we find that for large $v$, the curves of $M^2_{s1}t_a^{2\beta^{o}_{1}/\nu z}$ for different $t_a$ are very close, indicating that $M^2_{s1}\propto t_a^{-2\beta^{o}_1/\nu z}$. Moreover, $M^2_{s1}t_a^{2\beta_1^{o}/\nu z}$ is almost proportional to $v^{2(\beta_1-\beta^{o}_{1})/\nu r_J}$ for large $v$. Taking these factors into account, one finds that $M_{s1}^2\propto t_a^{-2\beta^{o}_{1}/\nu z}v^{2(\beta_1-\beta_{1}^{o})/\nu r_J}$. After rescaling $M_{s1}^2$ and $v$ according to Eq.~(\ref{opdisorderspeJ}), we find that the rescaled curves collapse onto each other as shown in Fig.~\ref{fig:inJ}(a2). The slight deviation of the rescaled curves implies potential scaling corrections. In addition, Eq.~(\ref{opdisorderspeJ}) is further verified for the case with $g\neq 0$ as shown in Figs.~\ref{fig:inJ}(b1) and (b2).

Some remarks on Eq.~(\ref{opdisorderspeJ}) are as follows: (i) For a finite-size system, after a large waiting time $t_a$, the initial state at the boundary is a equilibrium critical state characterized by $L$. In this case, $t_a$ in Eq.~(\ref{opdisorderspeJ}) should be replaced by $L^{z}$. Similar scaling forms are discussed in Refs.~\cite{Wangtl2025arx,Zeng2025nc,Zeng2025prb,Wang2025arx}.
(ii) The scaling form in Eq.~(\ref{opdisorderspeJ}) is also different from the case of driven dynamics from a nonequilibrium state near the critical point~\cite{Huangyy2016prb}. In the latter case, only one set of critical exponents should be included; while for Eq.~(\ref{opdisorderspeJ}), two sets of critical exponents for both ordinary and special transitions are included.

\section{\label{ext}Driven dynamics in the extraordinary transition}

In this section, we investigate the driven dynamics of the extraordinary transition. The extraordinary transition occurs for $J_s>J_{sc}$ and $T=T_b$. At $T_b$ and large $J_s$, the boundary is in the ordered phase. Here, we mainly focus on the driven dynamics near the ordered boundary.

Near the boundary, the order parameter is strongly influenced by the ordered boundary and expected to follow the behavior of $M_{sx}^2\propto x^{-2\beta/\nu}$. Thus, the full FTS form should be
\begin{equation}
\label{oporderext}
M_{sx}^2(g,L,v)=x^{-2\beta/\nu}f_{16}(vx^{r},gx^{-1/\nu},xL^{-1}).
\end{equation}
Without loss of generality and for conciseness, we verify Eq.~(\ref{oporderext}) for the heating dynamics driven from $T_0=T_b-2$ to $T_b$ at $J_{s}=2$, assuming a sufficiently large $L$ such that the finite-size effects are negligible.

\begin{figure}[tbp]
\centering
  \includegraphics[width=\linewidth,clip]{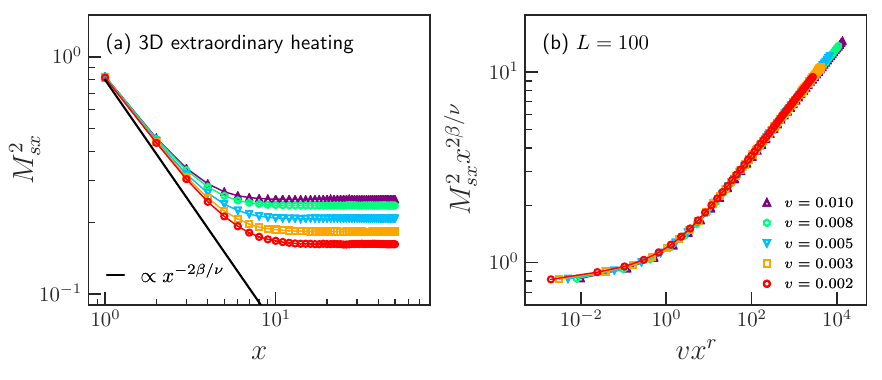}
  \vskip-3mm
  \caption{Heating dynamics in $3$D extraordinary transition at $T=T_b$ and $J_s=2$. The initial state is set at $T_0=T_b-2$.
    (a) Dependence of $M^{2}_{sx}$ on $x$ for $L=100$ with fixed $vL^{r}=50$ at $g=0$. The solid line represents a power law of $x^{-2\beta/\nu}$.
    (b) Rescaled curves of $M^{2}_{sx}x^{2\beta/\nu}$ versus $vx^r$. The good collapse confirms the scaling form in Eq.~(\ref{oporderext}).
  }
  \label{fig:fm3dextra}
\end{figure}

Figure~\ref{fig:fm3dextra}(a) shows the dependence of $M_{sx}^2$ on $x$ at $g=0$ with fixed $vL^r=50$. $M_{sx}^2$ asymptotically follows a power-law decay of $M_{sx}^2 \propto x^{-2\beta/\nu}$, consistent with the expectation that the ordered boundary dominates the scaling behavior near the surface. In Fig.~\ref{fig:fm3dextra}(b), the rescaled curves of $M_{sx}^2x^{2\beta/\nu}$ versus $vx^r$ collapse well, verifying the scaling form in Eq.~(\ref{oporderext}). These results demonstrate that the extraordinary transition exhibits scaling properties controlled by the bulk exponent $\beta$, with the boundary order providing the prefactor $x^{-2\beta/\nu}$ in Eq.~(\ref{oporderext}).

\section{\label{sum}Summary and Discussion}

In this paper, we have systematically investigated the driven critical dynamics of boundary universality classes in the $2$D and $3$D Ising models, including the ordinary, surface, special, and extraordinary transitions. By extending the FTS theory to the boundary, we developed the BFTS framework and examined its applicability under different driving protocols. For heating dynamics, all boundary universality classes are consistently described by the BFTS forms, with a clear crossover between bulk and boundary exponents. For cooling dynamics, however, the ordinary transition shows a breakdown of the normal BFTS and instead exhibits a logarithmic dependence on the driving rate, while surface, special, and extraordinary transitions remain consistent with BFTS. For the special transition, we further studied the case for changing the surface coupling. For decreasing $J_s$, we found the driven dynamics obeys BFTS with a modified driving dimension. For increasing $J_s$, we proposed the scaling form to incorporate the nonequilibrium state characterized by the waiting time. Finally, in the extraordinary transition, the ordered boundary enforces a power-law decay of the order parameter into the bulk, consistent with scaling theory.

Our results indicate that a core criterion for the applicability of the BFTS for cooling process is that whether the value of $2\beta_1/\nu-(d-1)$ is negative. For $2\beta_1/\nu-(d-1)<0$, the surface domain can form and the {\it normal} BFTS works; while for $2\beta_1/\nu-(d-1)\ge 0$, the surface fluctuations are strong enough to prevent the formation of domains and the {\it abnormal} logarithmic dynamic behaviors can appear.

Possible extensions of our present work include generalizations to quantum phase transitions, where non-trivial edge states may emerge in topological phases and their driven dynamics calls for further exploration~\cite{Zhang2017prl,Zhu2025arx,Wang2024prb,Liu2024prl,Liu2025arx,Toldin2025arx,Wu2020prb,Scaffidi2017prx,Ding2018prl,Weber2018prb,Verresen2021prx,Zhu2021prb,Yu2022prl}. It is also natural to consider boundary criticality in Dirac systems, where both exotic boundary phase transitions~\cite{Li2017prl,Grover2014sci} and novel critical phenomena near the boundary have been proposed~\cite{Ge2025arx,Jiang2025arx}. In addition, the recently uncovered extraordinary-log universality class in the O($N$) model~\cite{Metlitski2022sp} and other systems\cite{Toldin2021prl,Hu2021prl,Hu2025arx,Toldin2022prl,Padayasi2022sp,Sun2022prb,Zhang2022prb,Sun2022prb_log,Sun2025sp}, where the surface order parameter shows a logarithmic dependence, provides a further direction to explore the driven dynamics.

\section*{Acknowledgements}
We would like to thank Yi-Fan Jiang and Zi-Xiang Li for helpful discussions. S.Y. is supported by the National Natural Science Foundation of China (Grant No. 12222515) and the Science and Technology Projects in Guangdong Province (Grant No. 2021QN02X561) and Guangzhou City (Grant No. 2025A04J5408). Y.R.S. acknowledges support from the National Natural Science Foundation of China (Grant No. 12104109), the Science and Technology Projects in Guangzhou (Grant No. 2024A04J2092).

\bibliography{refzq}

\begin{thebibliography}{111}%
\makeatletter
\providecommand \@ifxundefined [1]{%
 \@ifx{#1\undefined}
}%
\providecommand \@ifnum [1]{%
 \ifnum #1\expandafter \@firstoftwo
 \else \expandafter \@secondoftwo
 \fi
}%
\providecommand \@ifx [1]{%
 \ifx #1\expandafter \@firstoftwo
 \else \expandafter \@secondoftwo
 \fi
}%
\providecommand \natexlab [1]{#1}%
\providecommand \enquote  [1]{``#1''}%
\providecommand \bibnamefont  [1]{#1}%
\providecommand \bibfnamefont [1]{#1}%
\providecommand \citenamefont [1]{#1}%
\providecommand \href@noop [0]{\@secondoftwo}%
\providecommand \href [0]{\begingroup \@sanitize@url \@href}%
\providecommand \@href[1]{\@@startlink{#1}\@@href}%
\providecommand \@@href[1]{\endgroup#1\@@endlink}%
\providecommand \@sanitize@url [0]{\catcode `\\12\catcode `\$12\catcode
  `\&12\catcode `\#12\catcode `\^12\catcode `\_12\catcode `\%12\relax}%
\providecommand \@@startlink[1]{}%
\providecommand \@@endlink[0]{}%
\providecommand \url  [0]{\begingroup\@sanitize@url \@url }%
\providecommand \@url [1]{\endgroup\@href {#1}{\urlprefix }}%
\providecommand \urlprefix  [0]{URL }%
\providecommand \Eprint [0]{\href }%
\providecommand \doibase [0]{http://dx.doi.org/}%
\providecommand \selectlanguage [0]{\@gobble}%
\providecommand \bibinfo  [0]{\@secondoftwo}%
\providecommand \bibfield  [0]{\@secondoftwo}%
\providecommand \translation [1]{[#1]}%
\providecommand \BibitemOpen [0]{}%
\providecommand \bibitemStop [0]{}%
\providecommand \bibitemNoStop [0]{.\EOS\space}%
\providecommand \EOS [0]{\spacefactor3000\relax}%
\providecommand \BibitemShut  [1]{\csname bibitem#1\endcsname}%
\let\auto@bib@innerbib\@empty
\bibitem [{\citenamefont {Kibble}(1976)}]{Kibble1976}%
  \BibitemOpen
  \bibfield  {author} {\bibinfo {author} {\bibfnamefont {T.~W.~B.}\
  \bibnamefont {Kibble}},\ }\bibfield  {title} {\enquote {\bibinfo {title}
  {{Topology of cosmic domains and strings}},}\ }\href {\doibase
  10.1088/0305-4470/9/8/029} {\bibfield  {journal} {\bibinfo  {journal}
  {Journal of Physics A: Mathematical and General}\ }\textbf {\bibinfo {volume}
  {9}},\ \bibinfo {pages} {1387} (\bibinfo {year} {1976})}\BibitemShut
  {NoStop}%
\bibitem [{\citenamefont {Zurek}(1985)}]{Zurek1985}%
  \BibitemOpen
  \bibfield  {author} {\bibinfo {author} {\bibfnamefont {W.~H.}\ \bibnamefont
  {Zurek}},\ }\bibfield  {title} {\enquote {\bibinfo {title} {{Cosmological
  experiments in superfluid helium?}}}\ }\href {\doibase 10.1038/317505a0}
  {\bibfield  {journal} {\bibinfo  {journal} {Nature}\ }\textbf {\bibinfo
  {volume} {317}},\ \bibinfo {pages} {505--508} (\bibinfo {year}
  {1985})}\BibitemShut {NoStop}%
\bibitem [{\citenamefont {Zurek}\ \emph {et~al.}(2005)\citenamefont {Zurek},
  \citenamefont {Dorner},\ and\ \citenamefont {Zoller}}]{Zoller2005prl}%
  \BibitemOpen
  \bibfield  {author} {\bibinfo {author} {\bibfnamefont {W.~H.}\ \bibnamefont
  {Zurek}}, \bibinfo {author} {\bibfnamefont {U.}~\bibnamefont {Dorner}}, \
  and\ \bibinfo {author} {\bibfnamefont {P.}~\bibnamefont {Zoller}},\
  }\bibfield  {title} {\enquote {\bibinfo {title} {{Dynamics of a Quantum Phase
  Transition}},}\ }\href {\doibase 10.1103/PhysRevLett.95.105701} {\bibfield
  {journal} {\bibinfo  {journal} {Phys. Rev. Lett.}\ }\textbf {\bibinfo
  {volume} {95}},\ \bibinfo {pages} {105701} (\bibinfo {year}
  {2005})}\BibitemShut {NoStop}%
\bibitem [{\citenamefont {Dziarmaga}(2005)}]{Dziarmaga2005prl}%
  \BibitemOpen
  \bibfield  {author} {\bibinfo {author} {\bibfnamefont {J.}~\bibnamefont
  {Dziarmaga}},\ }\bibfield  {title} {\enquote {\bibinfo {title} {{Dynamics of
  a Quantum Phase Transition: Exact Solution of the Quantum Ising Model}},}\
  }\href {\doibase 10.1103/PhysRevLett.95.245701} {\bibfield  {journal}
  {\bibinfo  {journal} {Phys. Rev. Lett.}\ }\textbf {\bibinfo {volume} {95}},\
  \bibinfo {pages} {245701} (\bibinfo {year} {2005})}\BibitemShut {NoStop}%
\bibitem [{\citenamefont {Polkovnikov}(2005)}]{PhysRevB.72.161201}%
  \BibitemOpen
  \bibfield  {author} {\bibinfo {author} {\bibfnamefont {A.}~\bibnamefont
  {Polkovnikov}},\ }\bibfield  {title} {\enquote {\bibinfo {title} {{Universal
  adiabatic dynamics in the vicinity of a quantum critical point}},}\ }\href
  {\doibase 10.1103/PhysRevB.72.161201} {\bibfield  {journal} {\bibinfo
  {journal} {Phys. Rev. B}\ }\textbf {\bibinfo {volume} {72}},\ \bibinfo
  {pages} {161201} (\bibinfo {year} {2005})}\BibitemShut {NoStop}%
\bibitem [{\citenamefont {Lin}\ \emph {et~al.}(2014)\citenamefont {Lin},
  \citenamefont {Wang}, \citenamefont {Kamiya}, \citenamefont {Chern},
  \citenamefont {Fan}, \citenamefont {Fan}, \citenamefont {Casas},
  \citenamefont {Liu}, \citenamefont {Kiryukhin}, \citenamefont {Zurek},
  \citenamefont {Batista},\ and\ \citenamefont {Cheong}}]{Lin2014natphy}%
  \BibitemOpen
  \bibfield  {author} {\bibinfo {author} {\bibfnamefont {S.-Z.}\ \bibnamefont
  {Lin}}, \bibinfo {author} {\bibfnamefont {X.}~\bibnamefont {Wang}}, \bibinfo
  {author} {\bibfnamefont {Y.}~\bibnamefont {Kamiya}}, \bibinfo {author}
  {\bibfnamefont {G.-W.}\ \bibnamefont {Chern}}, \bibinfo {author}
  {\bibfnamefont {F.}~\bibnamefont {Fan}}, \bibinfo {author} {\bibfnamefont
  {D.}~\bibnamefont {Fan}}, \bibinfo {author} {\bibfnamefont {B.}~\bibnamefont
  {Casas}}, \bibinfo {author} {\bibfnamefont {Y.}~\bibnamefont {Liu}}, \bibinfo
  {author} {\bibfnamefont {V.}~\bibnamefont {Kiryukhin}}, \bibinfo {author}
  {\bibfnamefont {W.~H.}\ \bibnamefont {Zurek}}, \bibinfo {author}
  {\bibfnamefont {C.~D.}\ \bibnamefont {Batista}}, \ and\ \bibinfo {author}
  {\bibfnamefont {S.-W.}\ \bibnamefont {Cheong}},\ }\bibfield  {title}
  {\enquote {\bibinfo {title} {{Topological defects as relics of emergent
  continuous symmetry and Higgs condensation of disorder in
  ferroelectrics}},}\ }\href {\doibase 10.1038/nphys3142} {\bibfield  {journal}
  {\bibinfo  {journal} {Nature Physics}\ }\textbf {\bibinfo {volume} {10}},\
  \bibinfo {pages} {970--977} (\bibinfo {year} {2014})}\BibitemShut {NoStop}%
\bibitem [{\citenamefont {Ko}\ \emph {et~al.}(2019)\citenamefont {Ko},
  \citenamefont {Park},\ and\ \citenamefont {Shin}}]{Ko2019}%
  \BibitemOpen
  \bibfield  {author} {\bibinfo {author} {\bibfnamefont {B.}~\bibnamefont
  {Ko}}, \bibinfo {author} {\bibfnamefont {J.~W.}\ \bibnamefont {Park}}, \ and\
  \bibinfo {author} {\bibfnamefont {Y.}~\bibnamefont {Shin}},\ }\bibfield
  {title} {\enquote {\bibinfo {title} {{Kibble-Zurek universality in a strongly
  interacting Fermi superfluid}},}\ }\href {\doibase 10.1038/s41567-019-0650-1}
  {\bibfield  {journal} {\bibinfo  {journal} {Nature Physics}\ }\textbf
  {\bibinfo {volume} {15}},\ \bibinfo {pages} {1227--1231} (\bibinfo {year}
  {2019})}\BibitemShut {NoStop}%
\bibitem [{\citenamefont {Keesling}\ \emph {et~al.}(2019)\citenamefont
  {Keesling}, \citenamefont {Omran}, \citenamefont {Levine}, \citenamefont
  {Bernien}, \citenamefont {Pichler}, \citenamefont {Choi}, \citenamefont
  {Samajdar}, \citenamefont {Schwartz}, \citenamefont {Silvi}, \citenamefont
  {Sachdev}, \citenamefont {Zoller}, \citenamefont {Endres}, \citenamefont
  {Greiner}, \citenamefont {Vuleti{\'{c}}},\ and\ \citenamefont
  {Lukin}}]{Keesling2019}%
  \BibitemOpen
  \bibfield  {author} {\bibinfo {author} {\bibfnamefont {A.}~\bibnamefont
  {Keesling}}, \bibinfo {author} {\bibfnamefont {A.}~\bibnamefont {Omran}},
  \bibinfo {author} {\bibfnamefont {H.}~\bibnamefont {Levine}}, \bibinfo
  {author} {\bibfnamefont {H.}~\bibnamefont {Bernien}}, \bibinfo {author}
  {\bibfnamefont {H.}~\bibnamefont {Pichler}}, \bibinfo {author} {\bibfnamefont
  {S.}~\bibnamefont {Choi}}, \bibinfo {author} {\bibfnamefont {R.}~\bibnamefont
  {Samajdar}}, \bibinfo {author} {\bibfnamefont {S.}~\bibnamefont {Schwartz}},
  \bibinfo {author} {\bibfnamefont {P.}~\bibnamefont {Silvi}}, \bibinfo
  {author} {\bibfnamefont {S.}~\bibnamefont {Sachdev}}, \bibinfo {author}
  {\bibfnamefont {P.}~\bibnamefont {Zoller}}, \bibinfo {author} {\bibfnamefont
  {M.}~\bibnamefont {Endres}}, \bibinfo {author} {\bibfnamefont
  {M.}~\bibnamefont {Greiner}}, \bibinfo {author} {\bibfnamefont
  {V.}~\bibnamefont {Vuleti{\'{c}}}}, \ and\ \bibinfo {author} {\bibfnamefont
  {M.~D.}\ \bibnamefont {Lukin}},\ }\bibfield  {title} {\enquote {\bibinfo
  {title} {{Quantum Kibble--Zurek mechanism and critical dynamics on a
  programmable Rydberg simulator}},}\ }\href {\doibase
  10.1038/s41586-019-1070-1} {\bibfield  {journal} {\bibinfo  {journal}
  {Nature}\ }\textbf {\bibinfo {volume} {568}},\ \bibinfo {pages} {207--211}
  (\bibinfo {year} {2019})}\BibitemShut {NoStop}%
\bibitem [{\citenamefont {Maegochi}\ \emph {et~al.}(2022)\citenamefont
  {Maegochi}, \citenamefont {Ienaga},\ and\ \citenamefont
  {Okuma}}]{PhysRevLett.129.227001}%
  \BibitemOpen
  \bibfield  {author} {\bibinfo {author} {\bibfnamefont {S.}~\bibnamefont
  {Maegochi}}, \bibinfo {author} {\bibfnamefont {K.}~\bibnamefont {Ienaga}}, \
  and\ \bibinfo {author} {\bibfnamefont {S.}~\bibnamefont {Okuma}},\ }\bibfield
   {title} {\enquote {\bibinfo {title} {{Kibble-Zurek Mechanism for Dynamical
  Ordering in a Driven Vortex System}},}\ }\href {\doibase
  10.1103/PhysRevLett.129.227001} {\bibfield  {journal} {\bibinfo  {journal}
  {Phys. Rev. Lett.}\ }\textbf {\bibinfo {volume} {129}},\ \bibinfo {pages}
  {227001} (\bibinfo {year} {2022})}\BibitemShut {NoStop}%
\bibitem [{\citenamefont {Ebadi}\ \emph {et~al.}(2021)\citenamefont {Ebadi},
  \citenamefont {Wang}, \citenamefont {Levine}, \citenamefont {Keesling},
  \citenamefont {Semeghini}, \citenamefont {Omran}, \citenamefont {Bluvstein},
  \citenamefont {Samajdar}, \citenamefont {Pichler}, \citenamefont {Ho},
  \citenamefont {Choi}, \citenamefont {Sachdev}, \citenamefont {Greiner},
  \citenamefont {Vuleti{\'{c}}},\ and\ \citenamefont {Lukin}}]{Ebadi2021}%
  \BibitemOpen
  \bibfield  {author} {\bibinfo {author} {\bibfnamefont {S.}~\bibnamefont
  {Ebadi}}, \bibinfo {author} {\bibfnamefont {T.~T.}\ \bibnamefont {Wang}},
  \bibinfo {author} {\bibfnamefont {H.}~\bibnamefont {Levine}}, \bibinfo
  {author} {\bibfnamefont {A.}~\bibnamefont {Keesling}}, \bibinfo {author}
  {\bibfnamefont {G.}~\bibnamefont {Semeghini}}, \bibinfo {author}
  {\bibfnamefont {A.}~\bibnamefont {Omran}}, \bibinfo {author} {\bibfnamefont
  {D.}~\bibnamefont {Bluvstein}}, \bibinfo {author} {\bibfnamefont
  {R.}~\bibnamefont {Samajdar}}, \bibinfo {author} {\bibfnamefont
  {H.}~\bibnamefont {Pichler}}, \bibinfo {author} {\bibfnamefont {W.~W.}\
  \bibnamefont {Ho}}, \bibinfo {author} {\bibfnamefont {S.}~\bibnamefont
  {Choi}}, \bibinfo {author} {\bibfnamefont {S.}~\bibnamefont {Sachdev}},
  \bibinfo {author} {\bibfnamefont {M.}~\bibnamefont {Greiner}}, \bibinfo
  {author} {\bibfnamefont {V.}~\bibnamefont {Vuleti{\'{c}}}}, \ and\ \bibinfo
  {author} {\bibfnamefont {M.~D.}\ \bibnamefont {Lukin}},\ }\bibfield  {title}
  {\enquote {\bibinfo {title} {{Quantum phases of matter on a 256-atom
  programmable quantum simulator}},}\ }\href {\doibase
  10.1038/s41586-021-03582-4} {\bibfield  {journal} {\bibinfo  {journal}
  {Nature}\ }\textbf {\bibinfo {volume} {595}},\ \bibinfo {pages} {227--232}
  (\bibinfo {year} {2021})}\BibitemShut {NoStop}%
\bibitem [{\citenamefont {Du}\ \emph {et~al.}(2023)\citenamefont {Du},
  \citenamefont {Fang}, \citenamefont {Won}, \citenamefont {De}, \citenamefont
  {Huang}, \citenamefont {Xu}, \citenamefont {You}, \citenamefont
  {Gómez-Ruiz}, \citenamefont {del Campo},\ and\ \citenamefont
  {Cheong}}]{Du2023}%
  \BibitemOpen
  \bibfield  {author} {\bibinfo {author} {\bibfnamefont {K.}~\bibnamefont
  {Du}}, \bibinfo {author} {\bibfnamefont {X.}~\bibnamefont {Fang}}, \bibinfo
  {author} {\bibfnamefont {C.}~\bibnamefont {Won}}, \bibinfo {author}
  {\bibfnamefont {C.}~\bibnamefont {De}}, \bibinfo {author} {\bibfnamefont
  {F.-T.}\ \bibnamefont {Huang}}, \bibinfo {author} {\bibfnamefont
  {W.}~\bibnamefont {Xu}}, \bibinfo {author} {\bibfnamefont {H.}~\bibnamefont
  {You}}, \bibinfo {author} {\bibfnamefont {F.~J.}\ \bibnamefont
  {Gómez-Ruiz}}, \bibinfo {author} {\bibfnamefont {A.}~\bibnamefont {del
  Campo}}, \ and\ \bibinfo {author} {\bibfnamefont {S.-W.}\ \bibnamefont
  {Cheong}},\ }\bibfield  {title} {\enquote {\bibinfo {title} {{Kibble–Zurek
  mechanism of Ising domains}},}\ }\href {\doibase 10.1038/s41567-023-02112-5}
  {\bibfield  {journal} {\bibinfo  {journal} {Nature Physics}\ }\textbf
  {\bibinfo {volume} {19}},\ \bibinfo {pages} {1495--1501} (\bibinfo {year}
  {2023})}\BibitemShut {NoStop}%
\bibitem [{\citenamefont {Qiu}\ \emph {et~al.}(2020)\citenamefont {Qiu},
  \citenamefont {Liang}, \citenamefont {Yang}, \citenamefont {Yang},
  \citenamefont {Tian}, \citenamefont {Xu},\ and\ \citenamefont
  {Duan}}]{sciadv.aba7292}%
  \BibitemOpen
  \bibfield  {author} {\bibinfo {author} {\bibfnamefont {L.-Y.}\ \bibnamefont
  {Qiu}}, \bibinfo {author} {\bibfnamefont {H.-Y.}\ \bibnamefont {Liang}},
  \bibinfo {author} {\bibfnamefont {Y.-B.}\ \bibnamefont {Yang}}, \bibinfo
  {author} {\bibfnamefont {H.-X.}\ \bibnamefont {Yang}}, \bibinfo {author}
  {\bibfnamefont {T.}~\bibnamefont {Tian}}, \bibinfo {author} {\bibfnamefont
  {Y.}~\bibnamefont {Xu}}, \ and\ \bibinfo {author} {\bibfnamefont {L.-M.}\
  \bibnamefont {Duan}},\ }\bibfield  {title} {\enquote {\bibinfo {title}
  {{Observation of generalized Kibble-Zurek mechanism across a first-order
  quantum phase transition in a spinor condensate}},}\ }\href {\doibase
  10.1126/sciadv.aba7292} {\bibfield  {journal} {\bibinfo  {journal} {Science
  Advances}\ }\textbf {\bibinfo {volume} {6}},\ \bibinfo {pages} {eaba7292}
  (\bibinfo {year} {2020})}\BibitemShut {NoStop}%
\bibitem [{\citenamefont {Ebadi}\ \emph {et~al.}(2022)\citenamefont {Ebadi},
  \citenamefont {Keesling}, \citenamefont {Cain}, \citenamefont {Wang},
  \citenamefont {Levine}, \citenamefont {Bluvstein}, \citenamefont {Semeghini},
  \citenamefont {Omran}, \citenamefont {Liu}, \citenamefont {Samajdar},
  \citenamefont {Luo}, \citenamefont {Nash}, \citenamefont {Gao}, \citenamefont
  {Barak}, \citenamefont {Farhi}, \citenamefont {Sachdev}, \citenamefont
  {Gemelke}, \citenamefont {Zhou}, \citenamefont {Choi}, \citenamefont
  {Pichler}, \citenamefont {Wang}, \citenamefont {Greiner}, \citenamefont
  {Vuletić},\ and\ \citenamefont {Lukin}}]{science.abo6587}%
  \BibitemOpen
  \bibfield  {author} {\bibinfo {author} {\bibfnamefont {S.}~\bibnamefont
  {Ebadi}}, \bibinfo {author} {\bibfnamefont {A.}~\bibnamefont {Keesling}},
  \bibinfo {author} {\bibfnamefont {M.}~\bibnamefont {Cain}}, \bibinfo {author}
  {\bibfnamefont {T.~T.}\ \bibnamefont {Wang}}, \bibinfo {author}
  {\bibfnamefont {H.}~\bibnamefont {Levine}}, \bibinfo {author} {\bibfnamefont
  {D.}~\bibnamefont {Bluvstein}}, \bibinfo {author} {\bibfnamefont
  {G.}~\bibnamefont {Semeghini}}, \bibinfo {author} {\bibfnamefont
  {A.}~\bibnamefont {Omran}}, \bibinfo {author} {\bibfnamefont {J.-G.}\
  \bibnamefont {Liu}}, \bibinfo {author} {\bibfnamefont {R.}~\bibnamefont
  {Samajdar}}, \bibinfo {author} {\bibfnamefont {X.-Z.}\ \bibnamefont {Luo}},
  \bibinfo {author} {\bibfnamefont {B.}~\bibnamefont {Nash}}, \bibinfo {author}
  {\bibfnamefont {X.}~\bibnamefont {Gao}}, \bibinfo {author} {\bibfnamefont
  {B.}~\bibnamefont {Barak}}, \bibinfo {author} {\bibfnamefont
  {E.}~\bibnamefont {Farhi}}, \bibinfo {author} {\bibfnamefont
  {S.}~\bibnamefont {Sachdev}}, \bibinfo {author} {\bibfnamefont
  {N.}~\bibnamefont {Gemelke}}, \bibinfo {author} {\bibfnamefont
  {L.}~\bibnamefont {Zhou}}, \bibinfo {author} {\bibfnamefont {S.}~\bibnamefont
  {Choi}}, \bibinfo {author} {\bibfnamefont {H.}~\bibnamefont {Pichler}},
  \bibinfo {author} {\bibfnamefont {S.-T.}\ \bibnamefont {Wang}}, \bibinfo
  {author} {\bibfnamefont {M.}~\bibnamefont {Greiner}}, \bibinfo {author}
  {\bibfnamefont {V.}~\bibnamefont {Vuletić}}, \ and\ \bibinfo {author}
  {\bibfnamefont {M.~D.}\ \bibnamefont {Lukin}},\ }\bibfield  {title} {\enquote
  {\bibinfo {title} {{Quantum optimization of maximum independent set using
  Rydberg atom arrays}},}\ }\href {\doibase 10.1126/science.abo6587} {\bibfield
   {journal} {\bibinfo  {journal} {Science}\ }\textbf {\bibinfo {volume}
  {376}},\ \bibinfo {pages} {1209--1215} (\bibinfo {year} {2022})}\BibitemShut
  {NoStop}%
\bibitem [{\citenamefont {Sunami}\ \emph {et~al.}(2023)\citenamefont {Sunami},
  \citenamefont {Singh}, \citenamefont {Garrick}, \citenamefont {Beregi},
  \citenamefont {Barker}, \citenamefont {Luksch}, \citenamefont {Bentine},
  \citenamefont {Mathey},\ and\ \citenamefont {Foot}}]{science.abq6753}%
  \BibitemOpen
  \bibfield  {author} {\bibinfo {author} {\bibfnamefont {S.}~\bibnamefont
  {Sunami}}, \bibinfo {author} {\bibfnamefont {V.~P.}\ \bibnamefont {Singh}},
  \bibinfo {author} {\bibfnamefont {D.}~\bibnamefont {Garrick}}, \bibinfo
  {author} {\bibfnamefont {A.}~\bibnamefont {Beregi}}, \bibinfo {author}
  {\bibfnamefont {A.~J.}\ \bibnamefont {Barker}}, \bibinfo {author}
  {\bibfnamefont {K.}~\bibnamefont {Luksch}}, \bibinfo {author} {\bibfnamefont
  {E.}~\bibnamefont {Bentine}}, \bibinfo {author} {\bibfnamefont
  {L.}~\bibnamefont {Mathey}}, \ and\ \bibinfo {author} {\bibfnamefont {C.~J.}\
  \bibnamefont {Foot}},\ }\bibfield  {title} {\enquote {\bibinfo {title}
  {{Universal scaling of the dynamic BKT transition in quenched 2D Bose
  gases}},}\ }\href {\doibase 10.1126/science.abq6753} {\bibfield  {journal}
  {\bibinfo  {journal} {Science}\ }\textbf {\bibinfo {volume} {382}},\ \bibinfo
  {pages} {443--447} (\bibinfo {year} {2023})}\BibitemShut {NoStop}%
\bibitem [{\citenamefont {Li}\ \emph {et~al.}(2023)\citenamefont {Li},
  \citenamefont {Wu}, \citenamefont {Mei}, \citenamefont {Yao}, \citenamefont
  {Lian}, \citenamefont {Cai}, \citenamefont {Wang}, \citenamefont {Qi},
  \citenamefont {Yao}, \citenamefont {He}, \citenamefont {Zhou},\ and\
  \citenamefont {Duan}}]{PRXQuantum}%
  \BibitemOpen
  \bibfield  {author} {\bibinfo {author} {\bibfnamefont {B.-W.}\ \bibnamefont
  {Li}}, \bibinfo {author} {\bibfnamefont {Y.-K.}\ \bibnamefont {Wu}}, \bibinfo
  {author} {\bibfnamefont {Q.-X.}\ \bibnamefont {Mei}}, \bibinfo {author}
  {\bibfnamefont {R.}~\bibnamefont {Yao}}, \bibinfo {author} {\bibfnamefont
  {W.-Q.}\ \bibnamefont {Lian}}, \bibinfo {author} {\bibfnamefont {M.-L.}\
  \bibnamefont {Cai}}, \bibinfo {author} {\bibfnamefont {Y.}~\bibnamefont
  {Wang}}, \bibinfo {author} {\bibfnamefont {B.-X.}\ \bibnamefont {Qi}},
  \bibinfo {author} {\bibfnamefont {L.}~\bibnamefont {Yao}}, \bibinfo {author}
  {\bibfnamefont {L.}~\bibnamefont {He}}, \bibinfo {author} {\bibfnamefont
  {Z.-C.}\ \bibnamefont {Zhou}}, \ and\ \bibinfo {author} {\bibfnamefont
  {L.-M.}\ \bibnamefont {Duan}},\ }\bibfield  {title} {\enquote {\bibinfo
  {title} {{Probing Critical Behavior of Long-Range Transverse-Field Ising
  Model through Quantum Kibble-Zurek Mechanism}},}\ }\href {\doibase
  10.1103/PRXQuantum.4.010302} {\bibfield  {journal} {\bibinfo  {journal} {PRX
  Quantum}\ }\textbf {\bibinfo {volume} {4}},\ \bibinfo {pages} {010302}
  (\bibinfo {year} {2023})}\BibitemShut {NoStop}%
\bibitem [{\citenamefont {Kang}\ \emph {et~al.}(2023)\citenamefont {Kang},
  \citenamefont {Gao}, \citenamefont {Guo}, \citenamefont {Zhu}, \citenamefont
  {Huang}, \citenamefont {Hong}, \citenamefont {Cheong},\ and\ \citenamefont
  {Wang}}]{Kang2023jap}%
  \BibitemOpen
  \bibfield  {author} {\bibinfo {author} {\bibfnamefont {J.}~\bibnamefont
  {Kang}}, \bibinfo {author} {\bibfnamefont {Z.}~\bibnamefont {Gao}}, \bibinfo
  {author} {\bibfnamefont {C.}~\bibnamefont {Guo}}, \bibinfo {author}
  {\bibfnamefont {W.}~\bibnamefont {Zhu}}, \bibinfo {author} {\bibfnamefont
  {H.}~\bibnamefont {Huang}}, \bibinfo {author} {\bibfnamefont
  {J.}~\bibnamefont {Hong}}, \bibinfo {author} {\bibfnamefont {S.-W.}\
  \bibnamefont {Cheong}}, \ and\ \bibinfo {author} {\bibfnamefont
  {X.}~\bibnamefont {Wang}},\ }\bibfield  {title} {\enquote {\bibinfo {title}
  {{A snapshot of domain evolution between topological vortex and stripe in
  ferroelectric hexagonal ErMnO3}},}\ }\href {\doibase 10.1063/5.0138096}
  {\bibfield  {journal} {\bibinfo  {journal} {Journal of Applied Physics}\
  }\textbf {\bibinfo {volume} {133}},\ \bibinfo {pages} {124102} (\bibinfo
  {year} {2023})}\BibitemShut {NoStop}%
\bibitem [{\citenamefont {Zeng}\ \emph {et~al.}(2023)\citenamefont {Zeng},
  \citenamefont {Xia},\ and\ \citenamefont {del Campo}}]{Zeng2023prl}%
  \BibitemOpen
  \bibfield  {author} {\bibinfo {author} {\bibfnamefont {H.-B.}\ \bibnamefont
  {Zeng}}, \bibinfo {author} {\bibfnamefont {C.-Y.}\ \bibnamefont {Xia}}, \
  and\ \bibinfo {author} {\bibfnamefont {A.}~\bibnamefont {del Campo}},\
  }\bibfield  {title} {\enquote {\bibinfo {title} {{Universal Breakdown of
  Kibble-Zurek Scaling in Fast Quenches across a Phase Transition}},}\ }\href
  {\doibase 10.1103/PhysRevLett.130.060402} {\bibfield  {journal} {\bibinfo
  {journal} {Phys. Rev. Lett.}\ }\textbf {\bibinfo {volume} {130}},\ \bibinfo
  {pages} {060402} (\bibinfo {year} {2023})}\BibitemShut {NoStop}%
\bibitem [{\citenamefont {Weinberg}\ \emph {et~al.}(2025)\citenamefont
  {Weinberg}, \citenamefont {Xu},\ and\ \citenamefont
  {Sandvik}}]{Weinberg2025arx}%
  \BibitemOpen
  \bibfield  {author} {\bibinfo {author} {\bibfnamefont {P.}~\bibnamefont
  {Weinberg}}, \bibinfo {author} {\bibfnamefont {N.}~\bibnamefont {Xu}}, \ and\
  \bibinfo {author} {\bibfnamefont {A.~W.}\ \bibnamefont {Sandvik}},\
  }\bibfield  {title} {\enquote {\bibinfo {title} {{Defects and their Time
  Scales in Quantum and Classical Annealing of the Two-Dimensional Ising
  Model}},}\ }\href {https://arxiv.org/abs/2507.09273} {\bibfield  {journal}
  {\bibinfo  {journal} {arXiv:2507.09273}\ } (\bibinfo {year}
  {2025})}\BibitemShut {NoStop}%
\bibitem [{\citenamefont {Zhong}\ and\ \citenamefont
  {Xu}(2005)}]{Zhongf2005prb}%
  \BibitemOpen
  \bibfield  {author} {\bibinfo {author} {\bibfnamefont {F.}~\bibnamefont
  {Zhong}}\ and\ \bibinfo {author} {\bibfnamefont {Z.}~\bibnamefont {Xu}},\
  }\bibfield  {title} {\enquote {\bibinfo {title} {{Dynamic Monte Carlo
  renormalization group determination of critical exponents with linearly
  changing temperature}},}\ }\href {\doibase 10.1103/PhysRevB.71.132402}
  {\bibfield  {journal} {\bibinfo  {journal} {Phys. Rev. B}\ }\textbf {\bibinfo
  {volume} {71}},\ \bibinfo {pages} {132402} (\bibinfo {year}
  {2005})}\BibitemShut {NoStop}%
\bibitem [{\citenamefont {Gong}\ \emph {et~al.}(2010)\citenamefont {Gong},
  \citenamefont {Zhong}, \citenamefont {Huang},\ and\ \citenamefont
  {Fan}}]{Gong2010njp}%
  \BibitemOpen
  \bibfield  {author} {\bibinfo {author} {\bibfnamefont {S.}~\bibnamefont
  {Gong}}, \bibinfo {author} {\bibfnamefont {F.}~\bibnamefont {Zhong}},
  \bibinfo {author} {\bibfnamefont {X.}~\bibnamefont {Huang}}, \ and\ \bibinfo
  {author} {\bibfnamefont {S.}~\bibnamefont {Fan}},\ }\bibfield  {title}
  {\enquote {\bibinfo {title} {{Finite-time scaling via linear driving}},}\
  }\href {\doibase 10.1088/1367-2630/12/4/043036} {\bibfield  {journal}
  {\bibinfo  {journal} {New Journal of Physics}\ }\textbf {\bibinfo {volume}
  {12}},\ \bibinfo {pages} {043036} (\bibinfo {year} {2010})}\BibitemShut
  {NoStop}%
\bibitem [{\citenamefont {Zhong}(2006)}]{Zhong2006pre}%
  \BibitemOpen
  \bibfield  {author} {\bibinfo {author} {\bibfnamefont {F.}~\bibnamefont
  {Zhong}},\ }\bibfield  {title} {\enquote {\bibinfo {title} {{Probing
  criticality with linearly varying external fields: Renormalization group
  theory of nonequilibrium critical dynamics under driving}},}\ }\href
  {\doibase 10.1103/PhysRevE.73.047102} {\bibfield  {journal} {\bibinfo
  {journal} {Phys. Rev. E}\ }\textbf {\bibinfo {volume} {73}},\ \bibinfo
  {pages} {047102} (\bibinfo {year} {2006})}\BibitemShut {NoStop}%
\bibitem [{\citenamefont {Fan}\ and\ \citenamefont {Zhong}(2007)}]{Fan2007pre}%
  \BibitemOpen
  \bibfield  {author} {\bibinfo {author} {\bibfnamefont {S.}~\bibnamefont
  {Fan}}\ and\ \bibinfo {author} {\bibfnamefont {F.}~\bibnamefont {Zhong}},\
  }\bibfield  {title} {\enquote {\bibinfo {title} {{Determination of the
  dynamic and static critical exponents of the two-dimensional three-state
  Potts model using linearly varying temperature}},}\ }\href {\doibase
  10.1103/PhysRevE.76.041141} {\bibfield  {journal} {\bibinfo  {journal} {Phys.
  Rev. E}\ }\textbf {\bibinfo {volume} {76}},\ \bibinfo {pages} {041141}
  (\bibinfo {year} {2007})}\BibitemShut {NoStop}%
\bibitem [{\citenamefont {Huang}\ \emph {et~al.}(2010)\citenamefont {Huang},
  \citenamefont {Gong}, \citenamefont {Zhong},\ and\ \citenamefont
  {Fan}}]{Huangxz2010pre}%
  \BibitemOpen
  \bibfield  {author} {\bibinfo {author} {\bibfnamefont {X.}~\bibnamefont
  {Huang}}, \bibinfo {author} {\bibfnamefont {S.}~\bibnamefont {Gong}},
  \bibinfo {author} {\bibfnamefont {F.}~\bibnamefont {Zhong}}, \ and\ \bibinfo
  {author} {\bibfnamefont {S.}~\bibnamefont {Fan}},\ }\bibfield  {title}
  {\enquote {\bibinfo {title} {{Finite-time scaling via linear driving:
  Application to the two-dimensional Potts model}},}\ }\href {\doibase
  10.1103/PhysRevE.81.041139} {\bibfield  {journal} {\bibinfo  {journal} {Phys.
  Rev. E}\ }\textbf {\bibinfo {volume} {81}},\ \bibinfo {pages} {041139}
  (\bibinfo {year} {2010})}\BibitemShut {NoStop}%
\bibitem [{\citenamefont {Yin}\ \emph {et~al.}(2014)\citenamefont {Yin},
  \citenamefont {Mai},\ and\ \citenamefont {Zhong}}]{Yin2014prb}%
  \BibitemOpen
  \bibfield  {author} {\bibinfo {author} {\bibfnamefont {S.}~\bibnamefont
  {Yin}}, \bibinfo {author} {\bibfnamefont {P.}~\bibnamefont {Mai}}, \ and\
  \bibinfo {author} {\bibfnamefont {F.}~\bibnamefont {Zhong}},\ }\bibfield
  {title} {\enquote {\bibinfo {title} {{Nonequilibrium quantum criticality in
  open systems: The dissipation rate as an additional indispensable scaling
  variable}},}\ }\href {\doibase 10.1103/PhysRevB.89.094108} {\bibfield
  {journal} {\bibinfo  {journal} {Phys. Rev. B}\ }\textbf {\bibinfo {volume}
  {89}},\ \bibinfo {pages} {094108} (\bibinfo {year} {2014})}\BibitemShut
  {NoStop}%
\bibitem [{\citenamefont {Huang}\ \emph {et~al.}(2014)\citenamefont {Huang},
  \citenamefont {Yin}, \citenamefont {Feng},\ and\ \citenamefont
  {Zhong}}]{Huang2014prb}%
  \BibitemOpen
  \bibfield  {author} {\bibinfo {author} {\bibfnamefont {Y.}~\bibnamefont
  {Huang}}, \bibinfo {author} {\bibfnamefont {S.}~\bibnamefont {Yin}}, \bibinfo
  {author} {\bibfnamefont {B.}~\bibnamefont {Feng}}, \ and\ \bibinfo {author}
  {\bibfnamefont {F.}~\bibnamefont {Zhong}},\ }\bibfield  {title} {\enquote
  {\bibinfo {title} {{Kibble-Zurek mechanism and finite-time scaling}},}\
  }\href {\doibase 10.1103/PhysRevB.90.134108} {\bibfield  {journal} {\bibinfo
  {journal} {Phys. Rev. B}\ }\textbf {\bibinfo {volume} {90}},\ \bibinfo
  {pages} {134108} (\bibinfo {year} {2014})}\BibitemShut {NoStop}%
\bibitem [{\citenamefont {Feng}\ \emph {et~al.}(2016)\citenamefont {Feng},
  \citenamefont {Yin},\ and\ \citenamefont {Zhong}}]{feng2016prb}%
  \BibitemOpen
  \bibfield  {author} {\bibinfo {author} {\bibfnamefont {B.}~\bibnamefont
  {Feng}}, \bibinfo {author} {\bibfnamefont {S.}~\bibnamefont {Yin}}, \ and\
  \bibinfo {author} {\bibfnamefont {F.}~\bibnamefont {Zhong}},\ }\bibfield
  {title} {\enquote {\bibinfo {title} {{Theory of driven nonequilibrium
  critical phenomena}},}\ }\href {\doibase 10.1103/PhysRevB.94.144103}
  {\bibfield  {journal} {\bibinfo  {journal} {Phys. Rev. B}\ }\textbf {\bibinfo
  {volume} {94}},\ \bibinfo {pages} {144103} (\bibinfo {year}
  {2016})}\BibitemShut {NoStop}%
\bibitem [{\citenamefont {Fan}\ and\ \citenamefont {Zhong}(2009)}]{Fan2009pre}%
  \BibitemOpen
  \bibfield  {author} {\bibinfo {author} {\bibfnamefont {S.}~\bibnamefont
  {Fan}}\ and\ \bibinfo {author} {\bibfnamefont {F.}~\bibnamefont {Zhong}},\
  }\bibfield  {title} {\enquote {\bibinfo {title} {{Critical dynamics of the
  two-dimensional random-bond Potts model with nonequilibrium Monte Carlo
  simulations}},}\ }\href {\doibase 10.1103/PhysRevE.79.011122} {\bibfield
  {journal} {\bibinfo  {journal} {Phys. Rev. E}\ }\textbf {\bibinfo {volume}
  {79}},\ \bibinfo {pages} {011122} (\bibinfo {year} {2009})}\BibitemShut
  {NoStop}%
\bibitem [{\citenamefont {Hu}\ \emph {et~al.}(2015)\citenamefont {Hu},
  \citenamefont {Yin},\ and\ \citenamefont {Zhong}}]{huqj2015prb}%
  \BibitemOpen
  \bibfield  {author} {\bibinfo {author} {\bibfnamefont {Q.}~\bibnamefont
  {Hu}}, \bibinfo {author} {\bibfnamefont {S.}~\bibnamefont {Yin}}, \ and\
  \bibinfo {author} {\bibfnamefont {F.}~\bibnamefont {Zhong}},\ }\bibfield
  {title} {\enquote {\bibinfo {title} {{Scaling of the entanglement spectrum in
  driven critical dynamics}},}\ }\href {\doibase 10.1103/PhysRevB.91.184109}
  {\bibfield  {journal} {\bibinfo  {journal} {Phys. Rev. B}\ }\textbf {\bibinfo
  {volume} {91}},\ \bibinfo {pages} {184109} (\bibinfo {year}
  {2015})}\BibitemShut {NoStop}%
\bibitem [{\citenamefont {Cao}\ \emph {et~al.}(2018)\citenamefont {Cao},
  \citenamefont {Hu},\ and\ \citenamefont {Zhong}}]{Cao2018prb}%
  \BibitemOpen
  \bibfield  {author} {\bibinfo {author} {\bibfnamefont {X.}~\bibnamefont
  {Cao}}, \bibinfo {author} {\bibfnamefont {Q.}~\bibnamefont {Hu}}, \ and\
  \bibinfo {author} {\bibfnamefont {F.}~\bibnamefont {Zhong}},\ }\bibfield
  {title} {\enquote {\bibinfo {title} {{Scaling theory of entanglement entropy
  in confinements near quantum critical points}},}\ }\href {\doibase
  10.1103/PhysRevB.98.245124} {\bibfield  {journal} {\bibinfo  {journal} {Phys.
  Rev. B}\ }\textbf {\bibinfo {volume} {98}},\ \bibinfo {pages} {245124}
  (\bibinfo {year} {2018})}\BibitemShut {NoStop}%
\bibitem [{\citenamefont {Shu}\ \emph {et~al.}(2025{\natexlab{a}})\citenamefont
  {Shu}, \citenamefont {Jian}, \citenamefont {Sandvik},\ and\ \citenamefont
  {Yin}}]{Shu2025nc}%
  \BibitemOpen
  \bibfield  {author} {\bibinfo {author} {\bibfnamefont {Y.-R.}\ \bibnamefont
  {Shu}}, \bibinfo {author} {\bibfnamefont {S.-K.}\ \bibnamefont {Jian}},
  \bibinfo {author} {\bibfnamefont {A.~W.}\ \bibnamefont {Sandvik}}, \ and\
  \bibinfo {author} {\bibfnamefont {S.}~\bibnamefont {Yin}},\ }\bibfield
  {title} {\enquote {\bibinfo {title} {{Equilibration of topological defects
  near the deconfined quantum multicritical point}},}\ }\href {\doibase
  10.1038/s41467-025-58477-z} {\bibfield  {journal} {\bibinfo  {journal}
  {Nature Communications}\ }\textbf {\bibinfo {volume} {16}},\ \bibinfo {pages}
  {3402} (\bibinfo {year} {2025}{\natexlab{a}})}\BibitemShut {NoStop}%
\bibitem [{\citenamefont {Huang}\ and\ \citenamefont
  {Yin}(2020)}]{Huang2020prr}%
  \BibitemOpen
  \bibfield  {author} {\bibinfo {author} {\bibfnamefont {R.-Z.}\ \bibnamefont
  {Huang}}\ and\ \bibinfo {author} {\bibfnamefont {S.}~\bibnamefont {Yin}},\
  }\bibfield  {title} {\enquote {\bibinfo {title} {{Kibble-Zurek mechanism for
  a one-dimensional incarnation of a deconfined quantum critical point}},}\
  }\href {\doibase 10.1103/PhysRevResearch.2.023175} {\bibfield  {journal}
  {\bibinfo  {journal} {Phys. Rev. Res.}\ }\textbf {\bibinfo {volume} {2}},\
  \bibinfo {pages} {023175} (\bibinfo {year} {2020})}\BibitemShut {NoStop}%
\bibitem [{\citenamefont {Zeng}\ \emph
  {et~al.}(2025{\natexlab{a}})\citenamefont {Zeng}, \citenamefont {Yu},
  \citenamefont {Li}, \citenamefont {Li},\ and\ \citenamefont
  {Yin}}]{Zeng2025nc}%
  \BibitemOpen
  \bibfield  {author} {\bibinfo {author} {\bibfnamefont {Z.}~\bibnamefont
  {Zeng}}, \bibinfo {author} {\bibfnamefont {Y.-K.}\ \bibnamefont {Yu}},
  \bibinfo {author} {\bibfnamefont {Z.-X.}\ \bibnamefont {Li}}, \bibinfo
  {author} {\bibfnamefont {Z.-X.}\ \bibnamefont {Li}}, \ and\ \bibinfo {author}
  {\bibfnamefont {S.}~\bibnamefont {Yin}},\ }\bibfield  {title} {\enquote
  {\bibinfo {title} {{Finite-time scaling beyond the Kibble-Zurek prerequisite
  in Dirac systems}},}\ }\href {\doibase 10.1038/s41467-025-61611-6} {\bibfield
   {journal} {\bibinfo  {journal} {Nature Communications}\ }\textbf {\bibinfo
  {volume} {16}},\ \bibinfo {pages} {6181} (\bibinfo {year}
  {2025}{\natexlab{a}})}\BibitemShut {NoStop}%
\bibitem [{\citenamefont {Zeng}\ \emph
  {et~al.}(2025{\natexlab{b}})\citenamefont {Zeng}, \citenamefont {Yu},
  \citenamefont {Li},\ and\ \citenamefont {Yin}}]{Zeng2025prb}%
  \BibitemOpen
  \bibfield  {author} {\bibinfo {author} {\bibfnamefont {Z.}~\bibnamefont
  {Zeng}}, \bibinfo {author} {\bibfnamefont {Y.-K.}\ \bibnamefont {Yu}},
  \bibinfo {author} {\bibfnamefont {Z.-X.}\ \bibnamefont {Li}}, \ and\ \bibinfo
  {author} {\bibfnamefont {S.}~\bibnamefont {Yin}},\ }\bibfield  {title}
  {\enquote {\bibinfo {title} {{Nonequilibrium critical dynamics with emergent
  supersymmetry}},}\ }\href {\doibase 10.1103/qtwr-8kth} {\bibfield  {journal}
  {\bibinfo  {journal} {Phys. Rev. B}\ }\textbf {\bibinfo {volume} {112}},\
  \bibinfo {pages} {L060301} (\bibinfo {year}
  {2025}{\natexlab{b}})}\BibitemShut {NoStop}%
\bibitem [{\citenamefont {Shu}\ \emph {et~al.}(2025{\natexlab{b}})\citenamefont
  {Shu}, \citenamefont {Yang},\ and\ \citenamefont {Yin}}]{Shu2025arx}%
  \BibitemOpen
  \bibfield  {author} {\bibinfo {author} {\bibfnamefont {Y.-R.}\ \bibnamefont
  {Shu}}, \bibinfo {author} {\bibfnamefont {L.-Y.}\ \bibnamefont {Yang}}, \
  and\ \bibinfo {author} {\bibfnamefont {S.}~\bibnamefont {Yin}},\ }\bibfield
  {title} {\enquote {\bibinfo {title} {{Finite-time scaling with two
  characteristic time scales: Driven critical dynamics with emergent
  symmetry}},}\ }\href {https://arxiv.org/abs/2503.16796} {\bibfield  {journal}
  {\bibinfo  {journal} {arXiv:2503.16796}\ } (\bibinfo {year}
  {2025}{\natexlab{b}})}\BibitemShut {NoStop}%
\bibitem [{\citenamefont {Shu}\ \emph {et~al.}(2024)\citenamefont {Shu},
  \citenamefont {Liao},\ and\ \citenamefont {Yin}}]{Shu2024prb}%
  \BibitemOpen
  \bibfield  {author} {\bibinfo {author} {\bibfnamefont {Y.-R.}\ \bibnamefont
  {Shu}}, \bibinfo {author} {\bibfnamefont {T.}~\bibnamefont {Liao}}, \ and\
  \bibinfo {author} {\bibfnamefont {S.}~\bibnamefont {Yin}},\ }\bibfield
  {title} {\enquote {\bibinfo {title} {{Relaxation critical dynamics with
  emergent symmetry}},}\ }\href {\doibase 10.1103/PhysRevB.110.134306}
  {\bibfield  {journal} {\bibinfo  {journal} {Phys. Rev. B}\ }\textbf {\bibinfo
  {volume} {110}},\ \bibinfo {pages} {134306} (\bibinfo {year}
  {2024})}\BibitemShut {NoStop}%
\bibitem [{\citenamefont {Li}\ \emph {et~al.}(2019)\citenamefont {Li},
  \citenamefont {Zeng},\ and\ \citenamefont {Zhong}}]{Liyh2019pre}%
  \BibitemOpen
  \bibfield  {author} {\bibinfo {author} {\bibfnamefont {Y.}~\bibnamefont
  {Li}}, \bibinfo {author} {\bibfnamefont {Z.}~\bibnamefont {Zeng}}, \ and\
  \bibinfo {author} {\bibfnamefont {F.}~\bibnamefont {Zhong}},\ }\bibfield
  {title} {\enquote {\bibinfo {title} {{Driving driven lattice gases to
  identify their universality classes}},}\ }\href {\doibase
  10.1103/PhysRevE.100.020105} {\bibfield  {journal} {\bibinfo  {journal}
  {Phys. Rev. E}\ }\textbf {\bibinfo {volume} {100}},\ \bibinfo {pages}
  {020105} (\bibinfo {year} {2019})}\BibitemShut {NoStop}%
\bibitem [{\citenamefont {Wang}\ \emph
  {et~al.}(2024{\natexlab{a}})\citenamefont {Wang}, \citenamefont {Liu},
  \citenamefont {Li}, \citenamefont {Zhang},\ and\ \citenamefont
  {Yin}}]{Yinarxiv2024}%
  \BibitemOpen
  \bibfield  {author} {\bibinfo {author} {\bibfnamefont {W.}~\bibnamefont
  {Wang}}, \bibinfo {author} {\bibfnamefont {S.}~\bibnamefont {Liu}}, \bibinfo
  {author} {\bibfnamefont {J.}~\bibnamefont {Li}}, \bibinfo {author}
  {\bibfnamefont {S.-X.}\ \bibnamefont {Zhang}}, \ and\ \bibinfo {author}
  {\bibfnamefont {S.}~\bibnamefont {Yin}},\ }\bibfield  {title} {\enquote
  {\bibinfo {title} {{}},}\ }\href {https://arxiv.org/abs/2411.06648}
  {\bibfield  {journal} {\bibinfo  {journal} {arXiv: 2411.06648}\ } (\bibinfo
  {year} {2024}{\natexlab{a}})}\BibitemShut {NoStop}%
\bibitem [{\citenamefont {Huang}\ \emph {et~al.}(2016)\citenamefont {Huang},
  \citenamefont {Yin}, \citenamefont {Hu},\ and\ \citenamefont
  {Zhong}}]{Huangyy2016prb}%
  \BibitemOpen
  \bibfield  {author} {\bibinfo {author} {\bibfnamefont {Y.}~\bibnamefont
  {Huang}}, \bibinfo {author} {\bibfnamefont {S.}~\bibnamefont {Yin}}, \bibinfo
  {author} {\bibfnamefont {Q.}~\bibnamefont {Hu}}, \ and\ \bibinfo {author}
  {\bibfnamefont {F.}~\bibnamefont {Zhong}},\ }\bibfield  {title} {\enquote
  {\bibinfo {title} {{Kibble-Zurek mechanism beyond adiabaticity: Finite-time
  scaling with critical initial slip}},}\ }\href {\doibase
  10.1103/PhysRevB.93.024103} {\bibfield  {journal} {\bibinfo  {journal} {Phys.
  Rev. B}\ }\textbf {\bibinfo {volume} {93}},\ \bibinfo {pages} {024103}
  (\bibinfo {year} {2016})}\BibitemShut {NoStop}%
\bibitem [{\citenamefont {Yin}\ \emph {et~al.}(2016)\citenamefont {Yin},
  \citenamefont {Lo},\ and\ \citenamefont {Chen}}]{Yin2016prb}%
  \BibitemOpen
  \bibfield  {author} {\bibinfo {author} {\bibfnamefont {S.}~\bibnamefont
  {Yin}}, \bibinfo {author} {\bibfnamefont {C.-Y.}\ \bibnamefont {Lo}}, \ and\
  \bibinfo {author} {\bibfnamefont {P.}~\bibnamefont {Chen}},\ }\bibfield
  {title} {\enquote {\bibinfo {title} {{Scaling in driven dynamics starting in
  the vicinity of a quantum critical point}},}\ }\href {\doibase
  10.1103/PhysRevB.94.064302} {\bibfield  {journal} {\bibinfo  {journal} {Phys.
  Rev. B}\ }\textbf {\bibinfo {volume} {94}},\ \bibinfo {pages} {064302}
  (\bibinfo {year} {2016})}\BibitemShut {NoStop}%
\bibitem [{\citenamefont {Deng}\ \emph {et~al.}(2009)\citenamefont {Deng},
  \citenamefont {Ortiz},\ and\ \citenamefont {Viola}}]{Deng2008}%
  \BibitemOpen
  \bibfield  {author} {\bibinfo {author} {\bibfnamefont {S.}~\bibnamefont
  {Deng}}, \bibinfo {author} {\bibfnamefont {G.}~\bibnamefont {Ortiz}}, \ and\
  \bibinfo {author} {\bibfnamefont {L.}~\bibnamefont {Viola}},\ }\bibfield
  {title} {\enquote {\bibinfo {title} {{Dynamical non-ergodic scaling in
  continuous finite-order quantum phase transitions}},}\ }\href {\doibase
  10.1209/0295-5075/84/67008} {\bibfield  {journal} {\bibinfo  {journal}
  {Europhysics Letters}\ }\textbf {\bibinfo {volume} {84}},\ \bibinfo {pages}
  {67008} (\bibinfo {year} {2009})}\BibitemShut {NoStop}%
\bibitem [{\citenamefont {Chandran}\ \emph {et~al.}(2012)\citenamefont
  {Chandran}, \citenamefont {Erez}, \citenamefont {Gubser},\ and\ \citenamefont
  {Sondhi}}]{chandran2012prb}%
  \BibitemOpen
  \bibfield  {author} {\bibinfo {author} {\bibfnamefont {A.}~\bibnamefont
  {Chandran}}, \bibinfo {author} {\bibfnamefont {A.}~\bibnamefont {Erez}},
  \bibinfo {author} {\bibfnamefont {S.~S.}\ \bibnamefont {Gubser}}, \ and\
  \bibinfo {author} {\bibfnamefont {S.~L.}\ \bibnamefont {Sondhi}},\ }\bibfield
   {title} {\enquote {\bibinfo {title} {{Kibble-Zurek problem: Universality and
  the scaling limit}},}\ }\href {\doibase 10.1103/PhysRevB.86.064304}
  {\bibfield  {journal} {\bibinfo  {journal} {Phys. Rev. B}\ }\textbf {\bibinfo
  {volume} {86}},\ \bibinfo {pages} {064304} (\bibinfo {year}
  {2012})}\BibitemShut {NoStop}%
\bibitem [{\citenamefont {De~Grandi}\ \emph {et~al.}(2011)\citenamefont
  {De~Grandi}, \citenamefont {Polkovnikov},\ and\ \citenamefont
  {Sandvik}}]{Grandi2011prb}%
  \BibitemOpen
  \bibfield  {author} {\bibinfo {author} {\bibfnamefont {C.}~\bibnamefont
  {De~Grandi}}, \bibinfo {author} {\bibfnamefont {A.}~\bibnamefont
  {Polkovnikov}}, \ and\ \bibinfo {author} {\bibfnamefont {A.~W.}\ \bibnamefont
  {Sandvik}},\ }\bibfield  {title} {\enquote {\bibinfo {title} {{Universal
  nonequilibrium quantum dynamics in imaginary time}},}\ }\href {\doibase
  10.1103/PhysRevB.84.224303} {\bibfield  {journal} {\bibinfo  {journal} {Phys.
  Rev. B}\ }\textbf {\bibinfo {volume} {84}},\ \bibinfo {pages} {224303}
  (\bibinfo {year} {2011})}\BibitemShut {NoStop}%
\bibitem [{\citenamefont {Kolodrubetz}\ \emph {et~al.}(2012)\citenamefont
  {Kolodrubetz}, \citenamefont {Clark},\ and\ \citenamefont
  {Huse}}]{Huse2012prl}%
  \BibitemOpen
  \bibfield  {author} {\bibinfo {author} {\bibfnamefont {M.}~\bibnamefont
  {Kolodrubetz}}, \bibinfo {author} {\bibfnamefont {B.~K.}\ \bibnamefont
  {Clark}}, \ and\ \bibinfo {author} {\bibfnamefont {D.~A.}\ \bibnamefont
  {Huse}},\ }\bibfield  {title} {\enquote {\bibinfo {title} {{Nonequilibrium
  Dynamic Critical Scaling of the Quantum Ising Chain}},}\ }\href {\doibase
  10.1103/PhysRevLett.109.015701} {\bibfield  {journal} {\bibinfo  {journal}
  {Phys. Rev. Lett.}\ }\textbf {\bibinfo {volume} {109}},\ \bibinfo {pages}
  {015701} (\bibinfo {year} {2012})}\BibitemShut {NoStop}%
\bibitem [{\citenamefont {Liu}\ \emph {et~al.}(2014)\citenamefont {Liu},
  \citenamefont {Polkovnikov},\ and\ \citenamefont {Sandvik}}]{Liu2014prb}%
  \BibitemOpen
  \bibfield  {author} {\bibinfo {author} {\bibfnamefont {C.-W.}\ \bibnamefont
  {Liu}}, \bibinfo {author} {\bibfnamefont {A.}~\bibnamefont {Polkovnikov}}, \
  and\ \bibinfo {author} {\bibfnamefont {A.~W.}\ \bibnamefont {Sandvik}},\
  }\bibfield  {title} {\enquote {\bibinfo {title} {{Dynamic scaling at
  classical phase transitions approached through nonequilibrium quenching}},}\
  }\href {\doibase 10.1103/PhysRevB.89.054307} {\bibfield  {journal} {\bibinfo
  {journal} {Phys. Rev. B}\ }\textbf {\bibinfo {volume} {89}},\ \bibinfo
  {pages} {054307} (\bibinfo {year} {2014})}\BibitemShut {NoStop}%
\bibitem [{\citenamefont {Francuz}\ \emph {et~al.}(2016)\citenamefont
  {Francuz}, \citenamefont {Dziarmaga}, \citenamefont {Gardas},\ and\
  \citenamefont {Zurek}}]{Zurek2016prb}%
  \BibitemOpen
  \bibfield  {author} {\bibinfo {author} {\bibfnamefont {A.}~\bibnamefont
  {Francuz}}, \bibinfo {author} {\bibfnamefont {J.}~\bibnamefont {Dziarmaga}},
  \bibinfo {author} {\bibfnamefont {B.}~\bibnamefont {Gardas}}, \ and\ \bibinfo
  {author} {\bibfnamefont {W.~H.}\ \bibnamefont {Zurek}},\ }\bibfield  {title}
  {\enquote {\bibinfo {title} {{Space and time renormalization in phase
  transition dynamics}},}\ }\href {\doibase 10.1103/PhysRevB.93.075134}
  {\bibfield  {journal} {\bibinfo  {journal} {Phys. Rev. B}\ }\textbf {\bibinfo
  {volume} {93}},\ \bibinfo {pages} {075134} (\bibinfo {year}
  {2016})}\BibitemShut {NoStop}%
\bibitem [{\citenamefont {Domb}\ and\ \citenamefont {Lebowitz}(1983)}]{PTCP8}%
  \BibitemOpen
  \bibinfo {editor} {\bibfnamefont {C.}~\bibnamefont {Domb}}\ and\ \bibinfo
  {editor} {\bibfnamefont {J.~L.}\ \bibnamefont {Lebowitz}},\ eds.,\ \href@noop
  {} {\emph {\bibinfo {title} {{Phase Transitions and Critical Phenomena}}}},\
  Vol.~\bibinfo {volume} {8}\ (\bibinfo  {publisher} {Academic Press},\
  \bibinfo {year} {1983})\BibitemShut {NoStop}%
\bibitem [{\citenamefont {Domb}\ and\ \citenamefont {Lebowitz}(1986)}]{PTCP10}%
  \BibitemOpen
  \bibinfo {editor} {\bibfnamefont {C.}~\bibnamefont {Domb}}\ and\ \bibinfo
  {editor} {\bibfnamefont {J.~L.}\ \bibnamefont {Lebowitz}},\ eds.,\ \href@noop
  {} {\emph {\bibinfo {title} {{Phase Transitions and Critical Phenomena}}}},\
  Vol.~\bibinfo {volume} {10}\ (\bibinfo  {publisher} {Academic Press},\
  \bibinfo {year} {1986})\BibitemShut {NoStop}%
\bibitem [{\citenamefont {Diehl}(1997)}]{Diehl1997book}%
  \BibitemOpen
  \bibfield  {author} {\bibinfo {author} {\bibfnamefont {H.~W.}\ \bibnamefont
  {Diehl}},\ }\bibfield  {title} {\enquote {\bibinfo {title} {{The Theory of
  Boundary Critical Phenomena}},}\ }\href {\doibase 10.1142/S0217979297001751}
  {\bibfield  {journal} {\bibinfo  {journal} {International Journal of Modern
  Physics B}\ }\textbf {\bibinfo {volume} {11}},\ \bibinfo {pages} {3503--3523}
  (\bibinfo {year} {1997})}\BibitemShut {NoStop}%
\bibitem [{\citenamefont {Pleimling}(2004)}]{Pleimling2004jpa}%
  \BibitemOpen
  \bibfield  {author} {\bibinfo {author} {\bibfnamefont {M.}~\bibnamefont
  {Pleimling}},\ }\bibfield  {title} {\enquote {\bibinfo {title} {{Critical
  phenomena at perfect and non-perfect surfaces}},}\ }\href {\doibase
  10.1088/0305-4470/37/19/R01} {\bibfield  {journal} {\bibinfo  {journal}
  {Journal of Physics A: Mathematical and General}\ }\textbf {\bibinfo {volume}
  {37}},\ \bibinfo {pages} {R79} (\bibinfo {year} {2004})}\BibitemShut
  {NoStop}%
\bibitem [{\citenamefont {Binder}\ and\ \citenamefont
  {Hohenberg}(1972)}]{Binder1972prb}%
  \BibitemOpen
  \bibfield  {author} {\bibinfo {author} {\bibfnamefont {K.}~\bibnamefont
  {Binder}}\ and\ \bibinfo {author} {\bibfnamefont {P.~C.}\ \bibnamefont
  {Hohenberg}},\ }\bibfield  {title} {\enquote {\bibinfo {title} {{Phase
  Transitions and Static Spin Correlations in Ising Models with Free
  Surfaces}},}\ }\href {\doibase 10.1103/PhysRevB.6.3461} {\bibfield  {journal}
  {\bibinfo  {journal} {Phys. Rev. B}\ }\textbf {\bibinfo {volume} {6}},\
  \bibinfo {pages} {3461--3487} (\bibinfo {year} {1972})}\BibitemShut {NoStop}%
\bibitem [{\citenamefont {Binder}\ and\ \citenamefont
  {Landau}(1984)}]{Binder1984prl}%
  \BibitemOpen
  \bibfield  {author} {\bibinfo {author} {\bibfnamefont {K.}~\bibnamefont
  {Binder}}\ and\ \bibinfo {author} {\bibfnamefont {D.~P.}\ \bibnamefont
  {Landau}},\ }\bibfield  {title} {\enquote {\bibinfo {title} {{Crossover
  Scaling and Critical Behavior at the ``Surface-Bulk'' Multicritical
  Point}},}\ }\href {\doibase 10.1103/PhysRevLett.52.318} {\bibfield  {journal}
  {\bibinfo  {journal} {Phys. Rev. Lett.}\ }\textbf {\bibinfo {volume} {52}},\
  \bibinfo {pages} {318--321} (\bibinfo {year} {1984})}\BibitemShut {NoStop}%
\bibitem [{\citenamefont {Landau}\ and\ \citenamefont
  {Binder}(1990)}]{Landau1990prb}%
  \BibitemOpen
  \bibfield  {author} {\bibinfo {author} {\bibfnamefont {D.~P.}\ \bibnamefont
  {Landau}}\ and\ \bibinfo {author} {\bibfnamefont {K.}~\bibnamefont
  {Binder}},\ }\bibfield  {title} {\enquote {\bibinfo {title} {{Monte Carlo
  study of surface phase transitions in the three-dimensional Ising model}},}\
  }\href {\doibase 10.1103/PhysRevB.41.4633} {\bibfield  {journal} {\bibinfo
  {journal} {Phys. Rev. B}\ }\textbf {\bibinfo {volume} {41}},\ \bibinfo
  {pages} {4633--4645} (\bibinfo {year} {1990})}\BibitemShut {NoStop}%
\bibitem [{\citenamefont {Ruge}\ and\ \citenamefont
  {Wagner}(1995)}]{Ruge1995prb}%
  \BibitemOpen
  \bibfield  {author} {\bibinfo {author} {\bibfnamefont {C.}~\bibnamefont
  {Ruge}}\ and\ \bibinfo {author} {\bibfnamefont {F.}~\bibnamefont {Wagner}},\
  }\bibfield  {title} {\enquote {\bibinfo {title} {{Critical parameters for the
  $d=3$ Ising model in a film geometry}},}\ }\href {\doibase
  10.1103/PhysRevB.52.4209} {\bibfield  {journal} {\bibinfo  {journal} {Phys.
  Rev. B}\ }\textbf {\bibinfo {volume} {52}},\ \bibinfo {pages} {4209--4216}
  (\bibinfo {year} {1995})}\BibitemShut {NoStop}%
\bibitem [{\citenamefont {Deng}\ \emph {et~al.}(2005)\citenamefont {Deng},
  \citenamefont {Bl\"ote},\ and\ \citenamefont {Nightingale}}]{Deng2005pre}%
  \BibitemOpen
  \bibfield  {author} {\bibinfo {author} {\bibfnamefont {Y.}~\bibnamefont
  {Deng}}, \bibinfo {author} {\bibfnamefont {H.~W.~J.}\ \bibnamefont
  {Bl\"ote}}, \ and\ \bibinfo {author} {\bibfnamefont {M.~P.}\ \bibnamefont
  {Nightingale}},\ }\bibfield  {title} {\enquote {\bibinfo {title} {{Surface
  and bulk transitions in three-dimensional $\mathrm{O}(n)$ models}},}\ }\href
  {\doibase 10.1103/PhysRevE.72.016128} {\bibfield  {journal} {\bibinfo
  {journal} {Phys. Rev. E}\ }\textbf {\bibinfo {volume} {72}},\ \bibinfo
  {pages} {016128} (\bibinfo {year} {2005})}\BibitemShut {NoStop}%
\bibitem [{\citenamefont {Deng}(2006)}]{Deng2006pre}%
  \BibitemOpen
  \bibfield  {author} {\bibinfo {author} {\bibfnamefont {Y.}~\bibnamefont
  {Deng}},\ }\bibfield  {title} {\enquote {\bibinfo {title} {{Bulk and surface
  phase transitions in the three-dimensional $O(4)$ spin model}},}\ }\href
  {\doibase 10.1103/PhysRevE.73.056116} {\bibfield  {journal} {\bibinfo
  {journal} {Phys. Rev. E}\ }\textbf {\bibinfo {volume} {73}},\ \bibinfo
  {pages} {056116} (\bibinfo {year} {2006})}\BibitemShut {NoStop}%
\bibitem [{\citenamefont {Metlitski}(2022)}]{Metlitski2022sp}%
  \BibitemOpen
  \bibfield  {author} {\bibinfo {author} {\bibfnamefont {M.~A.}\ \bibnamefont
  {Metlitski}},\ }\bibfield  {title} {\enquote {\bibinfo {title} {{Boundary
  criticality of the O($N$) model in $d=3$ critically revisited}},}\ }\href
  {\doibase 10.21468/SciPostPhys.12.4.131} {\bibfield  {journal} {\bibinfo
  {journal} {SciPost Phys.}\ }\textbf {\bibinfo {volume} {12}},\ \bibinfo
  {pages} {131} (\bibinfo {year} {2022})}\BibitemShut {NoStop}%
\bibitem [{\citenamefont {Parisen~Toldin}(2021)}]{Toldin2021prl}%
  \BibitemOpen
  \bibfield  {author} {\bibinfo {author} {\bibfnamefont {F.}~\bibnamefont
  {Parisen~Toldin}},\ }\bibfield  {title} {\enquote {\bibinfo {title}
  {{Boundary Critical Behavior of the Three-Dimensional Heisenberg Universality
  Class}},}\ }\href {\doibase 10.1103/PhysRevLett.126.135701} {\bibfield
  {journal} {\bibinfo  {journal} {Phys. Rev. Lett.}\ }\textbf {\bibinfo
  {volume} {126}},\ \bibinfo {pages} {135701} (\bibinfo {year}
  {2021})}\BibitemShut {NoStop}%
\bibitem [{\citenamefont {Hu}\ \emph {et~al.}(2021)\citenamefont {Hu},
  \citenamefont {Deng},\ and\ \citenamefont {Lv}}]{Hu2021prl}%
  \BibitemOpen
  \bibfield  {author} {\bibinfo {author} {\bibfnamefont {M.}~\bibnamefont
  {Hu}}, \bibinfo {author} {\bibfnamefont {Y.}~\bibnamefont {Deng}}, \ and\
  \bibinfo {author} {\bibfnamefont {J.-P.}\ \bibnamefont {Lv}},\ }\bibfield
  {title} {\enquote {\bibinfo {title} {{Extraordinary-Log Surface Phase
  Transition in the Three-Dimensional $XY$ Model}},}\ }\href {\doibase
  10.1103/PhysRevLett.127.120603} {\bibfield  {journal} {\bibinfo  {journal}
  {Phys. Rev. Lett.}\ }\textbf {\bibinfo {volume} {127}},\ \bibinfo {pages}
  {120603} (\bibinfo {year} {2021})}\BibitemShut {NoStop}%
\bibitem [{\citenamefont {Hu}\ and\ \citenamefont {Li}(2025)}]{Hu2025arx}%
  \BibitemOpen
  \bibfield  {author} {\bibinfo {author} {\bibfnamefont {R.}~\bibnamefont
  {Hu}}\ and\ \bibinfo {author} {\bibfnamefont {W.}~\bibnamefont {Li}},\
  }\bibfield  {title} {\enquote {\bibinfo {title} {{Boundary bootstrap for the
  three-dimensional O($N$) normal universality class}},}\ }\href
  {https://arxiv.org/abs/2508.20854} {\bibfield  {journal} {\bibinfo  {journal}
  {arXiv:2508.20854}\ } (\bibinfo {year} {2025})}\BibitemShut {NoStop}%
\bibitem [{\citenamefont {Parisen~Toldin}\ and\ \citenamefont
  {Metlitski}(2022)}]{Toldin2022prl}%
  \BibitemOpen
  \bibfield  {author} {\bibinfo {author} {\bibfnamefont {F.}~\bibnamefont
  {Parisen~Toldin}}\ and\ \bibinfo {author} {\bibfnamefont {M.~A.}\
  \bibnamefont {Metlitski}},\ }\bibfield  {title} {\enquote {\bibinfo {title}
  {{Boundary Criticality of the 3D O($N$) Model: From Normal to
  Extraordinary}},}\ }\href {\doibase 10.1103/PhysRevLett.128.215701}
  {\bibfield  {journal} {\bibinfo  {journal} {Phys. Rev. Lett.}\ }\textbf
  {\bibinfo {volume} {128}},\ \bibinfo {pages} {215701} (\bibinfo {year}
  {2022})}\BibitemShut {NoStop}%
\bibitem [{\citenamefont {Padayasi}\ \emph {et~al.}(2022)\citenamefont
  {Padayasi}, \citenamefont {Krishnan}, \citenamefont {Metlitski},
  \citenamefont {Gruzberg},\ and\ \citenamefont {Meineri}}]{Padayasi2022sp}%
  \BibitemOpen
  \bibfield  {author} {\bibinfo {author} {\bibfnamefont {J.}~\bibnamefont
  {Padayasi}}, \bibinfo {author} {\bibfnamefont {A.}~\bibnamefont {Krishnan}},
  \bibinfo {author} {\bibfnamefont {M.~A.}\ \bibnamefont {Metlitski}}, \bibinfo
  {author} {\bibfnamefont {I.~A.}\ \bibnamefont {Gruzberg}}, \ and\ \bibinfo
  {author} {\bibfnamefont {M.}~\bibnamefont {Meineri}},\ }\bibfield  {title}
  {\enquote {\bibinfo {title} {{The extraordinary boundary transition in the
  $3d$ O($N$) model via conformal bootstrap}},}\ }\href {\doibase
  10.21468/SciPostPhys.12.6.190} {\bibfield  {journal} {\bibinfo  {journal}
  {SciPost Phys.}\ }\textbf {\bibinfo {volume} {12}},\ \bibinfo {pages} {190}
  (\bibinfo {year} {2022})}\BibitemShut {NoStop}%
\bibitem [{\citenamefont {Sun}\ \emph {et~al.}(2022)\citenamefont {Sun},
  \citenamefont {Lyu},\ and\ \citenamefont {Lv}}]{Sun2022prb}%
  \BibitemOpen
  \bibfield  {author} {\bibinfo {author} {\bibfnamefont {Y.}~\bibnamefont
  {Sun}}, \bibinfo {author} {\bibfnamefont {J.}~\bibnamefont {Lyu}}, \ and\
  \bibinfo {author} {\bibfnamefont {J.-P.}\ \bibnamefont {Lv}},\ }\bibfield
  {title} {\enquote {\bibinfo {title} {{Classical-quantum correspondence of
  special and extraordinary-log criticality: Villain's bridge}},}\ }\href
  {\doibase 10.1103/PhysRevB.106.174516} {\bibfield  {journal} {\bibinfo
  {journal} {Phys. Rev. B}\ }\textbf {\bibinfo {volume} {106}},\ \bibinfo
  {pages} {174516} (\bibinfo {year} {2022})}\BibitemShut {NoStop}%
\bibitem [{\citenamefont {Zhang}\ \emph {et~al.}(2022)\citenamefont {Zhang},
  \citenamefont {Ding}, \citenamefont {Deng},\ and\ \citenamefont
  {Zhang}}]{Zhang2022prb}%
  \BibitemOpen
  \bibfield  {author} {\bibinfo {author} {\bibfnamefont {L.-R.}\ \bibnamefont
  {Zhang}}, \bibinfo {author} {\bibfnamefont {C.}~\bibnamefont {Ding}},
  \bibinfo {author} {\bibfnamefont {Y.}~\bibnamefont {Deng}}, \ and\ \bibinfo
  {author} {\bibfnamefont {L.}~\bibnamefont {Zhang}},\ }\bibfield  {title}
  {\enquote {\bibinfo {title} {{Surface criticality of the antiferromagnetic
  Potts model}},}\ }\href {\doibase 10.1103/PhysRevB.105.224415} {\bibfield
  {journal} {\bibinfo  {journal} {Phys. Rev. B}\ }\textbf {\bibinfo {volume}
  {105}},\ \bibinfo {pages} {224415} (\bibinfo {year} {2022})}\BibitemShut
  {NoStop}%
\bibitem [{\citenamefont {Sun}\ and\ \citenamefont
  {Lv}(2022)}]{Sun2022prb_log}%
  \BibitemOpen
  \bibfield  {author} {\bibinfo {author} {\bibfnamefont {Y.}~\bibnamefont
  {Sun}}\ and\ \bibinfo {author} {\bibfnamefont {J.-P.}\ \bibnamefont {Lv}},\
  }\bibfield  {title} {\enquote {\bibinfo {title} {{Quantum extraordinary-log
  universality of boundary critical behavior}},}\ }\href {\doibase
  10.1103/PhysRevB.106.224502} {\bibfield  {journal} {\bibinfo  {journal}
  {Phys. Rev. B}\ }\textbf {\bibinfo {volume} {106}},\ \bibinfo {pages}
  {224502} (\bibinfo {year} {2022})}\BibitemShut {NoStop}%
\bibitem [{\citenamefont {Sun}\ and\ \citenamefont {Jian}(2025)}]{Sun2025sp}%
  \BibitemOpen
  \bibfield  {author} {\bibinfo {author} {\bibfnamefont {X.}~\bibnamefont
  {Sun}}\ and\ \bibinfo {author} {\bibfnamefont {S.-K.}\ \bibnamefont {Jian}},\
  }\bibfield  {title} {\enquote {\bibinfo {title} {{Boundary operator expansion
  and extraordinary phase transition in the tricritical O($N$) model}},}\
  }\href {\doibase 10.21468/SciPostPhys.18.6.210} {\bibfield  {journal}
  {\bibinfo  {journal} {SciPost Phys.}\ }\textbf {\bibinfo {volume} {18}},\
  \bibinfo {pages} {210} (\bibinfo {year} {2025})}\BibitemShut {NoStop}%
\bibitem [{\citenamefont {Zhang}\ and\ \citenamefont
  {Wang}(2017)}]{Zhang2017prl}%
  \BibitemOpen
  \bibfield  {author} {\bibinfo {author} {\bibfnamefont {L.}~\bibnamefont
  {Zhang}}\ and\ \bibinfo {author} {\bibfnamefont {F.}~\bibnamefont {Wang}},\
  }\bibfield  {title} {\enquote {\bibinfo {title} {{Unconventional Surface
  Critical Behavior Induced by a Quantum Phase Transition from the
  Two-Dimensional Affleck-Kennedy-Lieb-Tasaki Phase to a N\'eel-Ordered
  Phase}},}\ }\href {\doibase 10.1103/PhysRevLett.118.087201} {\bibfield
  {journal} {\bibinfo  {journal} {Phys. Rev. Lett.}\ }\textbf {\bibinfo
  {volume} {118}},\ \bibinfo {pages} {087201} (\bibinfo {year}
  {2017})}\BibitemShut {NoStop}%
\bibitem [{\citenamefont {Zhu}\ \emph {et~al.}(2025)\citenamefont {Zhu},
  \citenamefont {Liu}, \citenamefont {Wang}, \citenamefont {Wang},\ and\
  \citenamefont {Yan}}]{Zhu2025arx}%
  \BibitemOpen
  \bibfield  {author} {\bibinfo {author} {\bibfnamefont {Y.}~\bibnamefont
  {Zhu}}, \bibinfo {author} {\bibfnamefont {Z.}~\bibnamefont {Liu}}, \bibinfo
  {author} {\bibfnamefont {Z.}~\bibnamefont {Wang}}, \bibinfo {author}
  {\bibfnamefont {Y.-C.}\ \bibnamefont {Wang}}, \ and\ \bibinfo {author}
  {\bibfnamefont {Z.}~\bibnamefont {Yan}},\ }\bibfield  {title} {\enquote
  {\bibinfo {title} {{Bipartite entanglement and surface criticality}},}\
  }\href {https://arxiv.org/abs/2508.07277} {\bibfield  {journal} {\bibinfo
  {journal} {arXiv:2508.07277}\ } (\bibinfo {year} {2025})}\BibitemShut
  {NoStop}%
\bibitem [{\citenamefont {Wang}\ \emph
  {et~al.}(2024{\natexlab{b}})\citenamefont {Wang}, \citenamefont {Ning},
  \citenamefont {Liu}, \citenamefont {Rong}, \citenamefont {Wang},
  \citenamefont {Yan},\ and\ \citenamefont {Guo}}]{Wang2024prb}%
  \BibitemOpen
  \bibfield  {author} {\bibinfo {author} {\bibfnamefont {Z.}~\bibnamefont
  {Wang}}, \bibinfo {author} {\bibfnamefont {S.-Q.}\ \bibnamefont {Ning}},
  \bibinfo {author} {\bibfnamefont {Z.}~\bibnamefont {Liu}}, \bibinfo {author}
  {\bibfnamefont {J.}~\bibnamefont {Rong}}, \bibinfo {author} {\bibfnamefont
  {Y.-C.}\ \bibnamefont {Wang}}, \bibinfo {author} {\bibfnamefont
  {Z.}~\bibnamefont {Yan}}, \ and\ \bibinfo {author} {\bibfnamefont
  {W.}~\bibnamefont {Guo}},\ }\bibfield  {title} {\enquote {\bibinfo {title}
  {{Surface phase transitions in a $(1+1)$-dimensional $SU{(2)}_{1}$ conformal
  field theory boundary coupled to a $(2+1)$-dimensional ${\mathcal{Z}}_{2}$
  bulk}},}\ }\href {\doibase 10.1103/PhysRevB.110.115122} {\bibfield  {journal}
  {\bibinfo  {journal} {Phys. Rev. B}\ }\textbf {\bibinfo {volume} {110}},\
  \bibinfo {pages} {115122} (\bibinfo {year} {2024}{\natexlab{b}})}\BibitemShut
  {NoStop}%
\bibitem [{\citenamefont {Liu}\ \emph {et~al.}(2024)\citenamefont {Liu},
  \citenamefont {Huang}, \citenamefont {Wang}, \citenamefont {Yan},\ and\
  \citenamefont {Yao}}]{Liu2024prl}%
  \BibitemOpen
  \bibfield  {author} {\bibinfo {author} {\bibfnamefont {Z.}~\bibnamefont
  {Liu}}, \bibinfo {author} {\bibfnamefont {R.-Z.}\ \bibnamefont {Huang}},
  \bibinfo {author} {\bibfnamefont {Y.-C.}\ \bibnamefont {Wang}}, \bibinfo
  {author} {\bibfnamefont {Z.}~\bibnamefont {Yan}}, \ and\ \bibinfo {author}
  {\bibfnamefont {D.-X.}\ \bibnamefont {Yao}},\ }\bibfield  {title} {\enquote
  {\bibinfo {title} {{Measuring the Boundary Gapless State and Criticality via
  Disorder Operator}},}\ }\href {\doibase 10.1103/PhysRevLett.132.206502}
  {\bibfield  {journal} {\bibinfo  {journal} {Phys. Rev. Lett.}\ }\textbf
  {\bibinfo {volume} {132}},\ \bibinfo {pages} {206502} (\bibinfo {year}
  {2024})}\BibitemShut {NoStop}%
\bibitem [{\citenamefont {Liu}\ \emph {et~al.}(2025{\natexlab{a}})\citenamefont
  {Liu}, \citenamefont {Sato}, \citenamefont {Hou}, \citenamefont {Wang},
  \citenamefont {Guo},\ and\ \citenamefont {Assaad}}]{Liu2025arx}%
  \BibitemOpen
  \bibfield  {author} {\bibinfo {author} {\bibfnamefont {Y.}~\bibnamefont
  {Liu}}, \bibinfo {author} {\bibfnamefont {T.}~\bibnamefont {Sato}}, \bibinfo
  {author} {\bibfnamefont {D.}~\bibnamefont {Hou}}, \bibinfo {author}
  {\bibfnamefont {Z.}~\bibnamefont {Wang}}, \bibinfo {author} {\bibfnamefont
  {W.}~\bibnamefont {Guo}}, \ and\ \bibinfo {author} {\bibfnamefont {F.~F.}\
  \bibnamefont {Assaad}},\ }\bibfield  {title} {\enquote {\bibinfo {title}
  {{Edge modes of topological Mott insulators and deconfined quantum critical
  points}},}\ }\href {https://arxiv.org/abs/2508.04455} {\bibfield  {journal}
  {\bibinfo  {journal} {arXiv:2508.04455}\ } (\bibinfo {year}
  {2025}{\natexlab{a}})}\BibitemShut {NoStop}%
\bibitem [{\citenamefont {Toldin}\ \emph {et~al.}(2025)\citenamefont {Toldin},
  \citenamefont {Assaad},\ and\ \citenamefont {Metlitski}}]{Toldin2025arx}%
  \BibitemOpen
  \bibfield  {author} {\bibinfo {author} {\bibfnamefont {F.~P.}\ \bibnamefont
  {Toldin}}, \bibinfo {author} {\bibfnamefont {F.~F.}\ \bibnamefont {Assaad}},
  \ and\ \bibinfo {author} {\bibfnamefont {M.~A.}\ \bibnamefont {Metlitski}},\
  }\bibfield  {title} {\enquote {\bibinfo {title} {{Extraordinary transition at
  the edge of a correlated topological insulator}},}\ }\href
  {https://arxiv.org/abs/2508.00999} {\bibfield  {journal} {\bibinfo  {journal}
  {arXiv:2508.00999}\ } (\bibinfo {year} {2025})}\BibitemShut {NoStop}%
\bibitem [{\citenamefont {Wu}\ \emph {et~al.}(2020)\citenamefont {Wu},
  \citenamefont {Xu}, \citenamefont {Geng}, \citenamefont {Jian},\ and\
  \citenamefont {Xu}}]{Wu2020prb}%
  \BibitemOpen
  \bibfield  {author} {\bibinfo {author} {\bibfnamefont {X.-C.}\ \bibnamefont
  {Wu}}, \bibinfo {author} {\bibfnamefont {Y.}~\bibnamefont {Xu}}, \bibinfo
  {author} {\bibfnamefont {H.}~\bibnamefont {Geng}}, \bibinfo {author}
  {\bibfnamefont {C.-M.}\ \bibnamefont {Jian}}, \ and\ \bibinfo {author}
  {\bibfnamefont {C.}~\bibnamefont {Xu}},\ }\bibfield  {title} {\enquote
  {\bibinfo {title} {{Boundary criticality of topological quantum phase
  transitions in two-dimensional systems}},}\ }\href {\doibase
  10.1103/PhysRevB.101.174406} {\bibfield  {journal} {\bibinfo  {journal}
  {Phys. Rev. B}\ }\textbf {\bibinfo {volume} {101}},\ \bibinfo {pages}
  {174406} (\bibinfo {year} {2020})}\BibitemShut {NoStop}%
\bibitem [{\citenamefont {Scaffidi}\ \emph {et~al.}(2017)\citenamefont
  {Scaffidi}, \citenamefont {Parker},\ and\ \citenamefont
  {Vasseur}}]{Scaffidi2017prx}%
  \BibitemOpen
  \bibfield  {author} {\bibinfo {author} {\bibfnamefont {T.}~\bibnamefont
  {Scaffidi}}, \bibinfo {author} {\bibfnamefont {D.~E.}\ \bibnamefont
  {Parker}}, \ and\ \bibinfo {author} {\bibfnamefont {R.}~\bibnamefont
  {Vasseur}},\ }\bibfield  {title} {\enquote {\bibinfo {title} {{Gapless
  Symmetry-Protected Topological Order}},}\ }\href {\doibase
  10.1103/PhysRevX.7.041048} {\bibfield  {journal} {\bibinfo  {journal} {Phys.
  Rev. X}\ }\textbf {\bibinfo {volume} {7}},\ \bibinfo {pages} {041048}
  (\bibinfo {year} {2017})}\BibitemShut {NoStop}%
\bibitem [{\citenamefont {Ding}\ \emph {et~al.}(2018)\citenamefont {Ding},
  \citenamefont {Zhang},\ and\ \citenamefont {Guo}}]{Ding2018prl}%
  \BibitemOpen
  \bibfield  {author} {\bibinfo {author} {\bibfnamefont {C.}~\bibnamefont
  {Ding}}, \bibinfo {author} {\bibfnamefont {L.}~\bibnamefont {Zhang}}, \ and\
  \bibinfo {author} {\bibfnamefont {W.}~\bibnamefont {Guo}},\ }\bibfield
  {title} {\enquote {\bibinfo {title} {{Engineering Surface Critical Behavior
  of ($2+1$)-Dimensional O(3) Quantum Critical Points}},}\ }\href {\doibase
  10.1103/PhysRevLett.120.235701} {\bibfield  {journal} {\bibinfo  {journal}
  {Phys. Rev. Lett.}\ }\textbf {\bibinfo {volume} {120}},\ \bibinfo {pages}
  {235701} (\bibinfo {year} {2018})}\BibitemShut {NoStop}%
\bibitem [{\citenamefont {Weber}\ \emph {et~al.}(2018)\citenamefont {Weber},
  \citenamefont {Parisen~Toldin},\ and\ \citenamefont {Wessel}}]{Weber2018prb}%
  \BibitemOpen
  \bibfield  {author} {\bibinfo {author} {\bibfnamefont {L.}~\bibnamefont
  {Weber}}, \bibinfo {author} {\bibfnamefont {F.}~\bibnamefont
  {Parisen~Toldin}}, \ and\ \bibinfo {author} {\bibfnamefont {S.}~\bibnamefont
  {Wessel}},\ }\bibfield  {title} {\enquote {\bibinfo {title} {{Nonordinary
  edge criticality of two-dimensional quantum critical magnets}},}\ }\href
  {\doibase 10.1103/PhysRevB.98.140403} {\bibfield  {journal} {\bibinfo
  {journal} {Phys. Rev. B}\ }\textbf {\bibinfo {volume} {98}},\ \bibinfo
  {pages} {140403} (\bibinfo {year} {2018})}\BibitemShut {NoStop}%
\bibitem [{\citenamefont {Verresen}\ \emph {et~al.}(2021)\citenamefont
  {Verresen}, \citenamefont {Thorngren}, \citenamefont {Jones},\ and\
  \citenamefont {Pollmann}}]{Verresen2021prx}%
  \BibitemOpen
  \bibfield  {author} {\bibinfo {author} {\bibfnamefont {R.}~\bibnamefont
  {Verresen}}, \bibinfo {author} {\bibfnamefont {R.}~\bibnamefont {Thorngren}},
  \bibinfo {author} {\bibfnamefont {N.~G.}\ \bibnamefont {Jones}}, \ and\
  \bibinfo {author} {\bibfnamefont {F.}~\bibnamefont {Pollmann}},\ }\bibfield
  {title} {\enquote {\bibinfo {title} {{Gapless Topological Phases and
  Symmetry-Enriched Quantum Criticality}},}\ }\href {\doibase
  10.1103/PhysRevX.11.041059} {\bibfield  {journal} {\bibinfo  {journal} {Phys.
  Rev. X}\ }\textbf {\bibinfo {volume} {11}},\ \bibinfo {pages} {041059}
  (\bibinfo {year} {2021})}\BibitemShut {NoStop}%
\bibitem [{\citenamefont {Zhu}\ \emph {et~al.}(2021)\citenamefont {Zhu},
  \citenamefont {Ding}, \citenamefont {Zhang},\ and\ \citenamefont
  {Guo}}]{Zhu2021prb}%
  \BibitemOpen
  \bibfield  {author} {\bibinfo {author} {\bibfnamefont {W.}~\bibnamefont
  {Zhu}}, \bibinfo {author} {\bibfnamefont {C.}~\bibnamefont {Ding}}, \bibinfo
  {author} {\bibfnamefont {L.}~\bibnamefont {Zhang}}, \ and\ \bibinfo {author}
  {\bibfnamefont {W.}~\bibnamefont {Guo}},\ }\bibfield  {title} {\enquote
  {\bibinfo {title} {{Surface critical behavior of coupled Haldane chains}},}\
  }\href {\doibase 10.1103/PhysRevB.103.024412} {\bibfield  {journal} {\bibinfo
   {journal} {Phys. Rev. B}\ }\textbf {\bibinfo {volume} {103}},\ \bibinfo
  {pages} {024412} (\bibinfo {year} {2021})}\BibitemShut {NoStop}%
\bibitem [{\citenamefont {Yu}\ \emph {et~al.}(2022)\citenamefont {Yu},
  \citenamefont {Huang}, \citenamefont {Song}, \citenamefont {Xu},
  \citenamefont {Ding},\ and\ \citenamefont {Zhang}}]{Yu2022prl}%
  \BibitemOpen
  \bibfield  {author} {\bibinfo {author} {\bibfnamefont {X.-J.}\ \bibnamefont
  {Yu}}, \bibinfo {author} {\bibfnamefont {R.-Z.}\ \bibnamefont {Huang}},
  \bibinfo {author} {\bibfnamefont {H.-H.}\ \bibnamefont {Song}}, \bibinfo
  {author} {\bibfnamefont {L.}~\bibnamefont {Xu}}, \bibinfo {author}
  {\bibfnamefont {C.}~\bibnamefont {Ding}}, \ and\ \bibinfo {author}
  {\bibfnamefont {L.}~\bibnamefont {Zhang}},\ }\bibfield  {title} {\enquote
  {\bibinfo {title} {{Conformal Boundary Conditions of Symmetry-Enriched
  Quantum Critical Spin Chains}},}\ }\href {\doibase
  10.1103/PhysRevLett.129.210601} {\bibfield  {journal} {\bibinfo  {journal}
  {Phys. Rev. Lett.}\ }\textbf {\bibinfo {volume} {129}},\ \bibinfo {pages}
  {210601} (\bibinfo {year} {2022})}\BibitemShut {NoStop}%
\bibitem [{\citenamefont {Dietrich}\ and\ \citenamefont
  {Diehl}(1983)}]{Dietrich1983zpb}%
  \BibitemOpen
  \bibfield  {author} {\bibinfo {author} {\bibfnamefont {S.}~\bibnamefont
  {Dietrich}}\ and\ \bibinfo {author} {\bibfnamefont {H.~W.}\ \bibnamefont
  {Diehl}},\ }\bibfield  {title} {\enquote {\bibinfo {title} {{The effects of
  surfaces on dynamic critical behavior}},}\ }\href {\doibase
  10.1007/BF01319217} {\bibfield  {journal} {\bibinfo  {journal} {Zeitschrift
  f{\"u}r Physik B Condensed Matter}\ }\textbf {\bibinfo {volume} {51}},\
  \bibinfo {pages} {343--354} (\bibinfo {year} {1983})}\BibitemShut {NoStop}%
\bibitem [{\citenamefont {Kikuchi}\ and\ \citenamefont
  {Okabe}(1985)}]{Kikuchi1985prl}%
  \BibitemOpen
  \bibfield  {author} {\bibinfo {author} {\bibfnamefont {M.}~\bibnamefont
  {Kikuchi}}\ and\ \bibinfo {author} {\bibfnamefont {Y.}~\bibnamefont
  {Okabe}},\ }\bibfield  {title} {\enquote {\bibinfo {title} {{Monte Carlo
  Study of Critical Relaxation near a Surface}},}\ }\href {\doibase
  10.1103/PhysRevLett.55.1220} {\bibfield  {journal} {\bibinfo  {journal}
  {Phys. Rev. Lett.}\ }\textbf {\bibinfo {volume} {55}},\ \bibinfo {pages}
  {1220--1222} (\bibinfo {year} {1985})}\BibitemShut {NoStop}%
\bibitem [{\citenamefont {Diehl}(1994)}]{Diehl1994prb}%
  \BibitemOpen
  \bibfield  {author} {\bibinfo {author} {\bibfnamefont {H.~W.}\ \bibnamefont
  {Diehl}},\ }\bibfield  {title} {\enquote {\bibinfo {title} {{Universality
  classes for the dynamic surface critical behavior of systems with
  relaxational dynamics}},}\ }\href {\doibase 10.1103/PhysRevB.49.2846}
  {\bibfield  {journal} {\bibinfo  {journal} {Phys. Rev. B}\ }\textbf {\bibinfo
  {volume} {49}},\ \bibinfo {pages} {2846--2860} (\bibinfo {year}
  {1994})}\BibitemShut {NoStop}%
\bibitem [{\citenamefont {Ritschel}\ and\ \citenamefont
  {Czerner}(1995)}]{Ritschel1995prl}%
  \BibitemOpen
  \bibfield  {author} {\bibinfo {author} {\bibfnamefont {U.}~\bibnamefont
  {Ritschel}}\ and\ \bibinfo {author} {\bibfnamefont {P.}~\bibnamefont
  {Czerner}},\ }\bibfield  {title} {\enquote {\bibinfo {title} {{Universal
  Short-Time Behavior in Critical Dynamics near Surfaces}},}\ }\href {\doibase
  10.1103/PhysRevLett.75.3882} {\bibfield  {journal} {\bibinfo  {journal}
  {Phys. Rev. Lett.}\ }\textbf {\bibinfo {volume} {75}},\ \bibinfo {pages}
  {3882--3885} (\bibinfo {year} {1995})}\BibitemShut {NoStop}%
\bibitem [{\citenamefont {Pleimling}\ and\ \citenamefont
  {Igl\'oi}(2004)}]{Pleimling2004prl}%
  \BibitemOpen
  \bibfield  {author} {\bibinfo {author} {\bibfnamefont {M.}~\bibnamefont
  {Pleimling}}\ and\ \bibinfo {author} {\bibfnamefont {F.}~\bibnamefont
  {Igl\'oi}},\ }\bibfield  {title} {\enquote {\bibinfo {title} {{Nonequilibrium
  Critical Dynamics at Surfaces: Cluster Dissolution and Nonalgebraic
  Correlations}},}\ }\href {\doibase 10.1103/PhysRevLett.92.145701} {\bibfield
  {journal} {\bibinfo  {journal} {Phys. Rev. Lett.}\ }\textbf {\bibinfo
  {volume} {92}},\ \bibinfo {pages} {145701} (\bibinfo {year}
  {2004})}\BibitemShut {NoStop}%
\bibitem [{\citenamefont {Pleimling}\ and\ \citenamefont
  {Igl\'oi}(2005)}]{Pleimling2005prb}%
  \BibitemOpen
  \bibfield  {author} {\bibinfo {author} {\bibfnamefont {M.}~\bibnamefont
  {Pleimling}}\ and\ \bibinfo {author} {\bibfnamefont {F.}~\bibnamefont
  {Igl\'oi}},\ }\bibfield  {title} {\enquote {\bibinfo {title} {{Nonequilibrium
  critical dynamics in inhomogeneous systems}},}\ }\href {\doibase
  10.1103/PhysRevB.71.094424} {\bibfield  {journal} {\bibinfo  {journal} {Phys.
  Rev. B}\ }\textbf {\bibinfo {volume} {71}},\ \bibinfo {pages} {094424}
  (\bibinfo {year} {2005})}\BibitemShut {NoStop}%
\bibitem [{\citenamefont {Lin}\ and\ \citenamefont {Zheng}(2008)}]{Lin2008pre}%
  \BibitemOpen
  \bibfield  {author} {\bibinfo {author} {\bibfnamefont {S.~Z.}\ \bibnamefont
  {Lin}}\ and\ \bibinfo {author} {\bibfnamefont {B.}~\bibnamefont {Zheng}},\
  }\bibfield  {title} {\enquote {\bibinfo {title} {{Short-time critical
  dynamics at perfect and imperfect surfaces}},}\ }\href {\doibase
  10.1103/PhysRevE.78.011127} {\bibfield  {journal} {\bibinfo  {journal} {Phys.
  Rev. E}\ }\textbf {\bibinfo {volume} {78}},\ \bibinfo {pages} {011127}
  (\bibinfo {year} {2008})}\BibitemShut {NoStop}%
\bibitem [{\citenamefont {Nightingale}\ and\ \citenamefont
  {Bl\"ote}(2000)}]{Nightingale2000prb}%
  \BibitemOpen
  \bibfield  {author} {\bibinfo {author} {\bibfnamefont {M.~P.}\ \bibnamefont
  {Nightingale}}\ and\ \bibinfo {author} {\bibfnamefont {H.~W.~J.}\
  \bibnamefont {Bl\"ote}},\ }\bibfield  {title} {\enquote {\bibinfo {title}
  {{Monte Carlo computation of correlation times of independent relaxation
  modes at criticality}},}\ }\href {\doibase 10.1103/PhysRevB.62.1089}
  {\bibfield  {journal} {\bibinfo  {journal} {Phys. Rev. B}\ }\textbf {\bibinfo
  {volume} {62}},\ \bibinfo {pages} {1089--1101} (\bibinfo {year}
  {2000})}\BibitemShut {NoStop}%
\bibitem [{\citenamefont {Ferrenberg}\ \emph {et~al.}(2018)\citenamefont
  {Ferrenberg}, \citenamefont {Xu},\ and\ \citenamefont
  {Landau}}]{Ferrenberg2018pre}%
  \BibitemOpen
  \bibfield  {author} {\bibinfo {author} {\bibfnamefont {A.~M.}\ \bibnamefont
  {Ferrenberg}}, \bibinfo {author} {\bibfnamefont {J.}~\bibnamefont {Xu}}, \
  and\ \bibinfo {author} {\bibfnamefont {D.~P.}\ \bibnamefont {Landau}},\
  }\bibfield  {title} {\enquote {\bibinfo {title} {{Pushing the limits of Monte
  Carlo simulations for the three-dimensional Ising model}},}\ }\href {\doibase
  10.1103/PhysRevE.97.043301} {\bibfield  {journal} {\bibinfo  {journal} {Phys.
  Rev. E}\ }\textbf {\bibinfo {volume} {97}},\ \bibinfo {pages} {043301}
  (\bibinfo {year} {2018})}\BibitemShut {NoStop}%
\bibitem [{\citenamefont {Campostrini}\ \emph {et~al.}(2002)\citenamefont
  {Campostrini}, \citenamefont {Pelissetto}, \citenamefont {Rossi},\ and\
  \citenamefont {Vicari}}]{Campostrini2002pre}%
  \BibitemOpen
  \bibfield  {author} {\bibinfo {author} {\bibfnamefont {M.}~\bibnamefont
  {Campostrini}}, \bibinfo {author} {\bibfnamefont {A.}~\bibnamefont
  {Pelissetto}}, \bibinfo {author} {\bibfnamefont {P.}~\bibnamefont {Rossi}}, \
  and\ \bibinfo {author} {\bibfnamefont {E.}~\bibnamefont {Vicari}},\
  }\bibfield  {title} {\enquote {\bibinfo {title} {{25th-order high-temperature
  expansion results for three-dimensional Ising-like systems on the
  simple-cubic lattice}},}\ }\href {\doibase 10.1103/PhysRevE.65.066127}
  {\bibfield  {journal} {\bibinfo  {journal} {Phys. Rev. E}\ }\textbf {\bibinfo
  {volume} {65}},\ \bibinfo {pages} {066127} (\bibinfo {year}
  {2002})}\BibitemShut {NoStop}%
\bibitem [{\citenamefont {Simmons-Duffin}(2017)}]{Simmons-Duffin2017}%
  \BibitemOpen
  \bibfield  {author} {\bibinfo {author} {\bibfnamefont {D.}~\bibnamefont
  {Simmons-Duffin}},\ }\bibfield  {title} {\enquote {\bibinfo {title} {{The
  lightcone bootstrap and the spectrum of the 3d Ising CFT}},}\ }\href
  {\doibase 10.1007/JHEP03(2017)086} {\bibfield  {journal} {\bibinfo  {journal}
  {Journal of High Energy Physics}\ }\textbf {\bibinfo {volume} {2017}},\
  \bibinfo {pages} {86} (\bibinfo {year} {2017})}\BibitemShut {NoStop}%
\bibitem [{\citenamefont {Hasenbusch}(2020)}]{Hasenbusch2020pre}%
  \BibitemOpen
  \bibfield  {author} {\bibinfo {author} {\bibfnamefont {M.}~\bibnamefont
  {Hasenbusch}},\ }\bibfield  {title} {\enquote {\bibinfo {title} {{Dynamic
  critical exponent $z$ of the three-dimensional Ising universality class:
  Monte Carlo simulations of the improved Blume-Capel model}},}\ }\href
  {\doibase 10.1103/PhysRevE.101.022126} {\bibfield  {journal} {\bibinfo
  {journal} {Phys. Rev. E}\ }\textbf {\bibinfo {volume} {101}},\ \bibinfo
  {pages} {022126} (\bibinfo {year} {2020})}\BibitemShut {NoStop}%
\bibitem [{\citenamefont {McCoy}\ and\ \citenamefont
  {Wu}(1973)}]{McCoy1973book}%
  \BibitemOpen
  \bibfield  {author} {\bibinfo {author} {\bibfnamefont {B.~M.}\ \bibnamefont
  {McCoy}}\ and\ \bibinfo {author} {\bibfnamefont {T.~T.}\ \bibnamefont {Wu}},\
  }\href@noop {} {\emph {\bibinfo {title} {{The Two-Dimensional Ising
  Model}}}}\ (\bibinfo  {publisher} {Harvard University Press},\ \bibinfo
  {address} {Cambridge, MA},\ \bibinfo {year} {1973})\BibitemShut {NoStop}%
\bibitem [{\citenamefont
  {Hasenbusch}(2011{\natexlab{a}})}]{Hasenbusch2011prb_3}%
  \BibitemOpen
  \bibfield  {author} {\bibinfo {author} {\bibfnamefont {M.}~\bibnamefont
  {Hasenbusch}},\ }\bibfield  {title} {\enquote {\bibinfo {title} {{Monte Carlo
  study of surface critical phenomena: The special point}},}\ }\href {\doibase
  10.1103/PhysRevB.84.134405} {\bibfield  {journal} {\bibinfo  {journal} {Phys.
  Rev. B}\ }\textbf {\bibinfo {volume} {84}},\ \bibinfo {pages} {134405}
  (\bibinfo {year} {2011}{\natexlab{a}})}\BibitemShut {NoStop}%
\bibitem [{\citenamefont
  {Hasenbusch}(2011{\natexlab{b}})}]{Hasenbusch2011prb_2}%
  \BibitemOpen
  \bibfield  {author} {\bibinfo {author} {\bibfnamefont {M.}~\bibnamefont
  {Hasenbusch}},\ }\bibfield  {title} {\enquote {\bibinfo {title}
  {{Thermodynamic Casimir force: A Monte Carlo study of the crossover between
  the ordinary and the normal surface universality class}},}\ }\href {\doibase
  10.1103/PhysRevB.83.134425} {\bibfield  {journal} {\bibinfo  {journal} {Phys.
  Rev. B}\ }\textbf {\bibinfo {volume} {83}},\ \bibinfo {pages} {134425}
  (\bibinfo {year} {2011}{\natexlab{b}})}\BibitemShut {NoStop}%
\bibitem [{\citenamefont {Binder}\ and\ \citenamefont
  {Heermann}(2010)}]{binderbook}%
  \BibitemOpen
  \bibfield  {author} {\bibinfo {author} {\bibfnamefont {K.}~\bibnamefont
  {Binder}}\ and\ \bibinfo {author} {\bibfnamefont {D.~W.}\ \bibnamefont
  {Heermann}},\ }\href@noop {} {\emph {\bibinfo {title} {{Monte Carlo
  Simulation in Statistical Physics}}}}\ (\bibinfo  {publisher} {Springer
  Berlin, Heidelberg},\ \bibinfo {year} {2010})\BibitemShut {NoStop}%
\bibitem [{\citenamefont {Hohenberg}\ and\ \citenamefont
  {Halperin}(1977)}]{Hohenberg1977rmp}%
  \BibitemOpen
  \bibfield  {author} {\bibinfo {author} {\bibfnamefont {P.~C.}\ \bibnamefont
  {Hohenberg}}\ and\ \bibinfo {author} {\bibfnamefont {B.~I.}\ \bibnamefont
  {Halperin}},\ }\bibfield  {title} {\enquote {\bibinfo {title} {{Theory of
  dynamic critical phenomena}},}\ }\href {\doibase 10.1103/RevModPhys.49.435}
  {\bibfield  {journal} {\bibinfo  {journal} {Rev. Mod. Phys.}\ }\textbf
  {\bibinfo {volume} {49}},\ \bibinfo {pages} {435--479} (\bibinfo {year}
  {1977})}\BibitemShut {NoStop}%
\bibitem [{\citenamefont {Folk}\ and\ \citenamefont
  {Moser}(2006)}]{Folk2006jpa}%
  \BibitemOpen
  \bibfield  {author} {\bibinfo {author} {\bibfnamefont {R.}~\bibnamefont
  {Folk}}\ and\ \bibinfo {author} {\bibfnamefont {G.}~\bibnamefont {Moser}},\
  }\bibfield  {title} {\enquote {\bibinfo {title} {{Critical dynamics: a
  field-theoretical approach}},}\ }\href {\doibase 10.1088/0305-4470/39/24/R01}
  {\bibfield  {journal} {\bibinfo  {journal} {Journal of Physics A:
  Mathematical and General}\ }\textbf {\bibinfo {volume} {39}},\ \bibinfo
  {pages} {R207} (\bibinfo {year} {2006})}\BibitemShut {NoStop}%
\bibitem [{\citenamefont {T\"auber}(2014)}]{Tauber2014book}%
  \BibitemOpen
  \bibfield  {author} {\bibinfo {author} {\bibfnamefont {U.~C.}\ \bibnamefont
  {T\"auber}},\ }\href {\doibase 10.1017/CBO9781139046213} {\emph {\bibinfo
  {title} {{Critical Dynamics: A Field Theory Approach to Equilibrium and
  Non-Equilibrium Scaling Behavior}}}}\ (\bibinfo  {publisher} {Cambridge
  University Press},\ \bibinfo {year} {2014})\BibitemShut {NoStop}%
\bibitem [{\citenamefont {Chae}\ \emph {et~al.}(2012)\citenamefont {Chae},
  \citenamefont {Lee}, \citenamefont {Horibe}, \citenamefont {Tanimura},
  \citenamefont {Mori}, \citenamefont {Gao}, \citenamefont {Carr},\ and\
  \citenamefont {Cheong}}]{Chae2012prl}%
  \BibitemOpen
  \bibfield  {author} {\bibinfo {author} {\bibfnamefont {S.~C.}\ \bibnamefont
  {Chae}}, \bibinfo {author} {\bibfnamefont {N.}~\bibnamefont {Lee}}, \bibinfo
  {author} {\bibfnamefont {Y.}~\bibnamefont {Horibe}}, \bibinfo {author}
  {\bibfnamefont {M.}~\bibnamefont {Tanimura}}, \bibinfo {author}
  {\bibfnamefont {S.}~\bibnamefont {Mori}}, \bibinfo {author} {\bibfnamefont
  {B.}~\bibnamefont {Gao}}, \bibinfo {author} {\bibfnamefont {S.}~\bibnamefont
  {Carr}}, \ and\ \bibinfo {author} {\bibfnamefont {S.-W.}\ \bibnamefont
  {Cheong}},\ }\bibfield  {title} {\enquote {\bibinfo {title} {{Direct
  Observation of the Proliferation of Ferroelectric Loop Domains and
  Vortex-Antivortex Pairs}},}\ }\href {\doibase 10.1103/PhysRevLett.108.167603}
  {\bibfield  {journal} {\bibinfo  {journal} {Phys. Rev. Lett.}\ }\textbf
  {\bibinfo {volume} {108}},\ \bibinfo {pages} {167603} (\bibinfo {year}
  {2012})}\BibitemShut {NoStop}%
\bibitem [{\citenamefont {Skj\ae{}rv\o{}}\ \emph {et~al.}(2019)\citenamefont
  {Skj\ae{}rv\o{}}, \citenamefont {Meier}, \citenamefont {Feygenson},
  \citenamefont {Spaldin}, \citenamefont {Billinge}, \citenamefont {Bozin},\
  and\ \citenamefont {Selbach}}]{Skjaervo2019prx}%
  \BibitemOpen
  \bibfield  {author} {\bibinfo {author} {\bibfnamefont {S.~H.}\ \bibnamefont
  {Skj\ae{}rv\o{}}}, \bibinfo {author} {\bibfnamefont {Q.~N.}\ \bibnamefont
  {Meier}}, \bibinfo {author} {\bibfnamefont {M.}~\bibnamefont {Feygenson}},
  \bibinfo {author} {\bibfnamefont {N.~A.}\ \bibnamefont {Spaldin}}, \bibinfo
  {author} {\bibfnamefont {S.~J.~L.}\ \bibnamefont {Billinge}}, \bibinfo
  {author} {\bibfnamefont {E.~S.}\ \bibnamefont {Bozin}}, \ and\ \bibinfo
  {author} {\bibfnamefont {S.~M.}\ \bibnamefont {Selbach}},\ }\bibfield
  {title} {\enquote {\bibinfo {title} {{Unconventional Continuous Structural
  Disorder at the Order-Disorder Phase Transition in the Hexagonal
  Manganites}},}\ }\href {\doibase 10.1103/PhysRevX.9.031001} {\bibfield
  {journal} {\bibinfo  {journal} {Phys. Rev. X}\ }\textbf {\bibinfo {volume}
  {9}},\ \bibinfo {pages} {031001} (\bibinfo {year} {2019})}\BibitemShut
  {NoStop}%
\bibitem [{\citenamefont {Griffin}\ \emph {et~al.}(2012)\citenamefont
  {Griffin}, \citenamefont {Lilienblum}, \citenamefont {Delaney}, \citenamefont
  {Kumagai}, \citenamefont {Fiebig},\ and\ \citenamefont
  {Spaldin}}]{Griffin2012prx}%
  \BibitemOpen
  \bibfield  {author} {\bibinfo {author} {\bibfnamefont {S.~M.}\ \bibnamefont
  {Griffin}}, \bibinfo {author} {\bibfnamefont {M.}~\bibnamefont {Lilienblum}},
  \bibinfo {author} {\bibfnamefont {K.~T.}\ \bibnamefont {Delaney}}, \bibinfo
  {author} {\bibfnamefont {Y.}~\bibnamefont {Kumagai}}, \bibinfo {author}
  {\bibfnamefont {M.}~\bibnamefont {Fiebig}}, \ and\ \bibinfo {author}
  {\bibfnamefont {N.~A.}\ \bibnamefont {Spaldin}},\ }\bibfield  {title}
  {\enquote {\bibinfo {title} {{Scaling Behavior and Beyond Equilibrium in the
  Hexagonal Manganites}},}\ }\href {\doibase 10.1103/PhysRevX.2.041022}
  {\bibfield  {journal} {\bibinfo  {journal} {Phys. Rev. X}\ }\textbf {\bibinfo
  {volume} {2}},\ \bibinfo {pages} {041022} (\bibinfo {year}
  {2012})}\BibitemShut {NoStop}%
\bibitem [{\citenamefont {Meier}\ \emph {et~al.}(2017)\citenamefont {Meier},
  \citenamefont {Lilienblum}, \citenamefont {Griffin}, \citenamefont {Conder},
  \citenamefont {Pomjakushina}, \citenamefont {Yan}, \citenamefont {Bourret},
  \citenamefont {Meier}, \citenamefont {Lichtenberg}, \citenamefont {Salje},
  \citenamefont {Spaldin}, \citenamefont {Fiebig},\ and\ \citenamefont
  {Cano}}]{Meier2017prx}%
  \BibitemOpen
  \bibfield  {author} {\bibinfo {author} {\bibfnamefont {Q.~N.}\ \bibnamefont
  {Meier}}, \bibinfo {author} {\bibfnamefont {M.}~\bibnamefont {Lilienblum}},
  \bibinfo {author} {\bibfnamefont {S.~M.}\ \bibnamefont {Griffin}}, \bibinfo
  {author} {\bibfnamefont {K.}~\bibnamefont {Conder}}, \bibinfo {author}
  {\bibfnamefont {E.}~\bibnamefont {Pomjakushina}}, \bibinfo {author}
  {\bibfnamefont {Z.}~\bibnamefont {Yan}}, \bibinfo {author} {\bibfnamefont
  {E.}~\bibnamefont {Bourret}}, \bibinfo {author} {\bibfnamefont
  {D.}~\bibnamefont {Meier}}, \bibinfo {author} {\bibfnamefont
  {F.}~\bibnamefont {Lichtenberg}}, \bibinfo {author} {\bibfnamefont
  {E.~K.~H.}\ \bibnamefont {Salje}}, \bibinfo {author} {\bibfnamefont {N.~A.}\
  \bibnamefont {Spaldin}}, \bibinfo {author} {\bibfnamefont {M.}~\bibnamefont
  {Fiebig}}, \ and\ \bibinfo {author} {\bibfnamefont {A.}~\bibnamefont
  {Cano}},\ }\bibfield  {title} {\enquote {\bibinfo {title} {{Global Formation
  of Topological Defects in the Multiferroic Hexagonal Manganites}},}\ }\href
  {\doibase 10.1103/PhysRevX.7.041014} {\bibfield  {journal} {\bibinfo
  {journal} {Phys. Rev. X}\ }\textbf {\bibinfo {volume} {7}},\ \bibinfo {pages}
  {041014} (\bibinfo {year} {2017})}\BibitemShut {NoStop}%
\bibitem [{\citenamefont {Jian}\ \emph {et~al.}(2021)\citenamefont {Jian},
  \citenamefont {Xu}, \citenamefont {Wu},\ and\ \citenamefont
  {Xu}}]{Jian2021sp}%
  \BibitemOpen
  \bibfield  {author} {\bibinfo {author} {\bibfnamefont {C.-M.}\ \bibnamefont
  {Jian}}, \bibinfo {author} {\bibfnamefont {Y.}~\bibnamefont {Xu}}, \bibinfo
  {author} {\bibfnamefont {X.-C.}\ \bibnamefont {Wu}}, \ and\ \bibinfo {author}
  {\bibfnamefont {C.}~\bibnamefont {Xu}},\ }\bibfield  {title} {\enquote
  {\bibinfo {title} {{Continuous Néel-VBS quantum phase transition in
  non-local one-dimensional systems with SO($3$) symmetry}},}\ }\href {\doibase
  10.21468/SciPostPhys.10.2.033} {\bibfield  {journal} {\bibinfo  {journal}
  {SciPost Phys.}\ }\textbf {\bibinfo {volume} {10}},\ \bibinfo {pages} {033}
  (\bibinfo {year} {2021})}\BibitemShut {NoStop}%
\bibitem [{\citenamefont {Hasenbusch}\ and\ \citenamefont
  {Vicari}(2011)}]{Hasenbusch2011prb}%
  \BibitemOpen
  \bibfield  {author} {\bibinfo {author} {\bibfnamefont {M.}~\bibnamefont
  {Hasenbusch}}\ and\ \bibinfo {author} {\bibfnamefont {E.}~\bibnamefont
  {Vicari}},\ }\bibfield  {title} {\enquote {\bibinfo {title} {{Anisotropic
  perturbations in three-dimensional O($N$)-symmetric vector models}},}\ }\href
  {\doibase 10.1103/PhysRevB.84.125136} {\bibfield  {journal} {\bibinfo
  {journal} {Phys. Rev. B}\ }\textbf {\bibinfo {volume} {84}},\ \bibinfo
  {pages} {125136} (\bibinfo {year} {2011})}\BibitemShut {NoStop}%
\bibitem [{\citenamefont {Wang}\ \emph
  {et~al.}(2025{\natexlab{a}})\citenamefont {Wang}, \citenamefont {Jiang},\
  and\ \citenamefont {Yin}}]{Wangtl2025arx}%
  \BibitemOpen
  \bibfield  {author} {\bibinfo {author} {\bibfnamefont {T.-L.}\ \bibnamefont
  {Wang}}, \bibinfo {author} {\bibfnamefont {Y.-F.}\ \bibnamefont {Jiang}}, \
  and\ \bibinfo {author} {\bibfnamefont {S.}~\bibnamefont {Yin}},\ }\bibfield
  {title} {\enquote {\bibinfo {title} {{Driven Critical Dynamics in Tricitical
  Point}},}\ }\href {https://arxiv.org/abs/2505.12595} {\bibfield  {journal}
  {\bibinfo  {journal} {arXiv:2505.12595}\ } (\bibinfo {year}
  {2025}{\natexlab{a}})}\BibitemShut {NoStop}%
\bibitem [{\citenamefont {Wang}\ \emph
  {et~al.}(2025{\natexlab{b}})\citenamefont {Wang}, \citenamefont {Li},\ and\
  \citenamefont {Li}}]{Wanght2025arx}%
  \BibitemOpen
  \bibfield  {author} {\bibinfo {author} {\bibfnamefont {H.}~\bibnamefont
  {Wang}}, \bibinfo {author} {\bibfnamefont {X.}~\bibnamefont {Li}}, \ and\
  \bibinfo {author} {\bibfnamefont {C.}~\bibnamefont {Li}},\ }\bibfield
  {title} {\enquote {\bibinfo {title} {{Tricritical Kibble-Zurek Scaling in
  Rydberg Atom Ladders}},}\ }\href {https://arxiv.org/abs/2505.12979}
  {\bibfield  {journal} {\bibinfo  {journal} {arXiv:2505.12979}\ } (\bibinfo
  {year} {2025}{\natexlab{b}})}\BibitemShut {NoStop}%
\bibitem [{\citenamefont {Liu}\ \emph {et~al.}(2025{\natexlab{b}})\citenamefont
  {Liu}, \citenamefont {Yin},\ and\ \citenamefont {Shu}}]{Liu2025cpb}%
  \BibitemOpen
  \bibfield  {author} {\bibinfo {author} {\bibfnamefont {J.-W.}\ \bibnamefont
  {Liu}}, \bibinfo {author} {\bibfnamefont {S.}~\bibnamefont {Yin}}, \ and\
  \bibinfo {author} {\bibfnamefont {Y.-R.}\ \bibnamefont {Shu}},\ }\bibfield
  {title} {\enquote {\bibinfo {title} {{Scaling corrections in driven critical
  dynamics: Application to the two-dimensional dimerized quantum Heisenberg
  model}},}\ }\href {\doibase 10.1088/1674-1056/adc672} {\bibfield  {journal}
  {\bibinfo  {journal} {Chinese Physics B}\ }\textbf {\bibinfo {volume} {34}},\
  \bibinfo {pages} {057502} (\bibinfo {year} {2025}{\natexlab{b}})}\BibitemShut
  {NoStop}%
\bibitem [{\citenamefont {Wang}\ \emph
  {et~al.}(2025{\natexlab{c}})\citenamefont {Wang}, \citenamefont {Yu},
  \citenamefont {Sun},\ and\ \citenamefont {Zhai}}]{Wang2025arx}%
  \BibitemOpen
  \bibfield  {author} {\bibinfo {author} {\bibfnamefont {X.-Y.}\ \bibnamefont
  {Wang}}, \bibinfo {author} {\bibfnamefont {W.-J.}\ \bibnamefont {Yu}},
  \bibinfo {author} {\bibfnamefont {Y.-M.}\ \bibnamefont {Sun}}, \ and\
  \bibinfo {author} {\bibfnamefont {L.-J.}\ \bibnamefont {Zhai}},\ }\bibfield
  {title} {\enquote {\bibinfo {title} {{Driven dynamics of localization phase
  transition in the Aubry-Andr\'{e} model with initial gapless extended
  states}},}\ }\href {https://arxiv.org/abs/2509.06358} {\bibfield  {journal}
  {\bibinfo  {journal} {arXiv:2509.06358}\ } (\bibinfo {year}
  {2025}{\natexlab{c}})}\BibitemShut {NoStop}%
\bibitem [{\citenamefont {Li}\ \emph {et~al.}(2017)\citenamefont {Li},
  \citenamefont {Jiang},\ and\ \citenamefont {Yao}}]{Li2017prl}%
  \BibitemOpen
  \bibfield  {author} {\bibinfo {author} {\bibfnamefont {Z.-X.}\ \bibnamefont
  {Li}}, \bibinfo {author} {\bibfnamefont {Y.-F.}\ \bibnamefont {Jiang}}, \
  and\ \bibinfo {author} {\bibfnamefont {H.}~\bibnamefont {Yao}},\ }\bibfield
  {title} {\enquote {\bibinfo {title} {{Edge Quantum Criticality and Emergent
  Supersymmetry in Topological Phases}},}\ }\href {\doibase
  10.1103/PhysRevLett.119.107202} {\bibfield  {journal} {\bibinfo  {journal}
  {Phys. Rev. Lett.}\ }\textbf {\bibinfo {volume} {119}},\ \bibinfo {pages}
  {107202} (\bibinfo {year} {2017})}\BibitemShut {NoStop}%
\bibitem [{\citenamefont {Grover}\ \emph {et~al.}(2014)\citenamefont {Grover},
  \citenamefont {Sheng},\ and\ \citenamefont {Vishwanath}}]{Grover2014sci}%
  \BibitemOpen
  \bibfield  {author} {\bibinfo {author} {\bibfnamefont {T.}~\bibnamefont
  {Grover}}, \bibinfo {author} {\bibfnamefont {D.~N.}\ \bibnamefont {Sheng}}, \
  and\ \bibinfo {author} {\bibfnamefont {A.}~\bibnamefont {Vishwanath}},\
  }\bibfield  {title} {\enquote {\bibinfo {title} {{Emergent Space-Time
  Supersymmetry at the Boundary of a Topological Phase}},}\ }\href {\doibase
  10.1126/science.1248253} {\bibfield  {journal} {\bibinfo  {journal}
  {Science}\ }\textbf {\bibinfo {volume} {344}},\ \bibinfo {pages} {280--283}
  (\bibinfo {year} {2014})}\BibitemShut {NoStop}%
\bibitem [{\citenamefont {Ge}\ \emph {et~al.}(2025)\citenamefont {Ge},
  \citenamefont {Yao},\ and\ \citenamefont {Jian}}]{Ge2025arx}%
  \BibitemOpen
  \bibfield  {author} {\bibinfo {author} {\bibfnamefont {Y.}~\bibnamefont
  {Ge}}, \bibinfo {author} {\bibfnamefont {H.}~\bibnamefont {Yao}}, \ and\
  \bibinfo {author} {\bibfnamefont {S.-K.}\ \bibnamefont {Jian}},\ }\bibfield
  {title} {\enquote {\bibinfo {title} {{Boundary criticality in two-dimensional
  interacting topological insulators}},}\ }\href
  {https://arxiv.org/abs/2504.12600} {\bibfield  {journal} {\bibinfo  {journal}
  {arXiv:2504.12600}\ } (\bibinfo {year} {2025})}\BibitemShut {NoStop}%
\bibitem [{\citenamefont {Jiang}\ \emph {et~al.}(2025)\citenamefont {Jiang},
  \citenamefont {Ge},\ and\ \citenamefont {Jian}}]{Jiang2025arx}%
  \BibitemOpen
  \bibfield  {author} {\bibinfo {author} {\bibfnamefont {H.}~\bibnamefont
  {Jiang}}, \bibinfo {author} {\bibfnamefont {Y.}~\bibnamefont {Ge}}, \ and\
  \bibinfo {author} {\bibfnamefont {S.-K.}\ \bibnamefont {Jian}},\ }\bibfield
  {title} {\enquote {\bibinfo {title} {{Boundary criticality for the
  Gross-Neveu-Yukawa models}},}\ }\href {https://arxiv.org/abs/2503.13247}
  {\bibfield  {journal} {\bibinfo  {journal} {arXiv:2503.13247}\ } (\bibinfo
  {year} {2025})}\BibitemShut {NoStop}%
\end{thebibliography}%

\end{document}